\documentclass[aps,twocolumn,prb,preprintnumbers, 10pt]{revtex4-2}
\usepackage[left=1.9cm, right=1.9cm, top=1.9cm, bottom=1.9cm]{geometry}
\usepackage{xpatch}
\makeatletter
\patchcmd{\@ssect@ltx}
    {\addcontentsline{toc}{#1}{\protect\numberline{}#8}}
    {}
    {}
    {}
\makeatother
\pdfoutput=1
\usepackage{color}
\usepackage{enumitem}
\usepackage[dvipsnames]{xcolor}
\usepackage{hyperref}
\definecolor{amaranth}{rgb}{0.9, 0.17, 0.31}
\definecolor{forestForestGreen(web)}{rgb}{0.13, 0.55, 0.13}
\definecolor{blue(munsell)}{HTML}{005567}
\definecolor{bblue}{rgb}{0.0, 0.58, 0.71}
\hypersetup{
pdfstartview={FitH}, % fits the width of the page to the window
pdftitle={}, % title
colorlinks=true, % false: boxed links; true: colored links
linkcolor=blue(munsell), % color of internal links (change box color with linkbordercolor)
citecolor=blue(munsell), % color of links to bibliography
filecolor=magenta, % color of file links
urlcolor=blue(munsell)% color of external links
}
\usepackage{pgfplots}
\pgfplotsset{compat=1.18}
\usepackage{amssymb}
\usepackage{amsfonts}
\usepackage{graphicx}
\usepackage{epstopdf}
\usepackage{dcolumn}
\usepackage{amsmath}
\usepackage{latexsym,bm}
\usepackage{amsthm}
\usepackage{slashed}
\usepackage{float}
\usepackage{color}
\usepackage{url}
\usepackage{longtable}
\usepackage{rotating}
\usepackage[normalem]{ulem}
\usepackage{faktor}
\usepackage{tikz-cd}
\usepackage{tensor}
\usepackage{array}
\usepackage{tabularx}
\usepackage{makecell}
\usepackage{changepage}

\usepackage{tikz}
\usepackage{diagbox}
\usetikzlibrary{positioning}
\usetikzlibrary{calc}
\usetikzlibrary{decorations.pathreplacing,calligraphy}

\usepackage{xstring}
\usetikzlibrary{decorations.pathmorphing} 
\usetikzlibrary{decorations.markings} 
\usetikzlibrary{arrows} 
\usetikzlibrary{shapes} 
\usetikzlibrary{matrix} 
\usetikzlibrary{positioning} 
\usepackage[english]{babel} 
\usepackage[autostyle]{csquotes}
\usepackage{pifont}
\usetikzlibrary{shapes.multipart}

\tikzset{->-/.style={decoration={
  markings,
  mark=at position .5 with {\arrow{>}}},postaction={decorate}}}

\newcommand{\bea}{\begin{eqnarray}}
\newcommand{\eea}{\end{eqnarray}}
\newcommand{\be}{\begin{equation}}
\newcommand{\ee}{\end{equation}}
\newcommand{\ba}{\begin{aligned}}
\newcommand{\ea}{\end{aligned}}
\newcommand{\bit}{\begin{itemize}}
\newcommand{\eit}{\end{itemize}}
\newcommand{\ben}{\begin{enumerate}}
\newcommand{\een}{\end{enumerate}}
\newcommand{\nn}{\nonumber}

\newcommand{\GS}{\text{GS}}

\newcommand{\Bsym}{\mathfrak{B}^{\text{sym}}}
\newcommand{\Bphys}{\mathfrak{B}^{\text{phys}}}

\newcommand{\lb}{\left(}
\newcommand{\rb}{\right)}
\newcommand{\lbb}{\left[}
\newcommand{\rbb}{\right]}

\newcommand{\wt}{\widetilde}

\newcommand{\ot}{\otimes}

\newcommand{\Z}{{\mathbb Z}}

\newcommand{\Q}{{\mathbb Q}}

\newcommand{\cA}{\mathcal{A}}

\newcommand{\cC}{\mathcal{C}}
\newcommand{\cD}{\mathcal{D}}

\newcommand{\cM}{\mathcal{M}}

\newcommand{\cO}{\mathcal{O}}

\newcommand{\cS}{\mathcal{S}}
\newcommand{\cT}{\mathcal{T}}

\newcommand{\ms}{\mathsf}

\newcommand{\Tr}{\text{Tr}}

\renewcommand{\Vec}{\mathsf{Vec}}
\newcommand{\Rep}{\mathsf{Rep}}

\renewcommand{\dim}{\text{dim}}

\newcommand{\Ising}{\mathsf{Ising}}
\newcommand{\TY}{\mathsf{TY}}
\renewcommand{\Q}{\bm{Q}}

\makeatother

\usepackage{physics}

\newcommand{\bbI}{\mathbb I}
\newcommand{\Swap}{\text{Swap}}
\newcommand{\diag}{\text{diag}}
\newcommand{\I}{\text{I}}
\newcommand{\II}{\text{II}}
\newcommand{\III}{\text{III}}
\usepackage{float}
\newcommand{\gr}[1]{{\color{Brown}  #1}}
\usepackage{makecell}
\makeatletter
\newcommand\xlabel[2][]{\phantomsection\def\@currentlabelname{#1}\label{#2}}
\makeatother

\makeatletter
\def\l@subsubsection#1#2{}
\makeatother
\begin{document}

\title{Categorical Symmetries in Spin Models with 
Atom Arrays}

 \author{Alison Warman$^{1}$}
 \email{warman@maths.ox.ac.uk}
 \author{Fan Yang$^{2,3}$}
 \email{F.Yang@uibk.ac.at}
 \author{Apoorv Tiwari$^{4}$}
 \author{Hannes Pichler$^{2,3}$}
 \author{Sakura Sch\"afer-Nameki$^{1}$}

\affiliation{$^{1}$Mathematical Institute, University
of Oxford, Woodstock Road, Oxford, OX2 6GG, United Kingdom}
\affiliation{$^{2}$Institute for Theoretical Physics, University of Innsbruck, Innsbruck 6020, Austria}
\affiliation{$^{3}$Institute for Quantum Optics and Quantum Information of the Austrian Academy of Sciences, Innsbruck 6020, Austria}
\affiliation{$^{4}$Center for Quantum Mathematics at IMADA, Southern Denmark University,
Campusvej 55, 5230 Odense, Denmark}

%\date{\today}

\begin{abstract} 
\noindent 
Categorical symmetries have recently been shown to generalize the classification of phases of matter, significantly broadening the traditional Landau paradigm. To test these predictions, we propose a simple spin chain model that encompasses all gapped phases and second-order phase transitions governed by the categorical symmetry $\mathsf{Rep}(D_8)$. This model not only captures the essential features of non-invertible phases but is also straightforward enough to enable practical realization. Specifically, we outline an implementation using neutral atoms trapped in optical tweezer arrays. Employing a dual-species setup and Rydberg blockade, we propose a digital simulation approach that can efficiently implement the many-body evolution in several nontrivial quantum phases.

\end{abstract}

%%%%%%%%%%%%%%%%%%%%%%%%%%%

\maketitle

\noindent{\bf Introduction.}
Symmetries are fundamental in studying and constraining phases of matter, and were traditionally limited to groups. This framework underpins a wide range of theoretical physics, from quantum mechanics, condensed matter, to quantum field theory (QFT). 
However, recent developments have revealed that symmetries in any dimension are far more general and form so-called (higher) fusion categories, characterized by the key property that not every symmetry transformation has an inverse. In recent years, these categorical or non-invertible symmetries have been recognized as ubiquitous, with far-reaching implications, including a categorical extension of the Landau paradigm of phases, new constraints on Standard Model physics, and advances in formal QFT (for reviews, see \cite{Schafer-Nameki:2023jdn, Shao:2023gho}). Our aim is to establish a framework that allows direct testing of certain predictions from categorical symmetries in quantum simulators realizing spin chains.

\noindent{\bf Phases with Categorical Symmetries.}
In (1+1)d the most general (finite and internal) symmetry structure is a fusion category, whose generators $\{a,b,c,...\}$ satisfy the composition $\otimes$ rule 
$a \otimes b = \sum_c N_{ab}^c c$, where  $N_{ab}^c$ are non-negative integers,
which in general is not an invertible operation. 
Phases with these symmetries are described by a categorical extension of the Landau paradigm, and feature novel symmetry-breaking patterns and new symmetry-protected critical models. 
A systematic characterization of all gapped and gapless phases was recently proposed \cite{Bhardwaj:2023ayw, Bhardwaj:2023fca, Bhardwaj:2023idu, Bhardwaj:2023bbf, Chatterjee:2022tyg, Wen:2023otf, Huang:2023pyk, Bhardwaj:2024qrf} using the so-called Symmetry Topological Field Theory (SymTFT)  \cite{Ji:2019jhk, Gaiotto:2020iye, Apruzzi:2021nmk, Freed:2022qnc,  Moradi:2022lqp}\footnote{For gapped phases an alternative approach uses module categories in \cite{Thorngren:2019iar}.}.
This computes all symmetric gapped phases, including the order parameters, ground states, and the action of the categorical symmetry on the phases. 
The theory furthermore predicts new second-order phase transitions, which are gapless phases with categorical symmetries.

\begin{figure}
\centering
\includegraphics[width= \linewidth]{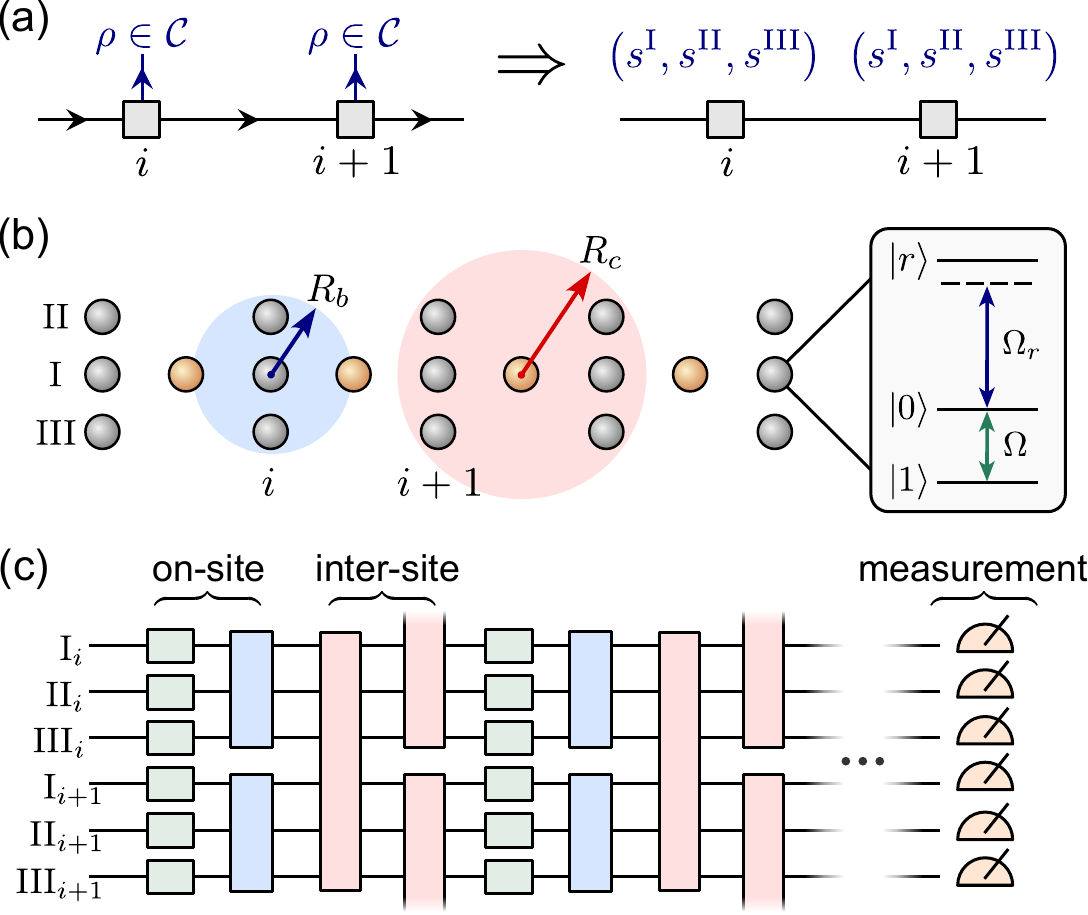}
\caption{(a) The anyon chain for a categorical symmetry (left) is used to construct lattice models, generically on a constrained Hilbert space. Here, we realize this on a spin-chain, realized on an unconstrained 3-qubit tensor product Hilbert space (right). (b) Lattice configuration and atomic level diagram used in our proposal for realizing the spin chain in (a) with categorical symmetry. The silver and the golden circles represent the data and the ancillary atomic qubits, respectively. (c) Trotterization scheme of the quantum circuit for simulating the many-body dynamics. On-site and inter-site terms can be realized by only  driving data atoms or ancillary atoms, respectively.} \label{fig:fig_scheme}
\end{figure}

\noindent{\bf Lattice Models and Cold Atom Implementation.}
This paper aims to explore these phenomena through simple spin-chains that realize categorical phases, while also  proposing an experimental protocol to probe them. This is particularly interesting for systems that may have phase transitions that are hard to study numerically, examples are precisely categorical symmetries, with e.g. the Haagerup symmetry being one of the most notorious examples (for recent progress see \cite{Huang:2021nvb, Vanhove:2021zop, Bottini:2025hri}). 
Lattice models with a categorical symmetry can be systematically constructed from the anyon chain \footnote{The input data that determines the Hilbert space is ($\cC$, $\cM$, $\rho\in\cC$), where $\cC$ is a fusion category, $\cM$ is a $\cC$-module category and $\cS= \cC^*_\cM$ is the actual symmetry category.}, which  generically results in a constrained Hilbert space \cite{Feiguin:2006ydp, Aasen:2016dop, Aasen:2020jwb, Cong:2017ffh,Buican:2017rxc, Inamura:2021wuo, Lootens:2021tet,  Bhardwaj:2024kvy}. This approach allows a systematic construction of commuting projector Hamiltonians for all gapped phases and second-order phase transitions of a categorical symmetry \cite{Bhardwaj:2024kvy}.

In this Letter, we present a benchmark spin-chain that demonstrates all the characteristic features of phases with non-invertible symmetries. Inspired by this general approach, the model has the distinctive advantage of being realized in an unconstrained tensor product Hilbert space based on qubits, as illustrated in Fig.~\ref{fig:fig_scheme}(a). Its Hamiltonians take the schematic form
\begin{equation}\label{genHam}
H =  \sum_i V_{i,i+1} +  \sum_i P_i\,, \vspace{-2.5mm}
\end{equation}
where $P_i$ denotes a projector acting on a few spins at a given site $i$, while $V_{i,i+1}$ is the interaction between neighboring sites $i$ and $i+1$. This system not only captures all the key non-invertible features but is also well-suited for experimental implementation.
For the latter, we focus on implementations with Rydberg-atom arrays, which have emerged as promising platforms for quantum simulation of spin models \cite{browaeys2020many,wu2020concise,kaufman2021quantum,morgado2021quantum}. These systems have already been used to experimentally probe quantum phase transitions breaking conventional discrete \cite{bernien2017probing,keeslingQuantumKibbleZurek2019b,scholl2021quantum,ebadi2021quantum} and continuous \cite{chen2023continuous} group symmetries, as well as topologically ordered phases \cite{Semeghini2021}. Here we discuss how these systems can be used to experimentally probe phases with non-invertible symmetries and the transitions between them. Specifically, our proposal is based on a digital evolution that can be induced by two driving lasers in a dual species Rydberg atom array \cite{singh2023mid, anand2024dual}, as shown in Fig.~\ref{fig:fig_scheme}(b).

\smallskip

\noindent{\bf Benchmark Model: $\Rep (D_8)$ Categorical Symmetry.}
The benchmark symmetry is $\Rep (D_8)$, i.e. the representations of the finite group $D_8 := \lb\Z_2^{a}\times \Z_2^{b}\rb \rtimes \Z_2^{c}$ of symmetries of the square (rotations and reflections). The generators are the irreducible  representations (irreps) \vspace{-2mm}
\be \label{eq:D8_irreps}
   \text{Generators of $\Rep (D_8)$}:\;\;  1,\;1_c,\;1_a,\;1_{ca},\;E \,, \vspace{-2mm}
\ee
with $\dim (1_k)=1$, $\dim (E)=2$, and their composition is the tensor product of irreps with $k=a, c, ca$:\vspace{-1mm}
\be\label{fusions}
\ba
    1_a\otimes1_c&=1_c\otimes1_a=1_{ca}\,,\qquad  1_k \otimes 1_k =1\,,\qquad     \\
    E\otimes 1_k&=1_k\otimes E=E\,,  \\
    E\otimes E&=1\oplus 1_a\oplus 1_c \oplus 1_{ca}\,,\vspace{-2.5mm}
\ea
\ee
the last line being a non-invertible composition rule. 
Using recent advancements in the representation theory of categorical symmetries, we can now demonstrate the full scope of the categorical Landau paradigm in action. This includes a comprehensive characterization of gapped phases and phase transitions \cite{Bhardwaj:2024qrf}. 
Lattice models for a subset of these phases have recently been realized using a complementary approach based on the cluster state \cite{Seifnashri:2024dsd, Li:2024gwx, Meng:2024nxx}. In this work, we show that all phases can be captured within a straightforward tensor product Hilbert space constructed from qubits.

The benchmark model provides a: 
(1) Derivation of all gapped and gapless phases governed by $\Rep(D_8)$, exhibiting the key features of categorical symmetries; 
(2) Realization in simple spin models, with few-body interactions; 
(3) Realistic near future implementation of the spin model in Rydberg atom arrays. 
{Our work furthermore opens up avenues for exploring dynamical questions, such as those related to quantum quenches, for which accurate and efficient analytic or numerical tools are currently lacking. }

\noindent{\bf Gapped Phases for $\Rep (D_8)$.} 
There are 11 distinct gapped phases for $\Rep (D_8)$, which can be labeled by 
 $(F,\beta)$, where $F$ is a representative of a conjugacy class of subgroups of $D_8$ and a cocycle $\beta\in H^2(F,U(1))$ \cite{Thorngren:2019iar,Bhardwaj:2024qrf}.
 
On general grounds \cite{Bhardwaj:2024kvy}, the gapped phases for this symmetry can be realized on a lattice model, where the states are labeled by group elements $\ms{g}, \ms{h} \in D_8$ and operators acting as 
\be\ba 
    L^{\ms g}\ket{\ms h} &=\ket{\ms g\ms h}\,,\quad R^{\ms g}\ket{\ms h}=\ket{\ms h\ms g}\,. \\     
\ea\ee
The commuting projector Hamiltonians for the gapped phases take the general form (\ref{genHam}), and are given by \footnote{For the expert reader,
%in the terminology of \cite{Bhardwaj:2024kvy}, 
we choose the categorical input data: $\cC=\Vec_{D_8},$ $\cM=\Vec,$ $\rho=\oplus_{g\in D_8} g,\,$
so that the symmetry of the model is $\cS=\cC^*_{\cM}=\Rep(D_8)$.}
\begin{equation} \label{eq:RepD8_Ham_main}
   H_{(F,\beta)}=-\tfrac{1}{|F|}\sum_i\sum_{\ms f\in F}(R_\beta^{\ms f^{-1}})_{i}(L_\beta^{\ms f})_{i+1} -\sum_i P^{(F)}_{i}\,.
\end{equation}
Here $ P^{(F)}$ is a projector onto the $F$ subgroup of $D_8$ \footnote{$H_{(F,\beta)}$ is Hermitian as the projectors are and  $(L_\beta^{\ms f})^\dagger=L_\beta^{\ms f^{-1}},\;(R_\beta^{\ms f})^\dagger=R_\beta^{\ms f^{-1}}$  and we sum over $f\in F$ where $F$ is a group.  One can write equivalent Hamiltonians by summing only over generators of $F$ and their inverses.}. %: since any element in $F$ can be written as a product of such generators, a state that minimizes the simplified Hamiltonian will also minimize the full one and vice-versa.}
For each $\Rep(D_8)$ irrep $\Gamma$, and $l, m \in\{1\,,\dots\,, \rm dim(\Gamma)\}$, we introduce a (diagonal) operator $Z_\Gamma^{l,m}$  that acts as \footnote{The operators $L_g,R_g,Z_\Gamma$ for dihedral groups presented as $\Z_N\rtimes\Z_2$ can be found in appendix B of \cite{Albert:2021vts}.}
\be
    Z_{\Gamma}^{l,m}|\ms g\rangle = \mathcal D_{\Gamma}^{l,m}(\ms g)|\ms g\rangle\,,
\ee
where $\cD_{\Gamma}(\ms g)$ represents $\ms g$ in the irrep $\Gamma$. 

The simplification, which is key to the potential realization in future experiments, is the identification of the state space with that of three qubits
\begin{equation}\label{ThreeQB}
    |s^{\I}\,s^{\II}\, s^{\III}\rangle\,, \qquad s^{\alpha}\in\{0,1\}\,,\vspace{-1.5mm}
\end{equation}
via the mapping:
\begin{equation}
\label{stollen}
\begin{alignedat}{3}
|1\rangle &\mapsto |000\rangle\,, \quad 
&|a\rangle &\mapsto |010\rangle\,, \quad 
&
|b\rangle &\mapsto |001\rangle\,, \\
  |c\rangle &\mapsto |100\rangle\,, \quad 
&|ca\rangle &\mapsto |101\rangle\,,
&|cb\rangle &\mapsto |110\rangle\,, \\ 
|ab\rangle &\mapsto |011\rangle\,, \quad  &|cab\rangle &\mapsto |111\rangle\,.
\end{alignedat}
\end{equation}
%we can map each lattice site to a three-qubit system 
The Hamiltonians  for each phase can thus be written in terms of Pauli operators $X_i^{\alpha}$ and $Z_i^{\alpha}$ acting on 
 qubit $\alpha= \I, \II, \III$  at site $i$:
\begin{align}  \label{eq:Loperators}
L^a&= X^{\II}\,, &R^{a}&=Q^{\I,+} X^{\II}+
Q^{\I,-} X^{\III}\,, \nn\\
L^b&= X^{\III}\,, &    R^{b}&=Q^{\I,-} X^{\II}+
    Q^{\I,+} X^{\III}\,, \\ 
L^c&= X^{\I}\,(\Swap)^{\II,\III} \,, &R^c&= X^{\I} \,,\nn
\end{align}
where  $Q^{\alpha,\pm} = \tfrac{1}{2}(1\pm Z^{\alpha})$ and $(\Swap)^{\II,\III}$ swaps $\II$ and $\III$.
The symmetry operators $\cS_{\Gamma}$ for each irrep $\Gamma$ are
\begin{align}
\cS_{1_a}&=\prod_{i} Z^{\I}_{i}\,,\;\;\; \cS_{1_c}=\prod_{i}  Z^{\II}_i Z^{\III}_{i}\,,\;\;\;\cS_{1_{ca}}=\prod_{i} Z^{\I}_{i} Z^{\II}_i Z^{\III}_{i}\,, \nn\\
    \cS_E&=Z_{E,\text{prod}}^{1,1}+Z_{E,\text{prod}}^{2,2}\,, 
    \hspace{-10pt}\label{eq:S_Gamma_main}
\end{align}
where 
for the two dimensional irrep $E$, with matrix representation $\cD_E(a)=Y,\,\cD_E(b)=-Y,\,\cD_E(c)=X$, 
$Z_{E,\text{prod}}=\prod_i(Z_E)_i$ 
is the product of the matrix
\be \label{eq:ZGammaE}
\ba
\vspace{-3mm}
Z_E^{1,1} & = Q^{1_{ca},+}(S^{\I})^\dagger  Z^{\III}\,, \\
Z_E^{1,2} & = Q^{1_{ca},-}\,(S^{\II})^{\dagger} S^{\III}(CZ)^{\II,\III}\,, \\
Z_E^{2,1} & =Q^{1_{ca},-}\, S^{\II} (S^{\III})^{\dagger}(CZ)^{\II,\III}\,, \\  
Z_E^{2,2} & = Q^{1_{ca},+}\,S^{\I}  Z^{\III} \,,
\ea
\ee
on all the lattice sites, with $Q^{1_{ca},\pm}=\tfrac{1}{2}(\bbI\pm Z^{\I} Z^{\II} Z^{\III})$,\\
$S={\rm diag}(1\,,i)$ and $CZ=\diag(1,1,1,-1)$\,.

The symmetry operators $\cS_{\Gamma}$ act on states as 
$\cS_{\Gamma}\ket{\Psi_{\ms g}}=\Tr(\cD_\Gamma(\ms g))\ket{\Psi_{\ms g}}$,
where ${\ms g}$ is the `holonomy' of the state, i.e. the product of all the group elements on the lattice sites. 
The diagonal operators $\cS_\Gamma$ commute with $ H_{(F,\beta)}$ since $ P^{(F)}$ is also diagonal and since $\cD_\Gamma(\ms f^{-1})\cD_\Gamma(\ms f)=\bbI$, hence the Hamiltonians \eqref{eq:RepD8_Ham_main} are $\Rep(D_8)$-symmetric. Their ground states are linear combinations of $\ket{\Psi_{[f]}}$ for $f\in F$, from which we can compute the $\Rep(D_8)$ symmetry action, as detailed in the supplemental material \cite{sm}.

The gapped phases, with their Hamiltonians, are summarized in the End Matter, table \ref{tab:phases}. They are either symmetry protected topological (SPT) or spontaneous symmetry breaking (SSB) phases for particular sub-symmetries of $\Rep (D_8)$. For the first five phases, one can first minimize the projector $-(\bbI+Z^\I)_i/2$ by setting $s^\I=0$ on all sites. On this subspace, $\cS_E$ simplifies to $\cS_E\simeq \prod_iZ^\II_i+\prod_iZ^\III_i$, and the Hamiltonians reduce to the black terms in table \ref{tab:phases}, involving up to four-body interactions between sites $\{\II_i,\III_i,\II_{i+1},\III_{i+1}\}$. As we will show below, this simplification can reduce the complexity of our experimental scheme.

%\vspace*{1mm}

%\clearpage
%\newpage

\noindent{\bf Phase Transitions.} 
One of the most recent advances is the systematic study of second-order phase transitions between two gapped phases with categorical symmetries, facilitated by the SymTFT framework \cite{Chatterjee:2022tyg, Bhardwaj:2024kvy,Bhardwaj:2024qrf, Wen:2023otf}. In simple spin-chain models, we can furthermore analyze the interpolation between two gapped phases with Hamiltonians $H_{(l)}$ or $H_{(m)}$ respectively, 
\begin{equation}
    H_{(l,m)}(\lambda)=\lambda H_{(l)}+(1-\lambda) H_{(m)} \,,\label{interpolation}
\end{equation}
where for some value between 0 and 1 we have the gapless Hamiltonian for the critical point describing the second-order phase transition between the two gapped phases. 
From the general continuum analysis in \cite{Bhardwaj:2024qrf} one can determine furthermore the CFTs that describe these transitions, which we corroborate from by a spin-chain analysis in the supplemental material \cite{sm}, where we show that the transitions can be realized within this simple framework. 

%which we corroborate from by a spin-chain analysis in appendix \ref{app:gapless}, where we show that the transitions can be realized within this simple framework. 

\noindent{\bf Hardware-efficient Simulation Scheme.} 
We now turn to a potential implementation of these spin models with Rydberg atom arrays. While these systems often realize spin models in an analog way \cite{scholl2021quantum,ebadi2021quantum,zeiher2017coherent,scholl2023erasure,steinert2023spatially,kim2024realization}, here we propose a digital approach \cite{levine2019parallel,ma2023high,cao2024multi,maskara2023programmable} that is versatile enough to accommodate the first five phases in Tab.~\ref{tab:phases}, and potentially can be extended to the remaining ones. 
%While arbitrary unitary operations can be realized in principle with the unitary gate set available in Rydberg atom arrays \cite{levine2019parallel,evered2023high,bluvstein2024logical}, we present a hardware-efficient, highly parallel scheme for implementing the above constructed spin models.
%In Rydberg-atom arrays, high-fidelity single-qubit rotations and two-qubit entangling gates have been achieved \cite{levine2019parallel,evered2023high,bluvstein2024logical}. While arbitrary unitary operations can be built from such a universal gate set, the required circuit depth can be prohibitively large and impractical. Here, we instead present a hardware-efficient, highly parallel scheme for implementing the above constructed spin model.

% \begin{figure}
% \centering
% \includegraphics[width=\linewidth]{Figurescheme.pdf}% Here is how to import EPS art
% \caption{(a) Lattice configuration and atomic level diagram for the scheme, where the silver and the golden circles represent the data and the ancillary atomic qubits, respectively. (b) Trotterization scheme of the quantum circuit for simulating the many-body dynamics.} \label{fig:fig_scheme}
% \end{figure} 

The configuration of the atom array is illustrated in Fig.~\ref{fig:fig_scheme}(b): for each lattice site $i$, a group of three atoms is aligned perpendicular to the direction of the chain, representing spins $\{\mathrm{I}_i, \mathrm{II}_i, \mathrm{III}_i\}$ as in Eq.~\eqref{ThreeQB}. These atomic data qubits are encoded by two long-living hyperfine ground states $|1\rangle$ (spin-down) and $|0\rangle$ (spin-up). The latter state can be  further coupled  to a Rydberg state $|r\rangle$ for inducing interactions between qubits via the Rydberg blockade mechanism \cite{jaksch2000fast}. We assume that certain subsets of the atoms on each site can be addressed by the Rydberg laser, e.g., by shifting the other atoms out of resonance using a few distinct tweezer patterns \cite{omranGenerationManipulationSchrodinger2019c}.  
In addition, an ancillary atom of a different species is placed between two neighboring sites $i$ and $i+1$ (see Fig.~\ref{fig:fig_scheme}(b)). These ancillary atoms have an analogous level scheme as our data atoms, but can be driven independently from them.
Moreover, this dual-species setup supports an asymmetric Rydberg blockade, in which the intra- and the inter-species blockade radius ($R_b$ and $R_c$) can be independently tuned, e.g., via F{\"o}rster resonance \cite{anand2024dual}. 
Here, we choose $R_c>R_b$ such that each data qubit can only be blockaded by its on-site nearest neighbors ($\mathrm{I}_i$ and $\mathrm{II}_i$; $\mathrm{I}_i$ and $\mathrm{III}_i$) or two closest ancillary qubits [see Fig.~\ref{fig:fig_scheme}(b)]. As we show below, such a dual-species configuration facilitates both parallel and efficient simulations of the desired multi-body interactions. %Insertion of ancillary qubits also benefits mid-circuit measurements and possible error mitigations \cite{singh2023mid}.

\begin{figure}
	\centering
	\includegraphics[width=\linewidth]{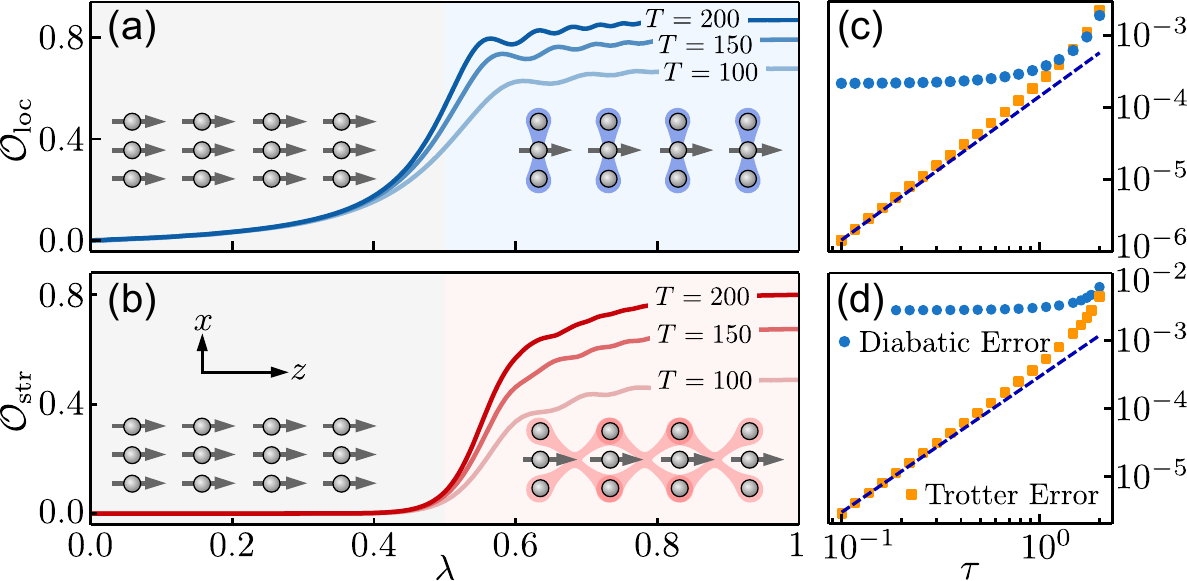}
	\caption{(a) and (b) show the evolution of the local and string order parameters $\mathcal{O}_\mathrm{loc}$ and $\mathcal{O}_\mathrm{str}$ for the annealing across the phase transition $\mathrm{Trivial}\rightarrow \Rep(D_8)/(\Z_2\times\Z_2)$ SSB and $\mathrm{Trivial}\rightarrow\mathrm{SPT}$, respectively. The results are obtained by time-evolving block decimation (TEBD) with $L=18$ sites ($54$ data qubits) and $\tau=1$. (c) and (d) show the errors in (a) and (b) as a function of the step size $\tau$ with $L=6$. The dashed curves are polynomial fits ($\propto\tau^2$) to the Trotter errors. In (a), symmetry-breaking defects are included at the boundaries to select one of the degenerate ground states.} \label{fig:fig_tebd}
\end{figure}

Simulation of the quantum evolution is then realized by a Trotterized sequence $e^{-\mathrm{i}H\tau}\approx \prod_pe^{-\mathrm{i}H_p\tau} + \mathcal{O}(\tau^2)$, where $H=\sum_p H_p$ and each $H_p$ is a collection of mutually commuting terms that can be executed in a parallel and system-size independent manner. With the physical system sketched above, arbitrary combinations of the simplified Hamiltonians $\{H_1,H_{\Z_2^{a}},H_{\Z_2^{ab}},H_{(\Z_2^a\times\Z_2^b)^\pm}\}$ in Table~\ref{tab:phases} (black terms) can be efficiently simulated. Expressing such a combination in the generic form Eq.~\eqref{genHam}, first, the on-site evolutions $e^{-\mathrm{i}P_i\tau}$ involving up to three-body interactions can be realized by three types of phase gates in few steps: single-qubit phase gate $U_\mathrm{P}(\phi)=e^{-\mathrm{i}\phi Q^\mathrm{I}_i}$, two-qubit controlled phase gate $U_\mathrm{CP}(\phi)=e^{-\mathrm{i}\phi Q^\mathrm{I}_i Q^\alpha_i}$ ($\alpha=\mathrm{II,III}$), and three-qubit controlled phase gate $U_\mathrm{CCP}(\phi)=e^{-\mathrm{i}\phi Q^\mathrm{I}_iQ^\mathrm{II}_iQ^\mathrm{III}_i}$, where $Q=(\mathbb{I}+Z)/2=|0\rangle\langle 0|$. As data qubits sitting on different sites are outside the blockade radius, the operation $e^{-\mathrm{i}\sum_i P_i\tau}$ can be parallelized \cite{evered2023high,bluvstein2024logical}. Second, the two-site evolution $e^{-\mathrm{i}V_{i,i+1}\tau}$ containing up to four-body plaquette interactions $U_\square(\phi)=\exp(-\mathrm{i}\phi \mathcal{O}_i^\mathrm{II}\mathcal{O}_{i+1}^\mathrm{II}\mathcal{O}_{i+1}^\mathrm{III}\mathcal{O}_{i}^\mathrm{III})$ with  $\mathcal{O}_j^\alpha\in\{\mathbb{I},X,Y,Z\}_j^\alpha$ can be constructed from a multi-qubit-$Z$ gate controlled by an ancillary qubit \cite{weimer2010rydberg,weimer2011digital,bohrdt2020multiparticle}. In such a construction, the ancillary qubit is only temporarily entangled with data qubits and can be traced out from the dynamics after completion of the gate operation {(see supplemental material \cite{sm})}.
%(see Appendix~\ref{app:appB}). 
Since each data qubit is only affected by its two closest ancillary qubits, this operation can be executed in just two steps $e^{-\mathrm{i}\sum_iV_{2i-1,2i}\tau}$ and $e^{-\mathrm{i}\sum_iV_{2i,2i+1}\tau}$.  The stroboscopic application of the on-site and the inter-site evolution then composes a Trotterization, realizing the desired many-body evolution [see Fig.~\ref{fig:fig_scheme}(c)]. %The detailed pulse sequences for the elementary gates are provided in {the supplemental material \cite{sm}}. %Appendix~\ref{app:appB}.

To show the performance of our scheme, we study a quantum annealing dynamics, in which the Hamiltonian is adiabatically varied in the form of Eq.~\eqref{interpolation} with a time-dependent $\lambda=t/T$. {The annealing time $T\sim 1/\Delta E\propto L$ (where $\Delta E$ is the energy gap and $L$ the number of lattice sites) scales linearly with the system size \cite{sm}, as expected for (1+1)d CFTs \cite{CARDY1986200,oshikawa2019universalfinitesizegapscaling}}. Here, we consider two interesting interpolations: $H_1$ to $H_{\Z_2^{ab}}$ and $H_1$ to $H_{(\Z_2^a\times\Z_2^b)^-}$. 
The former establishes the transition from a trivial paramagnetic-like phase with spins on different sites independently aligned, to a symmetry breaking phase where spins at $\{\mathrm{II}_i,\mathrm{III}_i\}$ form a Bell pair and are ordered in a correlated manner [see Fig.~\ref{fig:fig_tebd}(a)]. In this phase, the non-invertible symmetry $\mathcal{S}_E$ is spontaneously broken, while all group-like symmetries ($\cS_{1_a},\cS_{1_c},\cS_{1_{ca}}$) are preserved. The second interpolation realizes a transition from the trivial phase to an SPT phase featuring a cluster-state-like ground state [see Fig.~\ref{fig:fig_tebd}(b)]. When preparing the ground state with a digital annealing, there are two types of errors: the diabatic error induced by leakages from the ground-state manifold and the Trotter error caused by a finite step size $\tau$. As shown in Figs.~\ref{fig:fig_tebd}(c) and \ref{fig:fig_tebd}(d), the Trotter errors are small even for a very large Trotter step $\tau\sim 1$. To experimentally probe these transitions, one can measure local \cite{bernien2017probing} and string \cite{Semeghini2021,de2019observation} order parameters $\mathcal{O}_\mathrm{loc}$ and $\mathcal{O}_\mathrm{str}$ associated with each nontrivial phase {\footnote{The local and string order parameters here are given by $\mathcal{O}_\mathrm{loc}=\Sigma_i\langle{X^\mathrm{II}X^\mathrm{III}\rangle}_i/L$ and $\mathcal{O}_\mathrm{str}=\langle X^{\II}_{i} X^{\III}_{i+1} [\Pi_{k=1}^{l-2} (Z^{\II} Z^{\III})_{i+k}] X^{\II}_{i+l-2} X^{\III}_{i+l-1}\rangle$. In the TEBD simulation, we consider the string at the center of the system with a length $l=6$}} [see Figs.~\ref{fig:fig_tebd}(a) and \ref{fig:fig_tebd}(b)]. 
We note that by increasing the annealing time $T$ ($\propto$ total number of gates), the diabatic error can be reduced, and the order parameters exhibit sharper transitions.

\begin{figure}
	\centering
	\includegraphics[width=\linewidth]{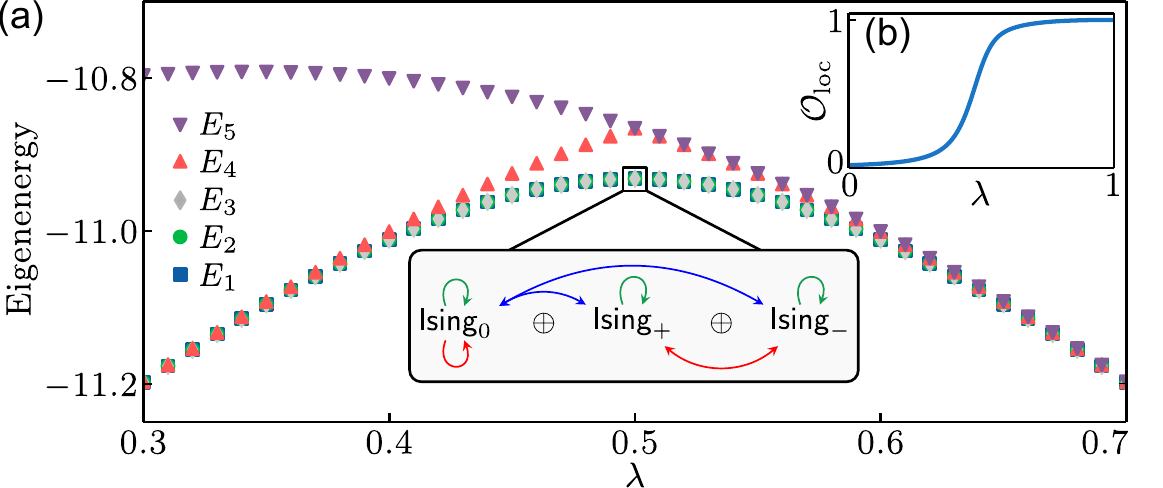}% Here is how to import EPS art
	\caption{Characterization of the quantum phase transition $(\Z_2\times \Z_2) \textrm{ SSB } \rightarrow\Rep(D_8) \textrm{ SSB}$. (a) and (b) show the spectrum (lowest five eigenenergies) and the order parameter $\mathcal{O}_\mathrm{loc}=\sum_iX_i^\mathrm{I}/L$ as a function of $\lambda$. The results are obtained by exact diagonalization with $L=6$ sites ($18$ qubits). In (b), a symmetry-breaking perturbation $-0.01\sum_i X_i^\mathrm{I}$ is applied.} \label{fig:fig_igssb}
\end{figure}

\noindent{\bf Generalization.} 
While the current configuration supports the exploration of the first five regimes in Table~\ref{tab:phases}, which already include an example of each type of phase except the $\Rep(D_8)$ SSB, a simple generalization can reach all possible phases emerging from the $\Rep(D_8)$ symmetry, 
{specifically the $\Rep(D_8)$ SSB. This is a gapped phase where the $\Rep(D_8)$ categorical symmetry is completely broken, and whose ground state degeneracy cannot be explained by any (UV or IR) group-like symmetry. The phase transition between a $\Z_2\times\Z_2$ SSB, e.g. $H_{(\Z_2^a\times\Z_2^b)^+}$, to the $\Rep(D_8)$ SSB $H_{(D_8)^+}$ is particularly noteworthy since it is an $\Ising$ intrinsically gapless SSB (igSSB) \cite{Bhardwaj:2024qrf}: it has 3 gapless ground states (also called ``universes''), each of which is the $\Ising$ CFT, that are related to each other by a spontaneously broken $\Ising$ categorical sub-symmetry of $\Rep(D_8)$ \footnote{There is $\sqrt{2}$ `relative Euler term' between $\Ising_0$ and each of $\Ising_\pm$, since $E$ has quantum dimension 2 whereas the $\Ising$ non-invertible symmetry has quantum dimension $\sqrt{2}$.} (see Fig. \ref{fig:fig_igssb}). It is ``intrinsically'' gapless since no gapped phases exhibit this symmetry-breaking pattern.}

To achieve the simulation for this general case, one can include a different Rydberg state $\ket{r^\prime}$ when implementing the two-site evolution. This Rydberg state should support a highly asymmetric blockade, in which data qubits themselves are completely non-interacting (even the nearest-neighbors), but can be blockaded by the neighboring ancillary atom, described by a constrained Hamiltonian $(\Omega_{r^\prime}/2)(\mathbb{I}_\mathrm{c}-\ketbra{r^\prime}_\mathrm{c})\otimes(\Sigma_n \ketbra{r^\prime}{0}_n+\mathrm{H.c.})$, where every data qubit ($n=1,\cdots,6$) in a given plaquette can be excited to the Rydberg state $\ket{r^\prime}_n$ if the ancillary qubit is not in the Rydberg state $\ket{r^\prime}_\mathrm{c}$. Then, using the same gate sequence \cite{sm} for the two-site evolution as before, arbitrary six-body interactions $U_\square^\prime(\phi)=\exp(-\mathrm{i}\phi \mathcal{O}_i^\mathrm{I}\mathcal{O}_{i+1}^\mathrm{I}\mathcal{O}_i^\mathrm{II}\mathcal{O}_{i+1}^\mathrm{II}\mathcal{O}_{i}^\mathrm{III}\mathcal{O}_{i+1}^\mathrm{III})$ with  $\mathcal{O}_j^\alpha\in\{\mathbb{I},X,Y,Z\}_j^\alpha$ within each plaquette can be engineered. Such a highly asymmetric regime can be achieved by utilizing the anisotropy of the Rydberg interaction \cite{PhysRevLett.112.183002} or microwave dressing \cite{young2021asymmetric}.

\noindent{\bf Conclusion and Outlook.} 
In summary, we have constructed a spin-chain model realizing all gapped phases and second-order phase transitions for the $\Rep(D_8)$ categorical symmetry, and provided a concrete proposal for implementing the model in a Rydberg quantum simulator. Based on the current scheme, it is possible to increase the circuit depth, as both the execution time and the robustness of each elementary gate can be improved by optimal control methods \cite{jandura2022time,fromonteil2024hamilton}. In addition to the dual-species scheme, one may also use a linear-chain configuration and coherently rearrange atomic qubits \cite{bluvstein2022quantum} to establish the multi-body interaction, e.g., using the globally driving scheme to realize the four-body interaction \cite{dlaska2022quantum}. While we focus on a digital implementation here, developing analog simulation schemes represents an interesting route as well.

The physical implementation presented here serves as a versatile framework for simulating a wide range of categorical symmetries, such as 
 $\Rep (S_3)$ non-invertible symmetries \cite{Eck:2023fqo, Bhardwaj:2024wlr, Chatterjee:2024ych}, and spin models in higher dimensions. Indeed, for higher categorical symmetries \cite{Kaidi:2021xfk, Choi:2021kmx, Bhardwaj:2022yxj} in  (2+1)d and (3+1)d, lattice models have been developed recently in \cite{Delcamp:2023kew, Inamura:2023qzl, Eck:2024myo, Gorantla:2024ocs, Inamura:2025cum}, giving rise to novel phases of matter e.g. \cite{Antinucci:2024ltv, Bhardwaj:2024qiv, Wen:2024qsg, Choi:2024rjm, Bhardwaj:2025piv, Bhardwaj:2025jtf}. Extending the framework discussed here to higher dimensions would naturally align with these developments and presents an exciting challenge for experimental realization in neutral-atom systems as well as other quantum simulation platforms, such as superconducting circuits \cite{huang2020superconducting} and trapped ions \cite{monroe2021programmable}.

\onecolumngrid
\vspace{-5mm}

\noindent
\begin{acknowledgments}
\vspace{-3mm}
    {We thank Lea Bottini, Luisa Eck, Paul Fendley, Yuhan Gai, Camillo Imbimbo, Dan Pajer, Zhongda Zeng, Klaus M{\o}lmer, and Torsten V. Zache for discussions at various stages of this work and  Lakshya Bhardwaj for initial collaboration. 
The work of S.S.N. and A.W. is supported by the UKRI Frontier Research Grant, underwriting the ERC Advanced Grant ``Generalized Symmetries in Quantum Field Theory and Quantum Gravity''.
A.T. is funded by Villum Fonden Grant no. VIL60714. Work in Innsbruck is supported by
the European Union’s Horizon Europe research and innovation program under Grant Agreement No. 101113690
(PASQuanS2.1), the ERC Starting grant QARA (Grant
No. 101041435), and the Austrian Science Fund (FWF) (Grant
No. DOI 10.55776/COE1).}
\end{acknowledgments}

The data that support the findings of this article are openly available \cite{data}.

\vspace{-4mm}

\twocolumngrid

\let\oldaddcontentsline\addcontentsline% Store \addcontentsline
\renewcommand{\addcontentsline}[3]{}% Make \addcontentsline a no-op
 \bibliographystyle{ytphys}
 \vspace{-10mm}
 \small 
 %\baselineskip=.9\baselineskip
 \let\bbb\bibitem\def\bibitem{\itemsep1.6pt\bbb}
\bibliography{ref.bib}
\let\addcontentsline\oldaddcontentsline% Restore \addcontentsline

\normalsize
\vspace{-6mm}
\section*{End Matter:
Summary Table of $\Rep(D_8)$-symmetric Gapped Phases}
\vspace{-4mm}
%Table \ref{tab:phases} summarizes all $\Rep(D_8)$-symmetric gapped phases, that are either symmetry protected topological (SPT) or spontaneous symmetry breaking (SSB) phases  for particular sub-symmetries of $\Rep (D_8)$, with their Hamiltonians, number of ground states and symmetry action, including the non-invertible $E$ satisfying (\ref{fusions}).
%They are either symmetry protected topological (SPT) or spontaneous symmetry breaking (SSB) phases for particular sub-symmetries of $\Rep (D_8)$. 
Table \ref{tab:phases} summarizes all $\Rep(D_8)$-symmetric gapped phases, with their Hamiltonians, number of ground states as well as the symmetry action, including the non-invertible generator $E$ satisfying (\ref{fusions}). They are either symmetry protected topological (SPT) or spontaneous symmetry breaking (SSB) phases for particular sub-symmetries of $\Rep (D_8)$. 
\onecolumngrid
\vspace{-1cm}
\begin{center}
\begin{minipage}{1.0\textwidth}
\begin{table}[H]
%\centering
\begin{tabular}{|c|c|}
\hline
Phase Type & \makecell[c]{Hamiltonian and Ground States with $\Rep (D_8)$ Action} \\
\hline \hline 
&\\[-3.8mm]
%%%%%%%%%%%%%%%%%%%%%%%%%%%%%%%%%%%%%%
 \makecell[c]{Trivial} & \makecell[c]{ $\hspace*{14.5mm} H_{1}=-\sum_i\lbb\bbI+\tfrac{1}{8}(\bbI+ Z^{\I})(\bbI+ Z^{\II})(\bbI+ Z^{\III})\rbb_{i}\simeq-\sum_i\lbb\bbI+\tfrac{1}{4}(\bbI+ Z^{\I})(Z^{\II}+Z^{\III})\rbb_{i}$ 
  \hspace{12.5mm}
\makecell[c]{\begin{tikzpicture}[baseline]
 \begin{scope}[shift={(-3,3)},scale=1]
 %\draw[gray,dotted] (-6.4,-2.2) to (9.4,-2.2);
     \node at (1.5,-2.75) {$\ket{\GS}$};
 \draw [-stealth](2,-2.9) .. controls (2.5,-3.2) and (2.5,-2.3) .. (2,-2.6);
 \end{scope}
 \end{tikzpicture} }
} \\[-1mm]
\hline
&\\[-2mm]
%%%%%%%%%%%%%%%%%%%%%%%%%%%%%%%%%%%%%%
\makecell[c]{$\Z_2$ SSB} & \makecell[c]{$H_{\Z_2^a}=-\tfrac{1}{2}\sum_i \left[\bbI_{i}\bbI_{i+1}+\gr{\tfrac{1}{2}(\bbI+Z^\I)_i}\, X^{\II}_iX_{i+1}^\II+\gr{\tfrac{1}{2}(\bbI-Z^\I)_i\, X^{\III}_i X^\II_{i+1}}\right]-\tfrac{1}{4}\sum_i\lbb(\bbI+ Z^{\I})(\bbI+ Z^{\III})\rbb_{i}$ \\[1mm]
\makecell[b]{\begin{tikzpicture}[baseline]
\begin{scope}[shift={(0,3)}]
%\draw[gray,dotted] (-6.4,-2.2) to (9.4,-2.2);
\end{scope}
\begin{scope}[shift={(1.5,3)}]
\node (1) at (-1.5,-3) {$\ket{\GS,+}$};
\node at (0,-3) {$\oplus$};
\node (2) at (1.5,-3) {$\ket{\GS,-}$};
\draw[red,stealth-stealth] (1) [bend left=20] to (2);
\draw[blue,stealth-stealth,transform canvas={xshift=0mm,yshift=1.5mm}] (1) [bend left=20] to (2);
\draw[blue,-stealth] (1) edge [in=60, out=120,looseness=5] (1);
\draw[ForestGreen,-stealth] (1) edge [in=70, out=110,looseness=5] (1);
\draw[blue,stealth-] (2) edge [in=60, out=120,looseness=5] (2);
\draw[ForestGreen,stealth-] (2) edge [in=70, out=110,looseness=5] (2);
\end{scope}
\end{tikzpicture} 
}}  \\
\hline
&\\[-2mm]
%%%%%%%%%%%%%%%%%%%%%%%%%%%%%%%%%%%%%% 
\makecell[c]{$\tfrac{\Rep(D_8)}{(\Z_2\times\Z_2)}$ SSB } & \makecell[c]{
$H_{\Z_2^{ab}}=-\tfrac{1}{2}\sum_i\left[\bbI_i\bbI_{i+1}+( X^{\II} X^{\III})_{i}(X^{\II} X^{\III})_{i+1}\right]-\tfrac{1}{4}\sum_i\lbb(\bbI+ Z^{\I})(\bbI+ Z^{\II} Z^{\III})\rbb_{i}$\\[1mm]
\makecell[b]{\begin{tikzpicture}[baseline]
\begin{scope}[shift={(0,3)}]
%\draw[gray,dotted] (-6.4,-2.2) to (9.4,-2.2);
\end{scope}
\begin{scope}[shift={(1.5,3)}]
\node (1) at (-1.5,-3) {$\ket{\GS,+}$};
\node at (0,-3) {$\oplus$};
\node (2) at (1.5,-3) {$\ket{\GS,-}$};
\draw[blue,stealth-stealth] (1) [bend left=20] to (2);
\draw[red,-stealth] (1) edge [in=60, out=120,looseness=5] (1);
\draw[ForestGreen,-stealth] (1) edge [in=70, out=110,looseness=5] (1);
\draw[red,stealth-] (2) edge [in=60, out=120,looseness=5] (2);
\draw[ForestGreen,stealth-] (2) edge [in=70, out=110,looseness=5] (2);
\end{scope}
\end{tikzpicture} 
}}  \\
\hline
&\\[-2mm]
%%%%%%%%%%%%%%%%%%%%%%%%%%%%%%%%%%%%%%
\makecell[c]{$\Z_2\times\Z_2$ SSB} & \makecell[c]
{$H_{(\Z_2^a\times\Z_2^b)^+}\simeq-\tfrac{1}{2}\sum_i \left[\gr{\tfrac{1}{2}(\bbI+Z^\I)_i}\, (X^{\II}_iX^\II_{i+1}+X^{\III}_iX^\III_{i+1})+\gr{\tfrac{1}{2}(\bbI-Z^\I)_i(X^{\III}_iX^\II_{i+1}+X^{\II}_iX^\III_{i+1})}\right]-\tfrac{1}{2}\sum_i[\bbI+Z^\I]_i$
\\[-1mm]
\makecell[b]{\begin{tikzpicture}[baseline]
\begin{scope}[shift={(0,3)}]
%\draw[gray,dotted] (-6.4,-2.2) to (9.4,-2.2);
\end{scope}
\begin{scope}[shift={(0,2.9)}]
\node (1) at (-1.5,-3) {$\ket{\GS,++}$};
\node at (-0.5,-3) {$\oplus$};
\node (2) at (0.5,-3) {$\ket{\GS,--}$};
\node at (1.5,-3) {$\oplus$};
\node (3) at (2.5,-3) {$\ket{\GS,+-}$};
\node at (3.5,-3) {$\oplus$};
\node (4) at (4.5,-3) {$\ket{\GS,-+}$};
\draw[blue,stealth-stealth] (1) [bend left=25] to (3);
\draw[blue,stealth-stealth] (1) [bend left=25] to (4);
\draw[blue,stealth-stealth] (2) [bend left=25] to (3);
\draw[blue,stealth-stealth] (2) [bend left=25] to (4);
\draw[red,stealth-stealth] (1) [bend right=30] to (2);
\draw[red,stealth-stealth] (3) [bend right=30] to (4);
\draw[ForestGreen,-stealth] (1) edge [in=70, out=110,looseness=5] (1);
\draw[ForestGreen,-stealth] (2) edge [in=70, out=110,looseness=5] (2);
\draw[ForestGreen,stealth-] (3) edge [in=70, out=110,looseness=5] (3);
\draw[ForestGreen,stealth-] (4) edge [in=70, out=110,looseness=5] (4);
%\oplus\ket{\GS,--}\oplus\ket{\GS,+-}\oplus\ket{\GS,-+}$};
\end{scope}
\end{tikzpicture} 
}} \\
&\\[-3mm]
\hline
&\\[-2mm]
%%%%%%%%%%%%%%%%%%%%%%%%%%%%%%%%%%%%%% 
\makecell[c]{SPT}  &  \makecell[c]{
$\hspace*{16mm}H_{(\Z_2^a\times\Z_2^b)^-}
    \simeq -\tfrac{1}{2}\sum_i \gr{\tfrac{1}{2}(\bbI+Z^\I)_i}\, \lbb X^{\II}_{i}(X^{\II}Z^{\III})_{i+1}+(Z^{\II}X^{\III})_{i}X^{\III}_{i+1}\rbb-\tfrac{1}{2}\sum_i[\bbI+Z^\I]_i$\\
    $\hspace*{16mm}\gr{-\tfrac{1}{2}\sum_i\tfrac{1}{2}(\bbI-Z^\I)_i(X^{\III}_iX^\II_{i+1}+(Z^\III X^{\II})_iX^\III_{i+1})}$ \hspace*{13mm} }\hspace*{10mm}  
\makecell[c]{\begin{tikzpicture}[baseline]
 \begin{scope}[shift={(-3,3)},scale=1]
 %\draw[gray,dotted] (-6.4,-2.2) to (9.4,-2.2);
     \node at (1.5,-2.75) {$\ket{\GS}$};
 \draw [-stealth](2,-2.9) .. controls (2.5,-3.2) and (2.5,-2.3) .. (2,-2.6);
 \end{scope}
 \end{tikzpicture} }
\\&\\[-3mm]
\hline
&\\[-2mm]
%%%%%%%%%%%%%%%%%%%%%%%%%%%%%%%%%%%%%%
$\Z_2$ SSB & \makecell[c]{$\gr{H_{\Z_2^c}=-\tfrac{1}{2}\sum_i\bbI_{i}\bbI_{i+1}-\tfrac{1}{2}\sum_i X^\I_i\left[\tfrac{1}{2}(\bbI+Z^\II Z^\III)\, X^{\I}+\tfrac{1}{2}(\bbI-Z^\II Z^\III)\, X^{\I}X^\II X^\III\right]_{i+1}
   -\tfrac{1}{4}\sum_i\lbb(\bbI+ Z^{\II})(\bbI+ Z^{\III})\rbb_{i}}$ \\[1mm]
\makecell[b]{\begin{tikzpicture}[baseline]
\begin{scope}[shift={(0,3)}]
%\draw[gray,dotted] (-6.4,-2.2) to (9.4,-2.2);
\end{scope}
\begin{scope}[shift={(1.5,3)}]
\node (1) at (-1.5,-3) {$\ket{\GS,+}$};
\node at (0,-3) {$\oplus$};
\node (2) at (1.5,-3) {$\ket{\GS,-}$};
\draw[ForestGreen,stealth-stealth] (1) [bend left=20] to (2);
\draw[blue,stealth-stealth,transform canvas={xshift=0mm,yshift=1.5mm}] (1) [bend left=20] to (2);
\draw[blue,-stealth] (1) edge [in=60, out=120,looseness=5] (1);
\draw[red,-stealth] (1) edge [in=70, out=110,looseness=5] (1);
\draw[blue,stealth-] (2) edge [in=60, out=120,looseness=5] (2);
\draw[red,stealth-] (2) edge [in=70, out=110,looseness=5] (2);
\end{scope}
\end{tikzpicture} 
}}  \\
\hline
&\\[-2mm]
%%%%%%%%%%%%%%%%%%%%%%%%%%%%%%%%%%%%%% 
$\Z_2\times\Z_2$ SSB & \makecell[c]{$\gr{H_{(\Z_2^c\times\Z_2^{ab})^+}\simeq -\tfrac{1}{2}\sum_i X^\I_i\left[\tfrac{1}{2}(\bbI+Z^\II Z^\III)\, X^{\I}+\tfrac{1}{2}(\bbI-Z^\II Z^\III)\, X^{\I}X^\II X^\III\right]_{i+1}}$\\\hspace*{13mm}$\gr{-\tfrac{1}{2}\sum_i\left[( X^{\II} X^{\III})_{i}(X^{\II} X^{\III})_{i+1}\right]-\tfrac{1}{2}\sum_i\left[\bbI+ (Z^{\II}Z^{\III})\right]_i}$ \\[-1mm]
\makecell[b]{\begin{tikzpicture}[baseline]
\begin{scope}[shift={(0,3)}]
%\draw[gray,dotted] (-6.4,-2.2) to (9.4,-2.2);
\end{scope}
\begin{scope}[shift={(0,2.9)}]
\node (1) at (-1.5,-3) {$\ket{\GS,++}$};
\node at (-0.5,-3) {$\oplus$};
\node (2) at (0.5,-3) {$\ket{\GS,--}$};
\node at (1.5,-3) {$\oplus$};
\node (3) at (2.5,-3) {$\ket{\GS,+-}$};
\node at (3.5,-3) {$\oplus$};
\node (4) at (4.5,-3) {$\ket{\GS,-+}$};
\draw[blue,stealth-stealth] (1) [bend left=25] to (3);
\draw[blue,stealth-stealth] (1) [bend left=25] to (4);
\draw[blue,stealth-stealth] (2) [bend left=25] to (3);
\draw[blue,stealth-stealth] (2) [bend left=25] to (4);
\draw[ForestGreen,stealth-stealth] (1) [bend right=30] to (2);
\draw[ForestGreen,stealth-stealth] (3) [bend right=30] to (4);
\draw[red,-stealth] (1) edge [in=70, out=110,looseness=5] (1);
\draw[red,-stealth] (2) edge [in=70, out=110,looseness=5] (2);
\draw[red,stealth-] (3) edge [in=70, out=110,looseness=5] (3);
\draw[red,stealth-] (4) edge [in=70, out=110,looseness=5] (4);
%\oplus\ket{\GS,--}\oplus\ket{\GS,+-}\oplus\ket{\GS,-+}$};
\end{scope}
\end{tikzpicture} 
}} \\
&\\[-3mm]
\hline
&\\[-2mm]
%%%%%%%%%%%%%%%%%%%%%%%%%%%%%%%%%%%%%% 
\makecell[c]{SPT}  &  \makecell[c]{
$\gr{H_{(\Z_2^c\times\Z_2^{ab})^-}
    \simeq -\tfrac{1}{8}\sum_i X^\I_i\left[(\bbI+Z^\II Z^\III)\, X^{\I}+(\bbI-Z^\II Z^\III)\, X^{\I}X^\II X^\III\right]_{i+1}[\bbI+Z^\II+Z^\III-Z^\II Z^\III]_{i+1}}$\\
    $\gr{-\tfrac{1}{2}\sum_i\left[( Z^\I X^{\II} X^{\III})_{i}(X^{\II} X^{\III})_{i+1}\right]-\tfrac{1}{2}\sum_i\left[\bbI+ (Z^{\II}Z^{\III})\right]_i}$ \hspace*{18mm} }\quad  
\makecell[c]{\begin{tikzpicture}[baseline]
 \begin{scope}[shift={(-4.8,3)},scale=1]
 %\draw[gray,dotted] (-6.4,-2.2) to (9.4,-2.2);
     \node at (1.5,-2.75) {$\ket{\GS}$};
 \draw [-stealth](2,-2.9) .. controls (2.5,-3.2) and (2.5,-2.3) .. (2,-2.6);
 \end{scope}
 \end{tikzpicture} }
\\&\\[-3mm]
%%%%%%%%%%%%%%%%%%%%%%%%%%%%%%%%%%%%%%
%\hline \hline \hline 
\hline
&\\[-2mm]
%%%%%%%%%%%%%%%%%%%%%%%%%%%%%%%%%%%%%%
\makecell[c]{$\Z_2$ SSB} & \makecell[c]{$\gr{H_{(D_8)^-}=-\tfrac{1}{8}\sum_i \left[(\bbI+Z^\I)\, X^{\II}+(\bbI-Z^\I)\, X^{\III}\right]_i X^\II_{i+1}\lbb\bbI-Z^{\I}+Z^{\III}+Z^{\I}Z^{\III}\rbb_{i+1} -\sum_i\bbI_i}$\\$\gr{  -\tfrac{1}{8}\sum_i\lbb\bbI+Z^{\II}+Z^{\III}-Z^{\II}Z^{\III}\rbb_i X^\I_i\left[(\bbI+Z^\II Z^\III)\, X^{\I}+(\bbI-Z^\II Z^\III)\, X^{\I}X^\II X^\III\right]_{i+1}}$\hspace*{-24mm}\\ 
\makecell[b]{\begin{tikzpicture}[baseline]
\begin{scope}[shift={(0,3)}]
%\draw[gray,dotted] (-6.4,-2.2) to (9.4,-2.2);
\end{scope}
\begin{scope}[shift={(1.5,3)}]
\node (1) at (-1.5,-3) {$\ket{\GS,+}$};
\node at (0,-3) {$\oplus$};
\node (2) at (1.5,-3) {$\ket{\GS,-}$};
\draw[ForestGreen,stealth-stealth] (1) [bend left=20,,transform canvas={xshift=0mm,yshift=-1.5mm}] to (2);
\draw[red,stealth-stealth] (1) [bend left=20] to (2);
\draw[blue,stealth-stealth,transform canvas={xshift=0mm,yshift=1.5mm}] (1) [bend left=20] to (2);
\draw[blue,-stealth] (1) edge [in=60, out=120,looseness=5] (1);
\draw[blue,stealth-] (2) edge [in=60, out=120,looseness=5] (2);
\end{scope}
\end{tikzpicture} 
}}  \\
\hline
&\\[-2mm]
%%%%%%%%%%%%%%%%%%%%%%%%%%%%%%%%%%%%%%
\makecell[c]{$\Z_2\times\Z_2$ SSB} & \makecell[c]{$\gr{ H_{\Z_4^{ca}}\simeq-\tfrac{1}{4}\sum_i\lbb(\bbI+ Z^{\I}) X^{\I}X^{\III}+
    (\bbI -Z^{\I}) X^{\I}X^{\II}\rbb_{i}\lbb X^{\I}X^{\III}(\Swap)^{\II,\III}\rbb_{i+1}}$\hspace*{45mm}\\
    $\gr{-\tfrac{1}{4}\sum_i\lbb(\bbI+ Z^{\I}) X^{\I}X^{\II}+
   (\bbI -Z^{\I}) X^{\I}X^{\III}\rbb_{i}\lbb X^{\I}X^{\II}(\Swap)^{\II,\III}\rbb_{i+1}-\tfrac{1}{2}\sum_i\left[\bbI+ Z^{\I} Z^{\II} Z^{\III}\right]_{i}}$\\[-1mm]
\makecell[b]{\begin{tikzpicture}[baseline]
\begin{scope}[shift={(0,3)}]
%\draw[gray,dotted] (-6.4,-2.2) to (9.4,-2.2);
\end{scope}
\begin{scope}[shift={(0,2.9)}]
\node (1) at (-1.5,-3) {$\ket{\GS,++}$};
\node at (-0.5,-3) {$\oplus$};
\node (2) at (0.5,-3) {$\ket{\GS,--}$};
\node at (1.5,-3) {$\oplus$};
\node (3) at (2.5,-3) {$\ket{\GS,+-}$};
\node at (3.5,-3) {$\oplus$};
\node (4) at (4.5,-3) {$\ket{\GS,-+}$};
\draw[blue,stealth-stealth] (1) [bend left=25] to (3);
\draw[blue,stealth-stealth] (1) [bend left=25] to (4);
\draw[blue,stealth-stealth] (2) [bend left=25] to (3);
\draw[blue,stealth-stealth] (2) [bend left=25] to (4);
\draw[ForestGreen,stealth-stealth] (1) [bend right=30] to (2);
\draw[red,stealth-stealth,transform canvas={xshift=0mm,yshift=-1.2mm}] (1) [bend right=30] to (2);
\draw[ForestGreen,stealth-stealth] (3) [bend right=30] to (4);
\draw[red,stealth-stealth,transform canvas={xshift=0mm,yshift=-1.2mm}] (3) [bend right=30] to (4);
\end{scope}
\end{tikzpicture} 
}} \\
&\\[-2mm]
\hline
&\\[-2mm]
%%%%%%%%%%%%%%%%%%%%%%%%%%%%%%%%%%%%%%
\makecell[c]{$\Rep(D_8)$ SSB} & \makecell[c]{$\gr{H_{(D_8)^+}\simeq  -\tfrac{1}{4}\sum_i \left[(\bbI+Z^\I)\, X^{\II}+(\bbI-Z^\I)\, X^{\III}\right]_i X^\II_{i+1} -\sum_i\bbI_i\;}$\\[1mm]
  \hspace{24mm}$\gr{-\tfrac{1}{4}\sum_i X^\I_i\left[(\bbI+Z^\II Z^\III)\, X^{\I}+(\bbI-Z^\II Z^\III)\, X^{\I}X^\II X^\III\right]_{i+1}}$\\
\makecell[b]{\begin{tikzpicture}[baseline]
\begin{scope}[shift={(0,3)}]
%\draw[gray,dotted] (-6.4,-2.2) to (9.4,-2.2);
\end{scope}
\begin{scope}[shift={(-1,2.8)}]
\node (1) at (-1.5,-3) {$\ket{\GS,1}$};
\node (1b) at (-1.55,-3.2) {};
\node at (-0.5,-3) {$\oplus$};
\node (2) at (0.5,-3) {$\ket{\GS,2}$};
\node (2b) at (0.45,-3.2) {};
\node at (1.5,-3) {$\oplus$};
\node (5) at (2.5,-3) {$\ket{\GS,5}$};
\node at (3.5,-3) {$\oplus$};
\node (3) at (4.5,-3) {$\ket{\GS,3}$};
\node (3b) at (4.55,-3.2) {};
\node at (5.5,-3) {$\oplus$};
\node (4) at (6.5,-3) {$\ket{\GS,4}$};
\node (4b) at (6.55,-3.2) {};
\draw[blue,stealth-stealth] (1) [bend left=30] to (5);
\draw[blue,stealth-stealth] (2) [bend left=30] to (5);
\draw[blue,stealth-stealth] (3) [bend right=30] to (5);
\draw[blue,stealth-stealth] (4) [bend right=30] to (5);
\draw[red,stealth-stealth,transform canvas={xshift=0mm,yshift=1mm}] (1) [bend right=30] to (2);
\draw[red,stealth-stealth,transform canvas={xshift=0mm,yshift=1mm}] (3) [bend right=30] to (4);
\draw[ForestGreen,stealth-stealth] (1b) [bend right=15] to (3b);
\draw[ForestGreen,stealth-stealth] (2b) [bend right=15] to (4b);
% \draw[red,stealth-stealth,transform canvas={xshift=3.5mm,yshift=0mm}] (1) [bend right=30] to (2);
% \draw[red,stealth-stealth,transform canvas={xshift=-3.5mm,yshift=0mm}] (3) [bend right=30] to (4);
%\oplus\ket{\GS,--}\oplus\ket{\GS,+-}\oplus\ket{\GS,-+}$};
\end{scope}
\end{tikzpicture} 
}}
\\
\hline
\end{tabular}
\caption{All gapped phases for the benchmark system with $\Rep (D_8)$ symmetry, showing the type of phase, i.e. SSB (spontaneously broken symmetry) and SPT (symmetry protected topological phase).  For each phase we write the  commuting projector Hamiltonian of the spin model, realized on $(\mathbb{C}^2_\I\otimes \mathbb{C}^2_\II \otimes \mathbb{C}^2_\III)^L$ and the number of ground states (GS) with the action of the  categorical symmetry.
We depict the $\Rep(D_8)$ symmetry action on the ground states as follows: $1_c$ in {\color{red!80!black}red}, $1_a$ in {\color{ForestGreen} green}, (their product is $1_{ca}=1_c\otimes 1_a$) and the non-invertible symmetry $E$ in {\color{blue}blue}. We use black for the full $\Rep(D_8)$. 
 %The first five phases are easier to implement in our experimental proposal based on dual-species atom arrays. 
 {The first five Hamiltonians can be simplified by first applying the $-(\bbI+Z^\I)_i /2$ projector: they then reduce to the black terms, which can be implemented in our experimental proposal based on dual-species atom arrays. The Hamiltonians written in brown require a slight generalization.}
\label{tab:phases}}
\end{table}
\end{minipage}
\end{center}
\twocolumngrid

\normalsize
\onecolumngrid
\begin{center}
    {\bf Supplemental material for Categorical Symmetries in Spin Models with 
Atom Arrays}\\[5mm]

{Alison Warman$^{1}$, Fan Yang$^{2,3}$, Apoorv Tiwari$^{4}$, Hannes Pichler$^{2,3}$ Sakura Sch\"afer-Nameki$^{1}$}\\[5mm]

\textit{\small $^{1}$Mathematical Institute, University
of Oxford, Woodstock Road, Oxford, OX2 6GG, United Kingdom,\\
$^{2}$Institute for Theoretical Physics, University of Innsbruck, Innsbruck 6020, Austria\\
$^{3}$Institute for Quantum Optics and Quantum Information of the Austrian Academy of Sciences, Innsbruck 6020, Austria,\\
$^{4}$Center for Quantum Mathematics at IMADA, Southern Denmark University,
Campusvej 55, 5230 Odense, Denmark}

\end{center}

\vspace{5mm}
\twocolumngrid
\appendix
\twocolumngrid

\tableofcontents 
\section{Lattice Model with $\Rep (D_8)$ symmetry} \label{app:appA}

\subsection{Preliminaries}
\noindent In this section, we describe some mathematical preliminaries related to the group $D_8$ and its representation theory, which play an important role in the model we will subsequently consider.
The dihedral group $D_8$ is the isometry group of a square and can be presented as
\begin{equation}
    D_8:=\lb\Z_2^{a}\times \Z_2^{b}\rb \rtimes \Z_2^{c}=\{1\,,a\,,b\,,ab\,,c\,,ca\,, cb\,, cab\}\,.
\end{equation}
It comprises of a normal subgroup \begin{equation}
    \Z_2^{a}\times \Z_2^{b}=\{1,\,a,\,b,\,ab\}\,.
\end{equation}
The adjoint action of $c\in\Z_2^{c}$  exchanges $a$ and $b$, i.e.,  
\begin{equation}
    cac=b\,,\quad cbc=a\,.
\end{equation}
$D_8$ has two order-4 elements, given by $ca=bc$ and its inverse $cb=ac$. All other non-identity elements are of order-2. The conjugacy classes of $D_8$ are:
\begin{equation}\label{eq:D8_conjclasses}
\begin{alignedat}{2}
    &[1]=\{1\}\,,\qquad &[ab]&=\{ab\}\,,\quad [ca]=\{ca,\,cb\}\,, \\
    &[a]=\{a,\,b\}\,,\qquad &[c]=&\{c,\,cab\}\,.    
\end{alignedat}
\end{equation}
The group $D_8$ has five irreducible representations labeled as $\{1,1_c,1_a,1_{ca},E\}$: the first four labeled as $1_k$ are 1-dimensional, while the $E$ representations is 2-dimensional.
The character table for $D_8$ is shown in table \ref{tab:D8_chars}.
\newpage
\begin{table}[H]
    \centering
    \begin{tabular}{|r|ccccc|}
    \hline
     & $[1]$ & $[ab]$ & $[ca]$ & $[c]$ & $[a]$ \\
     \hline
     1:\;& $1$ & $1$ & $1$ & $1$ & $1$ \\
     $1_c$:\;& $1$ & $1$ & $-1$ & $1$ & $-1$ \\
     $1_a$:\; & $1$ & $1$ & $-1$ & $-1$ & $1$ \\
     $1_{ca}$:\; & $1$ & $1$ & $1$ & $-1$ & $-1$ \\
     $E$:\; & $2$ & $-2$ & $0$ & $0$ & $0$ \\
     \hline
    \end{tabular}
    \caption{$D_8$ character table.}
    \label{tab:D8_chars}
\end{table}

\noindent The invertible, i.e., one dimensional, irreducible representations $\{1,1_c,1_a,1_{ca}\}$ can be read off from their characters while the 2-dimensional irreducible representation $E$ has the matrix representation
\be\label{eq:E matrix representation}
\begin{alignedat}{2}
    \cD_E(1)&=\bbI^{(2)}\,,\qquad %\sigma^{0}\,, \qquad 
    &\cD_E(c)&=X\,,\\ %\sigma^x\,, \\
    \cD_E(a)&=Y\,,\qquad %\sigma^{y}, \qquad
    &\cD_E(b)&=-Y\,. %-\sigma^{y}\,.
\end{alignedat} 
\ee

The fusion category $\Rep(D_8)$, which will be the symmetry category of our spin model, has five simple objects which correspond to the irreducible representations of $D_8$.
The fusion rules in $\Rep(D_8)$ are inherited from the tensor product of the representations.
Specifically, the 1 dimensional irreps $1_k$, for $k=c,a,ca$, fuse according to $\Z_2\times\Z_2$: 
\be
 1_a\otimes1_c=1_c\otimes1_a=1_{ca}\,,\qquad  1_k \otimes 1_k =1\,,\qquad     \\
\ee
whereas the two dimensional representation $E$ satisfies
\begin{equation}
\ba
   &E\otimes 1_k=1_k\otimes E=E\,, \\
    &E\otimes E=1\oplus 1_a\oplus 1_c \oplus 1_{ca}\,.
\ea  
\end{equation}
These fusion rules are those of a Tambara-Yamagami category $\Rep(D_8)=\TY(\Z_2\times\Z_2,\chi_1,+\tfrac{1}{2})$, defined in \cite{tamabara}, which also contains more refined data corresponding to the associator $
    \alpha_{R_1,R_2,R_3}:(R_1\ot R_2)\ot R_3 \rightarrow R_1\ot (R_2\ot R_3)$
encoding the $F$-symbols of the fusion category.

\subsection{SymTFT, Generalized Charges and Gapped Phases}
\label{app:SymTFT}

In this section we will briefly summarize the theoretical underpinning of the classification of phases in the categorical Landau paradigm (for more details see \cite{Bhardwaj:2023ayw, Bhardwaj:2023idu, Bhardwaj:2023fca, Bhardwaj:2023bbf, Bhardwaj:2024qrf}). The protagonist is the SymTFT for a categorical symmetry  $\cS$, which for a $d$-dimensional system is a $d+1$ dimensional topological field theory, which allows the separation of  symmetry from dynamics.

\onecolumngrid
\begin{center}
\begin{table}[H]
    \centering
    \begin{tabular}{|c|c|c|c|c|c|}
\hline
Label in \cite{Bhardwaj:2024qrf} & Lagrangian algebra & $(F)^{\beta}$ & \makecell[c]{Gapped phase \\ for $\cS=\Rep(D_8)$} & $n$ & Section\\
\hline
&&&&&\\[-3mm]
\nameref{eq:AD8_27} &  {$\textcolor{blue}{([1],1)} \oplus ([1],1_a) \oplus ([1],1_c) \oplus ([1],1_{ca}) \oplus 2 ([1],E)$}\xlabel[$\cA_{27}$]{eq:AD8_27} & $1$ & Trivial & 1 & \ref{sec:1} \\[1mm]
\nameref{eq:AD8_28} &  $\textcolor{blue}{([1],1)} \oplus ([ab],1_{ca}) \oplus ([1],1_{ca}) \oplus \textcolor{blue}{([ab],1)} \oplus  \textcolor{blue}{2 ([ca],1)}$\xlabel[$\cA_{28}$]{eq:AD8_28} & $\Z_4^{ca}$ & $\Z_2\times\Z_2$ SSB & 4 & \ref{sec:Z4ca}\\[1mm]
\nameref{eq:AD8_29} &  $\textcolor{blue}{([1],1)} \oplus ([ab],1_c) \oplus ([1],1_c) \oplus \textcolor{blue}{([ab],1)} \oplus \textcolor{blue}{2 ([c],1_{+,+})} $\xlabel[$\cA_{29}$]{eq:AD8_29} & $(\Z_2^c\times\Z_2^{ab})^+$ & $\Z_2\times\Z_2$ SSB & 4 & \ref{sec:Z2cZ2ab} \\[1mm]
\nameref{eq:AD8_30} &  $\textcolor{blue}{([1],1)} \oplus ([ab],1_{ca}) \oplus ([1],1_c) \oplus ([ab],1_a) \oplus 2 ([c],1_{+,-})$\xlabel[$\cA_{30}$]{eq:AD8_30} & $(\Z_2^c\times\Z_2^{ab})^-$ & SPT & 1 & \ref{sec:Z2cZ2ab-}\\[1mm]
\nameref{eq:AD8_31} &  {$\textcolor{blue}{([1],1)} \oplus ([1],1_a) \oplus ([ab],1_a) \oplus \textcolor{blue}{([ab],1)} \oplus\textcolor{blue}{2 ([a],1_{+,+})}$}\xlabel[$\cA_{31}$]{eq:AD8_31} & $(\Z_2^a\times\Z_2^{b})^+$ & $\Z_2\times\Z_2$ SSB & 4 & \ref{sec:Z2aZ2b}\\[1mm]
\nameref{eq:AD8_32} &  {$\textcolor{blue}{([1],1)} \oplus ([ab],1_{ca}) \oplus ([1],1_a) \oplus ([ab],1_c) \oplus 2 ([a],1_{+,-})$}\xlabel[$\cA_{32}$]{eq:AD8_32} & $(\Z_2^a\times\Z_2^{b})^-$ & SPT & 1 & \ref{sec:Z2aZ2b-}\\[1mm]
\hline
&&&&&\\[-3mm]
%%%%
\nameref{eq:AD8_33} &  $\textcolor{blue}{([1],1)} \oplus \textcolor{blue}{([ab],1)} \oplus \textcolor{blue}{([c],1_{+,+})} \oplus \textcolor{blue}{([a],1_{+,+})} 
\oplus \textcolor{blue}{([ca],1)}$\xlabel[$\cA_{33}$]{eq:AD8_33} & $D_8^{+}$ & $\Bsym$ and $\Rep(D_8)$ SSB & 5 & \ref{sec:D8}\\[0.5mm]
%%%%
\hline
&&&&&\\[-3mm]
\nameref{eq:AD8_34} &  $\textcolor{blue}{([1],1)} \oplus ([ab],1_{ca}) \oplus ([c],1_{+,-}) \oplus ([a],1_{+,-}) \oplus  \textcolor{blue}{([ca],1)}$\xlabel[$\cA_{34}$]{eq:AD8_34} & $D_8^{-}$ & $\Z_2$ SSB & 2 & \ref{sec:D8-}\\[1mm]
\nameref{eq:AD8_35} &  $\textcolor{blue}{([1],1)} \oplus ([1],1_c) \oplus ([1],E) \oplus ([c],1_{+,-}) \oplus \textcolor{blue}{([c],1_{+,+})} $\xlabel[$\cA_{35}$]{eq:AD8_35} & $\Z_2^c$ & $\Z_2$ SSB & 2 & \ref{sec:Z2c}\\[1mm]
\nameref{eq:AD8_36} &  {$\textcolor{blue}{([1],1)} \oplus ([1],1_a) \oplus ([1],E) \oplus ([a],1_{+,-}) \oplus \textcolor{blue}{([a],1_{+,+})}$}\xlabel[$\cA_{36}$]{eq:AD8_36} & $\Z_2^a $ & $\Z_2$ SSB & 2 & \ref{sec:Z2a}\\[1mm]
\nameref{eq:AD8_37} &  \makecell[c]{{$\textcolor{blue}{([1],1)} \oplus ([1],1_{c}) \oplus ([1],1_a) \oplus ([1],1_{ca}) \oplus $} \\ {$\textcolor{blue}{([ab],1)}\oplus([ab],1_c)\oplus ([ab],1_a) \oplus ([ab],1_{ca})$}} \xlabel[$\cA_{37}$]{eq:AD8_37} & $\Z_2^{ab} $ & $\tfrac{\Rep(D_8)}{(\Z_2\times\Z_2)}$ SSB & 2 & \ref{sec:Z2ab}\\[3mm]
\hline
\end{tabular}
    \caption{Lagrangian algebras for $\Rep (D_8)$ SymTFT, comprised of anyons, which are labeled by $\Q_{[g], R}$. These define gapped boundary conditions, where these anyons can condense. $(F)^\beta$ denotes the corresponding choice of (a representative of a) conjugacy class of subgroups $F$ and $\beta\in H^2(F,U(1))$; $n$ is the number of vacua in the gapped phase. The generalized charges that contain the local order parameters for the gapped phases are shown in blue: $\Q_{\textcolor{blue}{([g], R)}}$. The numbering of the Lagrangian algebras is that of reference \cite{Bhardwaj:2024qrf}.}
    \label{tab:D8_Lags}
\end{table}
\end{center}

\twocolumngrid
The SymTFT for a (1+1)d theory with symmetry $\cS$ is a (2+1)d TFT with two boundaries: $\Bsym$ specifies the symmetry, whereas $\Bphys$ encodes the dynamical information of a (1+1)d theory:
\begin{equation}
\begin{tikzpicture}[scale=2]
\begin{scope}[shift={(0,0)}]
\draw [cyan,  fill=cyan] 
(0,0) -- (0,1) -- (1,1) -- (1,0) -- (0,0) ; 
\draw [white] (0,0) -- (0,1) -- (1,1) -- (1,0) -- (0,0)  ; 
\draw [very thick] (0,0) -- (0,1) ;
\draw [very thick] (1,0) -- (1,1) ;
\node at (0.5,0.5) {SymTFT} ;
\node[above] at (0,1) {$\Bsym$}; 
\node[above] at (1,1) {$\Bphys$}; 
\end{scope}
\end{tikzpicture}
\end{equation}

The topological defects (anyons) of the SymTFT (which form the so-called Drinfeld center of the symmetry category) generalize the notion of representations to categorical symmetries:  
they give rise to generalized charges, which  form multiplets under the categorical symmetry. 
For $\Rep (D_8)$ the generalized charges are $\Q_{[g], R}$, where $[g]$ is a conjugacy class and $R$ a representation of the centralizer of $g\in[g]$. We will determine the generalized charges in detail and their lattice realization in section \ref{sec:Multiplets}. 

Gapped phases with symmetry $\cS$ are classified in the SymTFT by letting $\Bphys$ also be a  gapped boundary condition. Gapped boundary conditions are in 1-1 correspondence with so-called Lagrangian algebras, which specify the anyons that can end on the boundary of the SymTFT. For $\Rep (D_8)$ these are summarized in table \ref{tab:D8_Lags}. 
The simplicity of this classification of gapped phases is not the only advantage of the SymTFT approach: it encodes the order parameters for the phase, and the action of the categorical symmetry on these.

For $\Rep (D_8)$ there are 11 gapped phases, given by the pairing $\Bsym= \cA_{33}$ in table \ref{tab:D8_Lags} with $\Bphys$ taking value in each of the Lagrangian algebras in table \ref{tab:D8_Lags}. The order parameters are determined as the set of anyons that appear both in the set of anyons specifying the symmetry boundary and the physical boundary. The action of the symmetry on the order parameters is given by the linking in the (2+1)d SymTFT. In summary: 
\begin{itemize}
\item Gapped phases: SymTFT with choice of \\
$\Bsym = \cA_{33}$, \;\;$\Bphys = \cA_{i}$ in table \ref{tab:D8_Lags}.
\item Order parameters: $\cA_{33} \cap \cA_{i}$ (counted with multiplicities).
\item Number of ground states: $n=\sum_a n_{a}^{\cA_{33}} \,n_a^{\cA_{i}}$, where $n_a^{\cA_i}$ is the integer coefficient of the anyon $a$ in the algebra $\cA_i$.
\item Categorical symmetry action on ground states: linking of anyons in the SymTFT.
\end{itemize}

\subsection{Multiplets of the $\Rep(D_8)$ Symmetry}
\label{sec:Multiplets}

Symmetry representations play a central role in the Landau paradigm as phases and transitions are characterized by their condensation patterns.
The collection of symmetry representations condensed in a gapped phase serve as order parameters for that phase.
In the same spirit, generalized charges (representations) of non-invertible symmetries also characterize phases and transitions within the categorical Landau paradigm. \\

\noindent{\bf Generalized Charges for $\Rep (D_8)$.} 
We now turn to the symmetry representations of $\Rep(D_8)$.
We will deduce these representations in the following way.
We start with the representations of $D_8$ symmetry, which are more familiar.
We then gauge the $D_8$ symmetry (in two steps) to obtain the system with $\Rep(D_8)$ symmetry.
Since gauging is a topological manipulation, the multiplet of operators in a given $D_8$ representation after gauging become operators in a certain symmetry multiplet of the $\Rep(D_8)$ symmetry of the gauged model. 

\medskip {\bf Representations of $D_8$.} Local operators in a (1+1)d system transforming in representations of $D_8$ symmetry can either be genuine local operators or non-genuine local operators, i.e., attached to a symmetry defect $\ms g\in D_8$.
% \begin{equation}
% \includegraphics[width=0.24\textwidth]{LocalOps.png}
% \nonumber
% \end{equation}
Under the symmetry action by $\ms h\in D_8$, a $\ms g$-twisted sector operator transforms to an $\ms h\ms g \ms h^{-1}$ twisted sector operator.
Therefore $D_8$ representations are clubbed into twisted sectors labeled by conjugacy class.
Moreover, a non-genuine local operator in a certain twisted sector can transform meaningfully only in a representation of the centralizer of a chosen element in the conjugacy class.
To summarize, symmetry multiplets of $D_8$ are labeled as 
\begin{equation}\label{eq:SymTFT_labels}
 ([g],R)   
\end{equation}
where $[g]$ is a $D_8$ conjugacy class and $R$ is a representation of the centralizer for a representative $g\in[g]$. These are precisely the anyons in the SymTFT introduced in section \ref{app:SymTFT} (alternatively, they can  be labeled in terms of three copies of the toric code with non-trivial 3-cocycle \cite{Iqbal:2023wvm}).  
Among these, the self-local charges play a special role as they can condense or become topological in the infra-red. 
The self-local operators satisfy 
\begin{equation}\label{eq:T matrix}
    \frac{\Tr(\cD_R([g]))}{{\rm dim}(R)}=1\,.
\end{equation}
One can recognize that the labels \eqref{eq:SymTFT_labels} indeed also correspond to the labels of the $D_8$ SymTFT bulk topological lines \cite{Lin:2022dhv, Bhardwaj:2023wzd} and the condition \eqref{eq:T matrix} simply translates to the fact that the corresponding line is Bosonic and can therefore condense on a boundary. The sets of labels satisfying \eqref{eq:T matrix} are precisely the lines appearing in the Lagrangian algebras in Table~\ref{tab:D8_Lags}. 
These serve as order parameters for the different $\Rep(D_8)$ phases and transitions.

We now deduce the structure of the generalized charges $\Q_{[g], R}$, as multiplets under the $\Rep (D_8)$ symmetry. 
\begin{itemize}
    \item {\bf $\Q_{[1],\Gamma}$ generalized charge.} In $D_8$ symmetric systems, this corresponds to a ${\rm dim}(\Gamma)$ dimensional multiplet of genuine local operators that transform in the $\Gamma$ representation of $D_8$.
    After gauging the full $D_8$ symmetry, these $\Gamma$ charged untwisted operators map to uncharged $\Gamma$ twisted sector operators.
    Concretely an operators labeled $(\Gamma,\ell)$ with $\ell=1,\dots, \rm dim(\Gamma)$ acts by inserting a $\Gamma$ symmetry twist corresponding to the $\ell^{\rm th}$ vector in the underlying vector space of $\Gamma$.
    We denote such a twisted sector operator in the lattice model as 
    \begin{equation}
        \mathcal T_{(\Gamma,\ell)}\,.
    \end{equation}
    
\end{itemize}
To study the remaining generalized charges, we carry out the $D_8$ gauging in two steps.
We first gauge the normal subgroup $\Z_4^{ca}$ in $D_8$. 
Upon doing so, we obtain a new dual symmetry which is $\Rep(\Z_4^{ca})$.  
We denote the dual $\Rep(\Z_4^{ca})$ generators as $\rho_{\ms n}$ with $\ms n=0,1,2,3$. 
With the properties
\begin{equation}
    \rho_{\ms n}(ca)=i^{\ms n}\,.
\end{equation}
Additionally there is also the remaining $\Z_2^c$ symmetry which acts on $\Rep(\Z_4^{ca})$ by the outer automorphism
exchanging $\rho_{1}$ and $\rho_3$.
Therefore the full symmetry after gauging is 
\begin{equation}
    \Rep(\Z_4^{ca})\rtimes \Z_2^c\cong D_8\,.
\end{equation}
In the second step, we gauge $\Z_2^{c}$.
This causes  $\rho_1$ and $\rho_3$ to combine into a single $c$-invariant non-invertible symmetry of dimension 2
\begin{equation}
    \rho_1\oplus\rho_3 \cong E\,.
\end{equation}
Denoting the vector space underlying $\rho_{\ms n}$ as $v_{\ms n}$, it can immediately be seen that $D_8$ is represented on $v_1\oplus v_3$ precisely via the matrix representation in \eqref{eq:E matrix representation}.
The remaining symmetry generators are invertible. 
The symmetry $\rho_2$ survives the $\Z_2^c$ gauging and since $\rho_2(ca)=-1$ and $\rho_2(c)=1$, we may identify
\begin{equation}
    \rho_2\cong 1_c\,.
\end{equation}
Lastly, there is a $\Rep(\Z_2^{c})$ symmetry generated by $\rho_c$. Since $\rho_c(c)=-1$ and $\rho_c(ca)=-1$, we may identify 
\begin{equation}
    \rho_c\cong 1_a\,.
\end{equation}
Compatibility between fusions before and after gauging imposes that $1_c\,, 1_a$ and $E$ satisfy the $\Rep(D_8)$ fusion rules.
Similar partial gaugings were detailed in the context of 2-fusion categories in 2+1 dimensions in \cite{Bhardwaj:2022maz}. 
Let us now deduce the structure of the remaining self-local $\Rep(D_8)$ generalized charges.
\begin{itemize}
    \item {\bf $\Q_{[ab],1}$ generalized charge.} As a generalized charge for a $D_8$ symmetric system, this is a single uncharged operator in the $ab$ twisted sector.
    Upon gauging $\Z_4^{ca}$, it maps to a genuine local operator invariant under $\rho_2$ but transforming with a minus sign under $\rho_1$ and $\rho_3$.
    Indeed this agrees with 
    \begin{equation}
        \rho_{\ms n}=(-1)^{\ms n}\,.
    \end{equation}
    Since it is uncharged and untwisted with respect to $\Z_2^{c}$, it remains unaltered upon a subsequent gauging of $\Z_2^{c}$.
    Therefore, we obtain a single local operator uncharged under $\rho_2 \cong 1_c$, and with a linking $-2$ ($-1\times 2$ from $\rho_1$ and $\rho_3$) with $E$. We identify this operator as 
    \begin{equation}
        \Q_{([ab],1)}\cong X^{\II}X^{\III}\,.
    \end{equation}

    \item {\bf $\Q_{[ab],1_a}$ generalized charge.} As a generalized charge for a $D_8$ symmetric system, this is a single operator in the $ab$ twisted sector that is charged under $\rho_c$.
    Upon gauging $\Z_4^{ca}$, it maps to a genuine local operator that carries a charge $(-1)^{\ms n}$ under  $\rho_{\ms n}$ and $-1$ under $c$.
    Finally, upon the subsequent gauging of $\Z_2^c$, one obtains a $\rho_{c}\cong 1_a$-twisted operator with charge $-1$ under $\rho_1$ and $\rho_3$ or equivalently linking charge $-2$ under $E$. 
    We may identify this operator as 
    \begin{equation}
        \Q_{([ab],1_a)} \cong \mathcal T_{1_a}X^{\II}X^{\III}\,.
    \end{equation}
    \item {\bf $\Q_{[ab],1_c}$ generalized charge.} As a $D_8$ symmetry charge, this corresponds to an operator in the $ab$ twisted sector that transforms in the $1_c$ representation of $D_8$.
    We may decompose $1_c = (\rho_2\,, 1)\in \Rep(\Z_4^{ca}) \times \Rep(\Z_2^c)$.
    Therefore upon gauging $\Z_4^{ca}$, it maps to an operator in the $\rho_2$-twisted sector that carries a $(-1)^{\ms n}$ charge under $\rho_{\ms n}$.
    Since this operator is uncharged and untwisted with respect to $\Z_2^c$, it remains invariant under the $\Z_2^c$ gauging.
    As a $\Rep(D_8)$ multiplet, we obtain a single operator in the $1_c$ twisted sector that has a linking charge of $-2$  under $E$.
    We may identify this as
    \begin{equation}
        \Q_{[ab],1_c} \cong \mathcal T_{1_c}X^{\II}X^{\III}\,.
    \end{equation}
        \item {\bf $\Q_{[ca],1}$ generalized charge.} As a $D_8$ generalized charge, this is a doublet of twisted sectors operators $\cO_{ca}$ and $\cO_{cb}$ which are $ca$ and $cb$ twisted respectively.
        We first gauge the $\Z_4^{ca}$ symmetry to obtain a doublet of local operators in which $\cO_{ca}$ and $\cO_{cb}$ carry  $i^{\ms n}$ and $i^{- \ms n}$ charge under $\rho_{\ms n}$  respectively.
        The $\Z_2^{c}$ symmetry exchanges $\cO_{ca}$ and $\cO_{cb}$.
        We consider the linear combinations 
        \begin{equation}
        \cO_{[ca],\pm}=\cO_{ca}\pm \cO_{cb}\,.
        \end{equation}
        Among these $\cO_{[ca],+}$ is uncharged, while $\cO_{[ca],-}$ is charged under $c$.
        Upon gauging $\Z_2^c$, $\cO_{[ca],+}$ becomes an uncharged operator while $\cO_{[ca],-}$ becomes a $\rho_c\cong 1_a$ twisted operator.
        We identify this as the doublet 
       \begin{equation}
           \begin{split}
               \Q_{([ca],1)}& \cong\left\{\cO_{[ca],+}\sim X^{\I}X^{\II}+X^{\I}X^{\III}\,, \right. \\
          & \left.\cO_{[ca],-}\sim \mathcal T_{1_{a}}(X^{\I}X^{\II}-X^{\I}X^{\III})\right\}\,.  
           \end{split}
       \end{equation}
        
    \item {\bf $\Q_{[c],1_{++}}$ generalized charge.} As a $D_8$ generalized charge, this corresponds to a doublet of operators $\cO_{c}$ and $\cO_{cab}$ in the twisted sectors of $c$ and $cab$ respectively.
    Both $\cO_{c}$ and $\cO_{cab}$ are uncharged under the centralizer group $\mathbb Z_2^{c}\times \mathbb Z_2^{ab}$ and are exchanged under $\mathbb Z_2^{ca}$ (via conjugation).
    Consider the linear combinations
 \begin{equation}
        \cO_{[c],\pm}=\cO_{c}\pm \cO_{cab}\,.
        \end{equation}
  Among these, $\cO_{[c],+}$ is uncharged under $\Z_4^{ca}$ while $\cO_{[c],-}$ transforms under $\rho_2\in \Rep(\Z_4^{ca})$.
  Upon gauging $\Z_4^{ca}$, $\cO_{[c],+}$ becomes an uncharged $c$-twisted operator, while $\cO_{[c],-}$ becomes a $c\cdot \rho_2$ twisted uncharged operator.
  Upon the final gauging of $\Z_2^c$, $\cO_{[c],+}$ becomes a genuine local operator carrying a $\rho_c\cong 1_a$ charge while $\cO_{[c],-}$ becomes a $\rho_2\cong 1_c$ twisted sector operator carrying a $1_a$ charge. 
  On the lattice, we identify this doublet as
  \begin{equation}
  \begin{split}
    \Q_{[c],1_{++}}= &\left\{
     \cO_{[c],+}\sim (X^{\I}+X^{\I}X^{\II}X^{\III})\,,  \right.      \\
 & \ \left.  \cO_{[c],-}\sim (X^{\I}-X^{\I}X^{\II}X^{\III})\mathcal T_{1_{c}}
      \right\}\,.      
  \end{split}
  \end{equation}
\item {\bf $\Q_{[c], 1_{+-}}$ generalized charge.} As a $D_8$ generalized charge, this comprises of a doublet $\widetilde{\cO}_c$ and $\wt{\cO}_{cab}$, which transform in the representations $1_{+-}^{(c)}$ and $1_{+-}^{(cab)}$ respectively of the centralizer group $\Z_2^{c}\times \Z_2^{ab}$.
These representations are
\begin{equation}\label{eq:Qc1pm charges}
\begin{alignedat}{3}
   &1_{+-}^{(c)}(c)=1\,, \quad   \qquad 
   &1_{+-}^{(c)}(ab)&=-1\,, \\  
   &1_{+-}^{(cab)}(c)=-1\,, \quad   
   &1_{+-}^{(cab)}(ab)&=-1\,, \\  
\end{alignedat}
\end{equation}
The operators $\widetilde{\cO}_c$ and $\wt{\cO}_{cab}$ are exchanged under the action of $ca$. 
Moreover since $(ca)^2=ab$, we may pick the action
\begin{equation}
    ca: (\widetilde{\cO}_c\,,\wt{\cO}_{cab}) \longmapsto 
(i\widetilde{\cO}_{cab}\,,i\wt{\cO}_{c})\,.
\end{equation}
Consider the linear combinations 
\begin{equation}
\wt{\cO}_{[c],\pm}:=\wt{\cO}_c \pm \wt{\cO}_{cab}\,,
\end{equation}
which transform under $ca$ as
\begin{equation}
    ca:\wt{\cO}_{[c],\pm}\longmapsto  \pm i
    \wt{\cO}_{[c],\pm}\,.
\end{equation}
Upon gauging $\Z_4^{ca}$, these charged operators become twisted sector operators under the dual $\Rep(\Z_4^{ca})$ symmetry.
Specifically, $\wt{\cO}_{[c],+}$ is now $c\cdot \rho_1$ twisted while $\wt{\cO}_{[c],-}$ is $c\cdot \rho_3$ twisted.
These two twisted sector operators are exchanged under the $c$ action.
This follows from the fact that the twisted sectors $\rho_1$ and $\rho_3$ are exchanged under $c$ and is compatible with the charge assignment in \eqref{eq:Qc1pm charges}.
Hence we take odd and even combinations of $\wt{\cO}_{[c],\pm}$.
The even combination
\begin{equation}
    \wt{\cO}_{c}\propto \wt{\cO}_{[c],+} + \wt{\cO}_{[c],-}\,,
\end{equation}
is uncharged under $\Z_2^{c}$ and we therefore obtain a $E=\rho_1\oplus \rho_3$ twisted sector operator charged under $Z_{1a}$ upon gauging $\Z_2^c$. 
Instead the odd combination
\begin{equation}
    \wt{\cO}_{cab}\propto \wt{\cO}_{[c],+} - \wt{\cO}_{[c],-}\,,
\end{equation}
is charged under $c$ and therefore we obtain a line changing operator from $E$ to $1_a$ upon gauging $\Z_2^{c}$.
We propose the following lattice representatives for these operators
\begin{equation}
    \Q_{[c], 1_{+-}}=\{\cT_{(E,v_{+})}X^{\I}\,, 
    \cT_{(E,v_{+})\otimes 1_a}X^{\I}X^{\II}X^{\III}\}\,,
\end{equation}
where $v_+=v_1+v_2$ in the $E$ representation space is a $+1$ eigenvector of $\cD_E(c)=X$ and a $-1$ eigenvector of $\cD_E(ab)=-\bbI$.

\item {\bf $\Q_{[a], 1_{++}}$ generalized charge.} As a $D_8$ symmetry multiplet, this corresponds to a doublet of twisted sector operators $\cO_a$ and $\cO_b$, in the $a=c\times ca$ and $b=c\times (ca)^3$ twisted sectors respectively. 
Upon gauging $\Z_4^{ca}$, $\cO_a$ and $\cO_b$ both go to the twisted sector of $c$ and carry charges $(i)^{\ms n}$ and $(-i)^{\ms n}$ under $\rho_{\ms n}$ respectively.
Note that $\cO_a$ and $\cO_{b}$ are exchanged under the action of $c$ and therefore we must take odd and even combinations 
\begin{equation}
    \cO_{[a],\pm}:= \cO_{a}\pm \cO_{b}\,.
\end{equation}
The even and odd combinations are uncharged and charged under  $\Z_2^c$ respectively and therefore go to the untwisted and twisted sectors of $\rho_c\cong 1_a$ upon gauging $\Z_2^{c}$.
We identify these operators as
\begin{equation}
    \Q_{[a], 1_{++}}=\{X^{\II}+X^{\III}\,, 
    \cT_{ 1_a}(X^{\II}-X^{\III})\}\,.
\end{equation}

% \item {\bf $\Q_{[a], 1_{+-}}$ generalized charge.} As a $D_8$ symmetry multiplet this comprises of a doublet of operators $\wt{\cO}_a$ and $\wt{\cO}_b$ in the twisted sectors of $a$ and $b$ respectively.
% %
% $\wt{\cO}_a$ and $\wt{\cO}_b$ transform under representations $1_{+-}^{(a)}$ and $1_{+-}^{(b)}$ of the centralizer symmetry $\Z_2^a\times \Z_2^b$ respectively where 
% \begin{equation}
% \begin{split}
% &1_{+-}^{(a)}(a)=+1\,, \qquad 1_{+-}^{(a)}(b)=-1 \\   
% &1_{+-}^{(b)}(a)=-1\,, \qquad 1_{+-}^{(b)}(b)=+1\,.   
% \end{split}
% \end{equation}
% While under $ca$,
% \begin{equation}
%     ca:(\wt{\cO}_a\,, \wt{\cO}_b) \longmapsto 
%     (i\wt{\cO}_b\,, i\wt{\cO}_a)\,.
% \end{equation}
% Therefore $\wt{\cO}_{[a],\pm}:=\wt{\cO}_a \pm \wt{\cO}_b$ map under $ca$ as
% \begin{equation}
%     ca: \wt{\cO}_{[a],\pm} \longmapsto \pm i \wt{\cO}_{[a],\pm}\,.
% \end{equation}

\end{itemize}

\subsection{Solvable Hamiltonians for  Gapped Phases}
\noindent Gapped phases for $\Rep(D_8)$ symmetry \cite{Thorngren:2019iar,Bhardwaj:2024qrf} 
are classified by tuples $(F,\beta)$  where $F$ is a representative of a conjugacy class of subgroups of $D_8$ and $\beta$ is an element of the group cohomology $H^2(F,U(1))$ \cite{Ostrikmodule}.
A set of possible choices for $F$ is:
\be \label{eq:F_choices}
    F\in \{1\,, \Z_2^a\,, \Z_2^{ab}\,, \Z_2^c\,, \Z_4^{ca}\,, \Z_2^a\times\Z_2^b\,, \Z_2^c\times\Z_2^{ab}\,, D_8\}.
\ee
$\beta$ can be non-trivial only when $F\simeq\Z_2\times\Z_2$ or $D_8$, for which $\beta\in H^2(F,U(1))=\Z_2$. The fixed-point Hamiltonian for a gapped phase labeled as $(F,\beta)$ can be written as:
\begin{equation} \label{eq:RepD8_Ham}
   H_{(F,\beta)}=-\tfrac{1}{|F|}\sum_i\sum_{\ms f\in F}(R_\beta^{\ms f^{-1}})_{i}(L_\beta^{\ms f})_{i+1} -\sum_i P^{(F)}_{i}\,.
\end{equation}
Here $ P^{(F)}_{i}$ is an operator diagonal in the qubit or group basis which projects onto the sub-Hilbert space at site $i$ spanned by $|\ms f\rangle\text{ for }\ms f\in F$
\begin{equation}
     P^{(F)}_i=\sum_{\ms f\in F}|\ms f_i\rangle \langle \ms f_i|\,.
\end{equation}
These projectors have simple expressions in terms of Pauli operators $Z^{\alpha}_{i}$ that we will use in constructing concrete Hamiltonians with each phase implementable on atomic array setups. 
Meanwhile the operators $R^{\ms f^{-1}}_\beta$ and $L^{\ms f}_{\beta}$ serve as disordering operators within the projected subspace.
In the case where $\beta $ is trivial, these reduce to the usual left and right group multiplication operators defined in \eqref{eq:Loperators}.
Instead when $\beta$ is non-trivial, the operator action gets modified by a $U(1)$ phase controlled by the 2-cocycle $\beta$. Concretely,
\begin{equation}\label{eq:beta_on_ket}
    \begin{split}
         R^{\ms f^{-1}}_\beta|\ms f'\rangle &= \tfrac{1}{\beta(\ms f'\ms f^{-1}\,, \ms f)}|\ms f'\ms f^{-1}\rangle\,, \\
         L^{\ms f}_\beta|\ms f'\rangle &= {\beta(\ms f\,, \ms f')}|\ms f\ms f'\rangle\,, \\
    \end{split}
\end{equation}
The Hamiltonians \eqref{eq:RepD8_Ham}, have been normalized such that ground states of all phases have energy $-2L$ where $L$ is the number of lattice sites. {One can write equivalent Hamiltonians by summing only over generators of the subgroup $F$ and their inverses: since any element in $F$ can be written as a product of such generators, a state that minimizes the simplified Hamiltonian will also minimize the full one and vice-versa.}

\medskip \noindent {\bf Twisted sectors.} Typically when probing the characterization of phases with a certain symmetry $\mathcal S$, one also requires state spaces twisted by $\mathcal S$ action.
In the case of invertible symmetries, for example $\mathbb Z_2$, the symmetry twisted state space is  nothing but the familar space with antiperiodic boundary conditions.
Instead, state spaces twisted by non-invertible symmetries are generically not isomorphic to untwisted state spaces.
For instance twisting by a symmetry operator $\Gamma \in \Rep(G)$ at a link $(i,i+1)$ on the lattice involves inserting a defect at $(i,i+1)$ whose associated state space is the vector space underlying the representation $\Gamma$ which is $\mathbb C^{\rm dim(\Gamma)}$. 
For invertible representations which are one dimensional, this reduces to the familiar group-like case since the action on $\mathbb C$ is by multiplication of a complex number that can be absorbed into a coupling constant of the lattice model.
Concretely, the presence of a $\Gamma$ defect at link $(i_0,i_0+1)$ modifies the disordering term in \eqref{eq:RepD8_Ham} such that 
\begin{equation}
    (R_\beta^{\ms g^{-1}})_{i_0}(L_\beta^{\ms g})_{i_0+1}\,,
\end{equation}
is replaced with
\begin{equation}
    (R_\beta^{\ms g^{-1}})_{i_0}
    \lb\mathcal D_{\Gamma}(\ms g)\rb_{(i_0,i_0+1)}
    (L_\beta^{\ms g})_{i_0+1}
\end{equation}
where $\lb\cD_{\Gamma}(\ms g)\rb_{(i_0,i_0+1)}$ is an operator acting on the defect state space.

The symmetry operators act on the states as follows
\be \label{eq:SGamma}
\cS_{\Gamma}\ket{\Psi_{\ms g}}=\Tr(\cD_\Gamma(\ms g))\ket{\Psi_{\ms g}} \,,
\ee
where ${\ms g}$ is the `holonomy', i.e. the product of all the group elements on the lattice sites.

\subsection{Spin-Chain Realization of all $\Rep(D_8)$ Gapped Phases}
\noindent In this section, we provide a concrete qubit-based realization of commuting projector Hamiltonians within all $\Rep(D_8)$ symmetric gapped phases. 
These Hamiltonians have the form \eqref{eq:RepD8_Ham} for different choices of $(F,\beta)$.
\subsubsection{$\Rep(D_8)$ trivial phase for $F=1$} \label{sec:1}
Setting $F=1$ in eq. \eqref{eq:RepD8_Ham}, the Hamiltonian is:
\begin{equation}
\label{bambi}
\ba
    H_{1} &=-\sum_i\lbb\bbI+ P^{(1)}\rbb_{i} \\
    % = -\sum_{i}\left[\bbI+\tfrac{1}{8}\prod_{\alpha}(\bbI+Z_i^{\alpha})\right]\cr 
     &=-\sum_i\lbb\bbI+\tfrac{1}{8}(\bbI+ Z^{\I})(\bbI+ Z^{\II})(\bbI+ Z^{\III})\rbb_{i}
    \ea \,.
\end{equation}
It has a single ground state (which we write first in the $\Z_2^3$ basis and then the basis labeled by $g\in D_8$):
\begin{equation}
    \ket{\GS}= \bigotimes_i\ket{000}_{i}=\bigotimes_i\ket{1}_{i}\,.
\end{equation}
The $\Rep(D_8)$ symmetry \eqref{eq:SGamma} acts as  
\begin{equation}
    \cS_{\Gamma}\ket{\GS} ={\rm dim}(\Gamma)\ket{\GS}\,. 
\end{equation}
This is a trivial SPT phase for $\Rep(D_8)$ symmetry.

\subsubsection{$\Z_2$ SSB phase for $F=\Z_2^a$} \label{sec:Z2a}
Setting $F=\Z_2^a$ in eq. \eqref{eq:RepD8_Ham}, the Hamiltonian is:
\be
\ba
   H_{\Z_2^a}=&-\tfrac{1}{2}\sum_i\lb\bbI_{i}\bbI_{i+1}+R^{a}_{i}L^{a}_{i+1}\rb -\sum_i P^{(\Z_2^{a})}_{i}
   \\
   =&-\tfrac{1}{4}\sum_i\left[(\bbI+Z^\I)\, X^{\II}+(\bbI-Z^\I)\, X^{\III}\right]_i X^\II_{i+1}\\
    &-\tfrac{1}{2}\sum_i \bbI_{i}\bbI_{i+1}-\tfrac{1}{4}\sum_i\lbb(\bbI+ Z^{\I})(\bbI+ Z^{\III})\rbb_{i}\\
    %\approx & -\tfrac{1}{2}\sum_i\lb\bbI_{i}\bbI_{i+1}+X^{\II}_{i}X^{\II}_{i+1}\rb -\tfrac{1}{4}\sum_i\lbb(\bbI+ Z^{\I})(\bbI+ Z^{\III})\rbb_{i}\,.
\ea
\ee
The last term projects onto  $\Z_2^a=\{\ket{000}=\ket{1},\;\ket{010}=\ket{a}\}$ on each site $i$;
the first term, once projected to this subspace, is minimized by the $ X^{\II}$ eigenstates, there are therefore 2 linearly independent ground states:
\be \label{eq:GS_Z2a}
\ba
\ket{\GS,+}&= 
\bigotimes_i\tfrac{1}{\sqrt{2}}\left(\ket{000}+\ket{010}\right)_{i}\\
&=\bigotimes_i\tfrac{1}{\sqrt{2}}\left(\ket{1}+\ket{a}\right)_{i}\equiv \ket{\Psi_1}+\ket{\Psi_a}\,,\\
\ket{\GS,-}&=\bigotimes_i\tfrac{1}{\sqrt{2}}\left(\ket{000}-\ket{010}\right)_{i}\\
&=\bigotimes_i\tfrac{1}{\sqrt{2}}\left(\ket{1}-\ket{a}\right)_{i}\equiv \ket{\Psi_1}-\ket{\Psi_a}\,.
\ea
\ee
The action of the $\Rep(D_8)$ generators \eqref{eq:SGamma} on the ground states \eqref{eq:GS_Z2a} is:
\be
\ba
    \cS_1\ket{\GS,\pm}=\;\cS_{1_a}\ket{\GS,\pm}&=\ket{\GS,\pm}\,,\\
    \cS_{1_c}\ket{\GS,\pm}=\cS_{1_{ca}}\ket{\GS,\pm}&=\ket{\GS,\mp}\,,\\
    \cS_{E}\ket{\GS,\pm}&=\ket{\GS,+}+\ket{\GS,-}\,.
\ea
\ee
This is therefore a $\Z_2$ SSB phase for $\Rep(D_8)$: the $\Z_2$ symmetries exchanging the two ground states are $\cS_{1_c}$ and $\cS_{1_{ca}}$. \\
We note that $\cS_{E}$ sends each ground state to the sum of both: this a hallmark of non-invertible symmetries.

\noindent{\bf Order parameter.}
The local order parameter for this phase is the generalized charge $\Q_{([a], 1_{++})}$, which is realized on the lattice by the 
operator $X^{\II}_{i}$:
\be
    \bra{\GS,\pm}X^{\II}_{i}\ket{\GS,\pm} =\pm 1\,.
%     \cr  \bra{\GS,-}X^{\II}_{i}\ket{\GS,-}&=-1\,.
% \ea\ee
\ee

\subsubsection{$\Rep(D_8)/(\Z_2\times\Z_2)$ SSB phase for $F=\Z_2^{ab}$} \label{sec:Z2ab}
Setting $F=\Z_2^{ab}$ in  \eqref{eq:RepD8_Ham}, the Hamiltonian is:
\be
\ba
    H_{\Z_2^{ab}}=&-\tfrac{1}{2}\sum_i\lbb\bbI_i\bbI_{i+1}+R^{ab}_{i}L^{ab}_{i+1}\rbb -\sum_i P^{(\Z_2^{ab})}_{i}=\\
   =&-\tfrac{1}{2}\sum_i\left[\bbI_i\bbI_{i+1}+( X^{\II} X^{\III})_{i}(X^{\II} X^{\III})_{i+1}\right] \cr 
   &-\tfrac{1}{4}\sum_i\lbb(\bbI+ Z^{\I})(\bbI+ Z^{\II} Z^{\III})\rbb_{i}\,.
\ea
\ee
The second term projects onto  $\Z_2^{ab}=\{\ket{000}=\ket{1},\;\ket{011}=\ket{ab}\}$ on each site $i\in\{1,...,L\}$;
the first term is minimized by the $ X^{\II} X^{\III}$ eigenstates, there are therefore 2 linearly independent ground states:
\be \label{eq:GS_Z2ab}
\ba
\ket{\GS,+}&= 
\bigotimes_i\tfrac{1}{\sqrt{2}}\left(\ket{000}+\ket{011}\right)_{i}\\
&=\bigotimes_i\tfrac{1}{\sqrt{2}}\left(\ket{1}+\ket{ab}\right)_{i}\equiv \ket{\Psi_1}+\ket{\Psi_{ab}}\,,\\
\ket{\GS,-}&=\bigotimes_i\tfrac{1}{\sqrt{2}}\left(\ket{000}-\ket{011}\right)_{i}\\
&=\bigotimes_i\tfrac{1}{\sqrt{2}}\left(\ket{1}-\ket{ab}\right)_{i}\equiv \ket{\Psi_1}-\ket{\Psi_{ab}}\,.
\ea
\ee
Up to now, this discussion appears very similar to the previous case. However, once we take the $\Rep(D_8)$ symmetry into account, the phases are of different types. Indeed, the action of the $\Rep(D_8)$ generators on the ground states \eqref{eq:GS_Z2ab} is:
\be
\ba
\cS_\alpha\ket{\GS,\pm}&
=\ket{\GS,\pm}\,, \ \alpha= 1, 1_a, 1_c, 1_{ca}\\
    \cS_{E}\ket{\GS,\pm}&=2\ket{\GS,\mp}\,.
\ea
\ee
In this phase the symmetry generated by the 2-dimensional irreducible  representation $E$ is spontaneously broken: $\cS_E$ exchanges the ground states and also multiplies them by a factor of 2. Unlike previous case, the full $\Z_2\times\Z_2$ subsymmetry of $\Rep(D_8)$ is preserved, so this is a $\Rep(D_8)/(\Z_2\times\Z_2)$ SSB phase. 

\noindent{\bf Order parameter.}
The local order parameter for this phase comes the generalized charge $\Q_{[ab], 1}$ and is given by the operator $(X^{\II} X^{\III})_{i}$: 
\be
    \bra{\GS,\pm}(X^{\II} X^{\III})_{i}\ket{\GS,\pm}=\pm1 \,.
    % \,,\qquad \bra{\GS,-}(X^{\II} X^{\III})_{i}\ket{\GS,-}=-1\,.
\ee

\subsubsection{$\Z_2\times\Z_2$ SSB phase for $(F)^\beta=(\Z_2^a\times\Z_2^b)^+$} \label{sec:Z2aZ2b}
By choosing $F=\Z_2^a\times\Z_2^b$ and trivial $\beta$ in eq. \eqref{eq:RepD8_Ham}, we obtain the following effective Hamiltonian
\be\ba \label{eq:Ham_Z2aZ2b}
   H_{(\Z_2^a\times\Z_2^b)^+}\simeq
   &-\tfrac{1}{4}\sum_i (\bbI+Z^\I)_i\, (X^{\II}_iX^\II_{i+1}+X^{\III}_iX^\III_{i+1})\\
   &-\tfrac{1}{4}\sum_i(\bbI-Z^\I)_i(X^{\III}_iX^\II_{i+1}+X^{\II}_iX^\III_{i+1})\\
   &-\tfrac{1}{2}\sum_i[\bbI+Z^\I]_i
   %&-\tfrac{1}{2}\sum_i\lbb R^{a}_{i}L^{a}_{i+1}+R^{b}_{i}R^{b}_{i+1}\rbb -\sum_i P^{(\Z_2^{a}\times\Z_2^{b})}_{i} \cr 
   %=&-\tfrac{1}{4}\sum_i \left[(\bbI+Z^\I)\, X^{\II}+(\bbI-Z^\I)\, X^{\III}\right]_i X^\II_{i+1}\\
   %&-\tfrac{1}{4}\sum_i \left[(\bbI+Z^\I)\, X^{\III}+(\bbI-Z^\I)\, X^{\II}\right]_i X^\III_{i+1}\\
   %&-\tfrac{1}{2}\sum_i(\bbI_i+Z^\I_i)\\
    %\approx& -\tfrac{1}{2}\sum_i\lbb X^{\II}_{i}X^{\II}_{i+1}+X^{\III}_{i}X^{\III}_{i+1}\rbb \cr &-\tfrac{1}{2}\sum_i\left[\bbI+ Z^{\I}\right]_{i}\,.
\ea\ee
The projector term is minimized, on each site $i\in\{1,...,L\}$ by the states corresponding to the elements in $\Z_2^a\times\Z_2^{b}= \{\ket{000}=\ket{1},\;\ket{010}=\ket{a},\;\ket{001}=\ket{b},\;\ket{011}=\ket{ab}\}$. 
The remaining terms are minimized by the $ X^{\II}$ and $X^{\III}$ eigenstates independently, giving rise to 4 linearly independent ground states:
\be \label{eq:GS_Z2aZ2b}
\ba
\ket{\GS,++}&=\bigotimes_i\tfrac{1}{2}\left(\ket{1}+\ket{a}+\ket{b}+\ket{ab}\right)_{i}\\[-1mm]
&\equiv \ket{\Psi_1}+\ket{\Psi_{a}}+\ket{\Psi_{b}}+\ket{\Psi_{ab}}\,,\\ 
\ket{\GS,+-}&=\bigotimes_i\tfrac{1}{2}\left(\ket{1}+\ket{a}-\ket{b}-\ket{ab}\right)_{i}\\[-1mm]
&\equiv 
\ket{\Psi_1}+\ket{\Psi_{a}}-\ket{\Psi_{b}}-\ket{\Psi_{ab}}\,,\\
\ket{\GS,-+}&=\bigotimes_i\tfrac{1}{2}\left(\ket{1}-\ket{a}+\ket{b}-\ket{ab}\right)_{i}\\[-1mm] 
&\equiv 
\ket{\Psi_1}-\ket{\Psi_{a}}+\ket{\Psi_{b}}-\ket{\Psi_{ab}}\,,\\
\ket{\GS,--}&=\bigotimes_i\tfrac{1}{2}\left(\ket{1}-\ket{a}-\ket{b}+\ket{ab}\right)_{i}\\[-1mm] 
&\equiv \ket{\Psi_1}-\ket{\Psi_{a}}-\ket{\Psi_{b}}+\ket{\Psi_{ab}}\,.
\ea
\ee
From the character table \ref{tab:D8_chars}, we deduce the following action of $\Rep(D_8)$ generators on the ground states:
\be
\ba
\cS_1\ket{\GS,s,s'}&=\cS_{1_a}\ket{\GS,s,s'}=\ket{\GS,s,s'}, \\
\cS_{1_c}\ket{\GS,s,s'}&=\cS_{1_{ca}}\ket{\GS,s,s'}=\ket{\GS,-s,-s'}\\
    \cS_{E}\ket{\GS,++}&=\cS_{E}\ket{\GS,--}=\ket{\GS,+-}+\ket{\GS,-+}\,,\\
    \cS_{E}\ket{\GS,+-}&=\cS_{E}\ket{\GS,-+}=\ket{\GS,++}+\ket{\GS,--}\,,
\ea
\ee
for all $s, s'=\pm$.
The 4 ground states form 2 different $\Z_2$ orbits under the broken  generators $\cS_{1_c}$ and $\cS_{1_{ca}}$:
\be
    \cS_{1_c},\;\cS_{1_{ca}}:\quad 
    \ba
    &\ket{\GS,++}\leftrightarrow\ket{\GS,--}\cr &\ket{\GS,+-}\leftrightarrow\ket{\GS,-+}
    \ea
\ee
this is therefore a $\Z_2\times\Z_2$ SSB phase for $\Rep(D_8)$ symmetry. The local order parameters come from the generalized charges $\Q_{[ab], 1}$ and $2 \Q_{[a], 1_{+,+}}$, which are realized on the lattice by the operators
$X^{\II}_{i},\; X^{\III}_{i},\; (X^{\II} X^{\III})_{i}$
which have the following eigenvalues in the ground states:
\be
\ba
\bra{\GS,s,s'}X^{\II}_{i}\ket{\GS,s,s'}&=s,\\
\bra{\GS,s,s'}X^{\III}_{i}\ket{\GS,s,s'}&=s',\\
    \bra{\GS,s,s'}(X^{\II} X^{\III})_{i}\ket{\GS,s,s'}&=ss'\,.
\ea
\ee

\subsubsection{Non-trivial SPT phase for $(F)^\beta=(\Z_2^a\times\Z_2^b)^-$} \label{sec:Z2aZ2b-}
We now consider the same subgroup $F=\Z_2^a\times\Z_2^b$ as the previous case, but now with the non-identity
\be
    \beta\in H^2(\Z_2^a\times\Z_2^b,U(1))=\Z_2\,.
\ee
As a representative of $\beta$, we choose:
\be \label{eq:betaZ2Z2}
\ba
    \beta((p_1,q_1),(p_2,q_2))=(-1)^{p_1q_2}\,.
\ea
\ee
From the general discussion below equation \eqref{eq:RepD8_Ham}, this choice of $\beta$ is realized on the lattice by introducing the operators $Z^{\II},\,Z^{\III}$ in expression \eqref{eq:Ham_Z2aZ2b} as follows:
\be\ba \label{eq:Ham_Z2aZ2b_beta}
   &H_{(\Z_2^a\times\Z_2^b)^-}\simeq\\
    & -\tfrac{1}{4}\sum_i (\bbI+Z^\I)_i\, \lbb X^{\II}_{i}(X^{\II}Z^{\III})_{i+1}+(Z^{\II}X^{\III})_{i}X^{\III}_{i+1}\rbb\\
   &-\tfrac{1}{4}\sum_i(\bbI-Z^\I)_i(X^{\III}_iX^\II_{i+1}+(Z^\III X^{\II})_iX^\III_{i+1})\\
   &-\tfrac{1}{2}\sum_i(\bbI_i+Z^\I_i)\approx\\
     &-\tfrac{1}{2}\sum_i\lbb X^{\II}_{i}(X^{\II}Z^{\III})_{i+1}+(Z^{\II}X^{\III})_{i}X^{\III}_{i+1}+(\bbI_i+ Z^{\I}_i)\rbb\,.
\ea\ee
We note that, although $X^\alpha$ anti-commutes with $Z^\alpha$ for fixed $\alpha\in\{\I,\II,\III\}$, all the terms in square brackets commute with each other hence they can be minimized independently. Like for the previous case, the projector term is minimized for the group elements in $\Z_2^a\times\Z_2^{b}= \{\ket{000}=\ket{1},\;\ket{010}=\ket{a},\;\ket{001}=\ket{b},\;\ket{011}=\ket{ab}\}$.
A state $\ket{\GS}$ minimizing the remaining terms must furthermore satisfy:
\be \label{eq:conditions_ab}
\ba
    X^{\II}_{i}(X^{\II}Z^{\III})_{i+1}\ket{\GS}&=\ket{\GS}, \quad \forall\, i\in\{1,...,L\}\,, \\
    (Z^{\II}X^{\III})_{i}X^{\III}_{i+1}\ket{\GS}&=\ket{\GS}, \quad \forall i\in\{1,...,L\}\,.
\ea
\ee
The ground state is unique because the above are $2L$ independent equations for the $2L$ dimensional Hilbert space of qubits $\II$ and $\III$ on each lattice site (bit $\I$ is already fixed to 0). Furthermore, by taking the product over lattice sites $i$ of the constraints \eqref{eq:conditions_ab}, one has:
\be \label{eq:GS_SPT1_hol1}
\ba
     \prod_iZ^{\III}_{i}\ket{\GS}=\prod_iZ^{\II}_{i}\ket{\GS}=\ket{\GS}, \\
\ea
\ee
which implies that the holonomy of the ground state must be the identity. Since the terms in the Hamiltonian mutually commute, we can write the ground state as:
\be \label{eq:GS_non-triv_SPT}
\ba
    \ket{\GS}&=\prod_i\tfrac{1}{\sqrt{2}}\lbb\bbI_i\bbI_{i+1}+(Z^{\II}X^{\III})_{i}X^{\III}_{i+1}\rbb\bigotimes_i\ket{0,+,0}_{i}\,.
\ea
\ee
where $\ket{0,+,0}\equiv\tfrac{1}{\sqrt{2}}(\ket{000}+\ket{010})$ is a state that satisfies the first condition in \eqref{eq:conditions_ab}, whereas the second condition holds thanks to the operator acting on $\ket{0,+,0}$. 

We can write the symmetry generators $\cS_\Gamma$, for $\Gamma\in \Rep(D_8)$, on the $\Z_2^a\times\Z_2^b$ subspace as follows: 
\be
    \ba
\cS_1=\cS_{1_a}|_{\Z_2^a\times\Z_2^b}&=\prod_{i=1}^{L} \bbI_{i} \cr 
\cS_{1_c}=\cS_{1_{ca}}|_{\Z_2^a\times\Z_2^b} &=\prod_{i=1}^{L} (Z^{\II}_{i} Z^{\III}_{i}) \cr 
       \cS_E|_{\Z_2^a\times\Z_2^b}&=\prod_{i=1}^{L}  Z^{\II}_{i}+\prod_{i=1}^{L} Z^{\III}_{i} \,.
    \ea
\ee
Recalling eq. \eqref{eq:GS_SPT1_hol1}, we therefore see that: 
\be
    \cS_\Gamma\ket{\GS}=
    \dim(\Gamma)\ket{\GS}\,,
\ee
and this is a non-trivial $\Rep(D_8)$ SPT phase. 

\medskip\noindent{\bf String order parameters.} The SPT phase cannot be diagnosed by genuine local order parameters.
Instead we either require twisted sector operators or string order parameters that that are essentially the product of two twisted sector operators separated by a finite (and typically large) distance.
The twisted sector operators
map between ground states in different twisted sectors.
In the present work, for simplicity we restrict ourselves to untwisted sector ground states, which can be straightforwardly extended to twisted sectors using the approach in \cite{Bhardwaj:2024kvy}.
We may however detect the SPT using string order parameters \cite{Perez-Garcia:2008hhq, Pollmann:2012xdn}.
These are 
\be
\ba
\Q_{([1],1_a)}^{(i_0,i_0+N)}&=    \prod_{i=i_0}^{i_{0}+N}Z^{\I}_i \,, \\
\Q_{2([a],1_{+-})}^{(i_0,i_0+N)}&=
\begin{cases}
    X^\II_{i_0} ( \prod_{k=1}^{N-1} Z^{\III}_{i0+k} ) Z^{\III}_{i0+N} X^{\II}_{i0+N}\,, \\[2mm]
     (Z^{\II} X^{\III})_{i_0} (\prod_{k=1}^{N-1}   Z^{\II}_{i0+k}  )  X^{\III}_{i0+N}\,, 
\end{cases}\\
\Q_{([ab],1_c)}^{(i_0,i_0+N)}&=
     X^{\II}_{i0} X^{\III}_{i0+1} ( \prod_{k=1}^{N} (Z^{\II} Z^{\III})_{i0+k} ) X^{\II}_{i0+N} X^{\III}_{i0+N+1}\,,\\
\Q_{([ab],1_{ca})}^{(i_0,i_0+N)}&=\Q_{([1],1_a)}^{(i_0,i_0+N)}\Q_{([ab],1_c)}^{(i_0,i_0+N)}
\ea
\ee
%\Q_{([ab],1_c)}^{(i_0,i_0+N)}&=    \prod_{i=i_0}^{i_{0}+N}(X^{\II}X^{\III})_{i_0}(Z^{\II}_iZ^{\III}_i)(X^{\II}X^{\III})_{i_0+N}

which correspond to the condensed charges $([1],1_a),2([a],1_{+-}),([ab],1_c)$ and $([ab],1_{ca})$.
All these have a unit expectation value on the fixed-point SPT ground state.

\subsubsection{$\Z_2$ SSB phase for $F=\Z_2^c$} \label{sec:Z2c}
Setting $F=\Z_2^c$ in eq. \eqref{eq:RepD8_Ham} the Hamiltonian is:
\be\ba
   &H_{\Z_2^c}=-\tfrac{1}{2}\sum_i\lb\bbI+R^{c}_{i}L^{c}_{i+1}\rb -\sum_i P^{(\Z_2^{c})}_{i+1}=\\
  &  -\tfrac{1}{4}\sum_i X^\I_i\left[(\bbI+Z^\II Z^\III)\, X^{\I}+(\bbI-Z^\II Z^\III)\, X^{\I}X^\II X^\III\right]_{i+1}\\
  &-\tfrac{1}{2}\sum_i\bbI_{i}\bbI_{i+1}  -\tfrac{1}{4}\sum_i\lbb(\bbI+ Z^{\II})(\bbI+ Z^{\III})\rbb_{i}\,.
   %\approx &-\tfrac{1}{2}\sum_i\lb\bbI_{i}\bbI_{i+1}+X^{\I}_{i}X^{\I}_{i+1}\rb \cr &  -\tfrac{1}{4}\sum_i\lbb(\bbI+ Z^{\II})(\bbI+ Z^{\III})\rbb_{i}\,.
\ea\ee
The second term projects onto  $\Z_2^c=\{\ket{000}=\ket{1},\;\ket{100}=\ket{c}\}$ on each site $i$;
the first term on this subspace is then minimized by the $ X^{\I}$ eigenstates, there are therefore 2 linearly independent ground states:
\be \label{eq:GS_Z2c}
\ba
\ket{\GS,+}&= 
\bigotimes_i\tfrac{1}{\sqrt{2}}\left(\ket{000}+\ket{100}\right)_{i}\\
&=\bigotimes_i\tfrac{1}{\sqrt{2}}\left(\ket{1}+\ket{c}\right)_{i}\equiv \ket{\Psi_1}+\ket{\Psi_c}\,,\\
\ket{\GS,-}&=\bigotimes_i\tfrac{1}{\sqrt{2}}\left(\ket{000}-\ket{100}\right)_{i}\\
&=\bigotimes_i\tfrac{1}{\sqrt{2}}\left(\ket{1}-\ket{c}\right)_{i}\equiv \ket{\Psi_1}-\ket{\Psi_c}\,.
\ea
\ee
The action of the $\Rep(D_8)$ generators on the ground states \eqref{eq:GS_Z2c} is:
\be
\ba \cS_1\ket{\GS,\pm}=\;\cS_{1_c}\ket{\GS,\pm}&=\ket{\GS,\pm}\,,\\
    \cS_{1_a}\ket{\GS,\pm}=\cS_{1_{ca}}\ket{\GS,\pm}&=\ket{\GS,\mp}\,,\\
    \cS_{E}\ket{\GS,\pm}&=\ket{\GS,+}+\ket{\GS,-}\,.
\ea
\ee
The operators exchanging the ground states are $\cS_{1_a}$ and $\cS_{1_{ca}}$: this is therefore a $\Z_2$ SSB phase for $\Rep(D_8)$,  very similar to the one discussed in section \ref{sec:Z2a} (if we exchange $a$ and $c$). Also in this phase $\cS_{E}$ sends each ground state to the sum of both, which is a hallmark of non-invertible symmetries. \\
The local order parameter for this phase is  $\Q_{[c], 1_{++}}$, given by the operator $X^{\I}_{i}$: 
\be
    \bra{\GS,\pm}X^{\I}_{i}\ket{\GS,\pm}=\pm1\,.
\ee

\subsubsection{$\Z_2\times\Z_2$ SSB phase for $(F)^\beta=(\Z_2^c\times\Z_2^{ab})^+$} \label{sec:Z2cZ2ab}
By choosing $F=\Z_2^c\times\Z_2^{ab}$ and trivial $\beta$ in eq. \eqref{eq:RepD8_Ham}, we obtain the following effective Hamiltonian:
\be\ba \label{eq:Ham_Z2cZ2ab}
   &H_{(\Z_2^c\times\Z_2^{ab})^+}\simeq-\tfrac{1}{2}\sum_i \lbb R^{c}_{i}L^{c}_{i+1}
   +R^{ab}_{i}R^{ab}_{i+1}\rbb -\sum_i P^{(\Z_2^{c}\times\Z_2^{ab})}_{i}=\\
    &-\tfrac{1}{4}\sum_i X^\I_i\left[(\bbI+Z^\II Z^\III)\, X^{\I}+(\bbI-Z^\II Z^\III)\, X^{\I}X^\II X^\III\right]_{i+1}\\
    &-\tfrac{1}{2}\sum_i\left[( X^{\II} X^{\III})_{i}(X^{\II} X^{\III})_{i+1}+(\bbI_i+ Z^{\II}_iZ^{\III}_i)\right]
    %\approx& -\tfrac{1}{2}\sum_i\lbb X^{\I}_{i}X^{\I}_{i+1}+ (X^{\II}X^{\III})_{i}(X^{\II}X^{\III})_{i+1}\rbb \cr 
    %& -\tfrac{1}{2}\sum_i\left[\bbI+ Z^{\II}Z^{\III}\right]_{i}\,.
\ea\ee
The projector term is minimized, on each site $i$, by the states corresponding to the elements in $\Z_2^c\times\Z_2^{ab}=\{\ket{000}=\ket{1},\, \ket{100}=\ket{c},\,\ket{011}=\ket{ab},\, \ket{111}=\ket{cab}\}$.
The remaining terms are minimized by the $ X^{\I}$ and $X^{\II}X^{\III}$ eigenstates independently, giving rise to 4 linearly independent ground states:
\be \label{eq:GS_Z2cZ2ab}
\ba
\ket{\GS,++}&=\bigotimes_i\tfrac{1}{2}\left(\ket{1}+\ket{c}+\ket{cab}+\ket{ab}\right)_{i}\\[-1mm]
&\equiv \ket{\Psi_1}+\ket{\Psi_{c}}+\ket{\Psi_{cab}}+\ket{\Psi_{ab}}\,,\\ 
\ket{\GS,+-}&=\bigotimes_i\tfrac{1}{2}\left(\ket{1}+\ket{c}-\ket{cab}-\ket{ab}\right)_{i}\\[-1mm]
&\equiv 
\ket{\Psi_1}+\ket{\Psi_{c}}-\ket{\Psi_{cab}}-\ket{\Psi_{ab}}\,,\\
\ket{\GS,-+}&=\bigotimes_i\tfrac{1}{2}\left(\ket{1}-\ket{c}+\ket{cab}-\ket{ab}\right)_{i}\\[-1mm] 
&\equiv 
\ket{\Psi_1}-\ket{\Psi_{c}}+\ket{\Psi_{cab}}-\ket{\Psi_{ab}}\,,\\
\ket{\GS,--}&=\bigotimes_i\tfrac{1}{2}\left(\ket{1}-\ket{c}-\ket{cab}+\ket{ab}\right)_{i}\\[-1mm] 
&\equiv \ket{\Psi_1}-\ket{\Psi_{c}}-\ket{\Psi_{cab}}+\ket{\Psi_{ab}}\,.
\ea
\ee
The action of $\Rep(D_8)$ generators on the ground states is 
\be
\ba
\cS_1\ket{\GS,s,s'}&=\cS_{1_c}\ket{\GS,s,s'}=\ket{\GS,s,s'} \cr    \cS_{1_a}\ket{\GS,s,s'}&=\cS_{1_{ca}}\ket{\GS,s,s'}
=\ket{\GS,-s,-s'} \cr 
    \cS_{E}\ket{\GS,++}&=\cS_{E}\ket{\GS,--}=\ket{\GS,+-}+\ket{\GS,-+}\cr 
    \cS_{E}\ket{\GS,+-}&=\cS_{E}\ket{\GS,-+}=\ket{\GS,++}+\ket{\GS,--}\,.
\ea
\ee
The 4 ground states form 2 different $\Z_2$ orbits under the broken  generators $\cS_{1_a}$ and $\cS_{1_{ca}}$:
\be
    \cS_{1_a},\;\cS_{1_{ca}}:\quad 
    \ba & \ket{\GS,++}\leftrightarrow\ket{\GS,--}\cr 
    & \ket{\GS,+-}\leftrightarrow\ket{\GS,-+},
\ea
\ee
this is therefore a $\Z_2\times\Z_2$ SSB phase for $\Rep(D_8)$ symmetry. We note that this phase is similar to the $\Z_2\times\Z_2$ SSB phase discussed in section 
\ref{sec:Z2aZ2b}. The local order parameters are the operators
$\Q_{[ab], 1}$ and $2 \Q_{[c], 1_{++}}$. On the lattice they are realized by 
$X^{\I}_{i},\;(X^{\I}X^{\II} X^{\III})_{i},\; (X^{\II} X^{\III})_{i},\; $
which have the following eigenvalues in the ground states:
\be
\ba
    \bra{\GS,s,s'}X^{\I}_{i}\ket{\GS,s,s'}&=s,\\
    \bra{\GS,s,s'}(X^{\I}X^{\II} X^{\III})_{i}\ket{\GS,s,s'}&=s',\\
    \bra{\GS,s,s'}(X^{\II} X^{\III})_{i}\ket{\GS,s,s'}&=ss'.
\ea
\ee

\subsubsection{Non-trivial SPT phase for $(F)^\beta=(\Z_2^c\times\Z_2^{ab})^-$} \label{sec:Z2cZ2ab-}
We now consider the same subgroup $F=\Z_2^c\times\Z_2^{ab}$ as the previous case, but now with the non-identity
\be
    \beta\in H^2(\Z_2^c\times\Z_2^{ab},U(1))=\Z_2\,,
\ee
for whom we chose a representative given by equation \eqref{eq:betaZ2Z2}. Similarly to the case discussed in section \ref{sec:Z2aZ2b-}, the non-trivial $\beta$ implies that we modify the Hamiltonian \eqref{eq:Ham_Z2cZ2ab}, by introducing $Z$ operators as follows:
\be\ba \label{eq:Ham_Z2cZ2ab_beta}
   &H_{(\Z_2^c\times\Z_2^{ab})^-}\simeq \\
   &-\tfrac{1}{4}\sum_i X^\I_i\left[(\bbI+Z^\II Z^\III)\, X^{\I}+(\bbI-Z^\II Z^\III)\, X^{\I}X^\II X^\III\right]_{i+1}\\[-3mm]
   &\hspace{12mm}\times\tfrac{1}{2}[\bbI+Z^\II+Z^\III-Z^\II Z^\III]_{i+1}\\
    &-\tfrac{1}{2}\sum_i\left[( Z^\I X^{\II} X^{\III})_{i}(X^{\II} X^{\III})_{i+1}+(\bbI_i+ Z^{\II}_iZ^{\III}_i)\right]
    \approx \\
    &-\tfrac{1}{2}\sum_i X^{\I}_{i}(X^{\I}Z^{\II})_{i+1}-\tfrac{1}{2}\sum_i(Z^{\I}X^{\II}X^{\III})_{i}(X^{\II}X^{\III})_{i+1}\\
    &-\tfrac{1}{2}\sum_i\left[\bbI+ Z^{\II}Z^{\III}\right]_{i}\,.
\ea\ee
Like for the previous case, the projector term is minimized for the group elements in $\Z_2^c\times\Z_2^{ab}=\{\ket{000}=\ket{1},\, \ket{100}=\ket{c},\,\ket{011}=\ket{ab},\, \ket{111}=\ket{cab}\}$.
A state $\ket{\GS}$ minimizing the remaining terms must furthermore satisfy:
\be \label{eq:conditions_c_ab}
\ba
    X^{\I}_{i}(X^{\I}Z^{\II})_{i+1}\ket{\GS}&=
    \ket{\GS}, \quad \forall i=1,...,L \\
    (Z^{\I}X^{\II}X^{\III})_{i}(X^{\II}X^{\III})_{i+1}\ket{\GS}&=\ket{\GS}, \quad \forall i=1,\cdots,L.
\ea
\ee
Similarly to the phase discussed in section \ref{sec:Z2aZ2b-}, the ground state is unique because the above are $2L$ independent equations for the $2L$ dimensional Hilbert space of group elements in $\Z_2^c\times\Z_2^{ab}$. 
Furthermore, by taking the product over lattice sites $i$ of the constraints \eqref{eq:conditions_c_ab}, one has:
\be \label{eq:GS_SPT2_hol1}
\ba
     \prod_iZ^{\II}_{i}\ket{\GS}=\prod_iZ^{\I}_{i}\ket{\GS}=\ket{\GS}, \\
\ea
\ee
which implies that the holonomy of the ground state must be $1$.
Since the terms in the Hamiltonian mutually commute, we can write the ground state as:
\be \label{eq:GS_non-triv_SPT2}
\ba
        \ket{\GS}&=\prod_i\tfrac{1}{\sqrt{2}}\lbb\bbI_i\bbI_{i+1}+X^{\I}_{i}(X^{\I}Z^{\II})_{i+1}\rbb\bigotimes_i\ket{0,+}_{i}\,,
\ea
\ee
where $\ket{0,+}\equiv\tfrac{\ket{000}+\ket{011}}{\sqrt{2}}$ is a state that satisfies the second condition in \eqref{eq:conditions_c_ab}, whereas the first condition holds thanks to the operator acting on $\ket{0,+}$. 

We can write the symmetry generators $\cS_\Gamma$, for $\Gamma\in \Rep(D_8)$, on the $\Z_2^c\times\Z_2^{ab}$ subspace as follows: 
\be
    \ba
         \cS_1&=\cS_{1_c}|_{\Z_2^c\times\Z_2^{ab}}=\prod_{i=1}^{L} \bbI_{i}\cr 
         \cS_{1_a}&=\cS_{1_{ca}}|_{\Z_2^c\times\Z_2^{ab}}=\prod_{i=1}^{L} Z^{\I}_{i}\cr 
       \cS_E|_{\Z_2^c\times\Z_2^{ab}}&=\prod_{i=1}^{L} Z^{\II}_{i}+\prod_{i=1}^{L} (Z^{\I}_{i} Z^{\II}_{i})\,.
    \ea
\ee
Recalling eq. \eqref{eq:GS_SPT2_hol1}, we therefore see that: 
\be
    \cS_\Gamma\ket{\GS}=
    \dim(\Gamma)\ket{\GS}\,,
\ee
and this is a non-trivial $\Rep(D_8)$ SPT phase. 

\medskip\noindent{\bf String order parameter.} As for the SPT labeled by $(F)^{\beta}=(\Z_2^a\times \Z_2^b)^{-}$, this SPT can be detected by strong order parameters, for example
\begin{equation}
\begin{split}
\Q_{([1],1_c)}^{(i_0,i_0+N)}&=    \prod_{i=i_0}^{i_{0}+N}(Z^{\II}Z^{\III})_i \,, \\
%\Q_{([ab],1_a)}^{(i_0,i_0+N)}&=X^{\II}_{i0} X^{\III}_{i0+1} ( \prod_{k=1}^{N} Z^{\I}_{i0+k} ) X^{\II}_{i0+N} X^{\III}_{i0+N+1}\,,\\
\Q_{([ab],1_a)}^{(i_0,i_0+N)}&= (X^{\II}X^{\III})_{i_0}(   \prod_{i=i_0}^{i_{0}+N}Z^{\I}_i)(X^{\II}X^{\III})_{i_0+N}\,,
\end{split}
\end{equation}
which correspond to the condensed charges $([1],1_c)$ and $([ab],1_a)$ and have a unit expectation value on the SPT ground state.

\subsubsection{$\Z_2\times\Z_2$ SSB phase for $F=\Z_4^{ca}$} \label{sec:Z4ca}
Setting $F=\Z_4^{ca}$ in eq. \eqref{eq:RepD8_Ham}, the Hamiltonian is:
%\begin{widetext}
\be\ba \label{eq:Ham_Z4}
    &H_{\Z_4^{ca}}\simeq-\tfrac{1}{2}\sum_i\lb
    R^{cb}_{i}L^{ca}_{i+1}+
    R^{ca}_{i}L^{cb}_{i+1}\rb-\sum_i P^{(\Z_4^{ca})}_{i}\,.\\
    &-\tfrac{1}{4}\sum_i\lbb(\bbI+ Z^{\I}) X^{\I}X^{\III}+
    (\bbI -Z^{\I}) X^{\I}X^{\II}\rbb_{i}\times\\[-3mm]
    &\hspace{10mm}\times\lbb X^{\I}X^{\III}(\Swap)^{\II,\III}\rbb_{i+1}\\
    &-\tfrac{1}{4}\sum_i\lbb(\bbI+ Z^{\I}) X^{\I}X^{\II}+
   (\bbI -Z^{\I}) X^{\I}X^{\III}\rbb_{i}\times\\[-3mm]
   &\hspace{10mm}\times\lbb X^{\I}X^{\II}(\Swap)^{\II,\III}\rbb_{i+1}\\
   &-\tfrac{1}{2}\sum_i\left[\bbI+ Z^{\I} Z^{\II} Z^{\III}\right]_{i}\,.
\ea\ee
%\end{widetext}
The projector term is minimized, on each site $i$, by the states corresponding to the elements in $\Z_4=\{\ket{000}=\ket{1},\;\ket{101}=\ket{ca},\;\ket{011}=\ket{ab},\;\ket{110}=\ket{cb}\}$.
The remaining terms are then minimized by the following 4 linearly independent ground states:
\be \label{eq:GS_Z4}
\ba
\ket{\GS,0}&=\bigotimes_i\tfrac{1}{2}\left(\ket{1}+\ket{ca}+\ket{ab}+\ket{cb}\right)_{i} \\[-2mm]
&\equiv \ket{\Psi_1}+\ket{\Psi_{ca}}+\ket{\Psi_{ab}}+\ket{\Psi_{cb}},\\
\ket{\GS,1}&=\bigotimes_i\tfrac{1}{2}\left(\ket{1}+\zeta_4\ket{ca}-\ket{ab}-\zeta_4\ket{cb}\right)_{i}\\[-2mm]
&\equiv \ket{\Psi_1}+\zeta_4\ket{\Psi_{ca}}-\ket{\Psi_{ab}}-\zeta_4\ket{\Psi_{cb}},\\
\ket{\GS,2}&=\bigotimes_i\tfrac{1}{2}\left(\ket{1}-\ket{ca}+\ket{ab}-\ket{cb}\right)_{i}\\[-2mm]
&\equiv \ket{\Psi_1}-\ket{\Psi_{ca}}+\ket{\Psi_{ab}}-\ket{\Psi_{cb}},\\
\ket{\GS,3}&=\bigotimes_i\tfrac{1}{2}\left(\ket{1}-\zeta_4\ket{ca}-\ket{ab}+\zeta_4\ket{cb}\right)_{i}\\[-2mm] 
&\equiv \ket{\Psi_1}-\zeta_4\ket{\Psi_{ca}}-\ket{\Psi_{ab}}+\zeta_4\ket{\Psi_{cb}}\,,
\ea
\ee
where $\zeta_4=e^{\tfrac{2\pi i}{4}}$. The action of $\Rep(D_8)$ on the ground states is for $p\in \Z_4$:
\be
\ba
    \cS_1\ket{\GS,p}&=\cS_{1_{ca}}\ket{\GS,p}=\ket{\GS,p}, \\
     \cS_{1_a}\ket{\GS,p}&=\;\cS_{1_{c}}\ket{\GS,p}=\ket{\GS,p+2},\\
    \cS_{E}\ket{\GS,0}&=\cS_{E}\ket{\GS,2}=\ket{\GS,1}+\ket{\GS,3}\,,\\
    \cS_{E}\ket{\GS,1}&=\cS_{E}\ket{\GS,3}=\ket{\GS,0}+\ket{\GS,2}\,.\\
\ea
\ee
The 4 vacua form 2 different $\Z_2$ orbits under the broken  generators $\cS_{1_a}$ and $\cS_{1_{c}}$:
\be
    \cS_{1_a},\;\cS_{1_{c}}:\quad \ket{\GS,0}\leftrightarrow\ket{\GS,2},\quad \ket{\GS,1}\leftrightarrow\ket{\GS,3},
\ee
this is therefore a $\Z_2\times\Z_2$ SSB phase for $\Rep(D_8)$ symmetry. The order parameters are $\Q_{[ab], 1}$, $2\Q_{[ca], 1}$ which are realized by the operators
\be
\ba
    L^{ca}_{i} & =\lb X^{\I}X^{\III}(\Swap)^{\II,\III}\rb_{i}\,, \ 
    L^{ab}_{i}=(L^{ca})^2_{i}=\lb X^{\II} X^{\III}\rb_{i}\cr 
   L^{cb}_{i}&=(L^{ca})^3_{i}=\lb X^{\I}X^{\II}(\Swap)^{\II,\III}\rb_{i}\,,
\ea
\ee
which have the following eigenvalues in the ground states:
\be
\ba
    \bra{\GS,p}(L^{ca})^q_{i}\ket{\GS,p}&=(-\zeta_4)^{pq}, &&\quad\forall\;p,q,\in\{0,1,2,3\}\,.
\ea
\ee

\subsubsection{$\Rep(D_8)$ SSB phase for $(F)^\beta=(D_8)^+$} \label{sec:D8}
Setting $F=D_8$ with trivial $\beta$ in eq. \eqref{eq:RepD8_Ham}, the Hamiltonian is:
\be\ba \label{eq:Ham_D8}
& H_{(D_8)^+}\simeq-\tfrac{1}{2}\sum_i \left( R^a_iL^a_{i+1}+R^c_iL^c_{i+1}\right)-\sum_i\bbI_i\simeq \\ 
&-\tfrac{1}{4}\sum_i \left[(\bbI+Z^\I)\, X^{\II}+(\bbI-Z^\I)\, X^{\III}\right]_i X^\II_{i+1} -\sum_i\bbI_i\cr 
&  -\tfrac{1}{4}\sum_i X^\I_i\left[(\bbI+Z^\II Z^\III)\, X^{\I}+(\bbI-Z^\II Z^\III)\, X^{\I}X^\II X^\III\right]_{i+1}\,.
% H_{(D_8)^+}=&-\tfrac{1}{8}\sum_i\sum_{g\in D_8}R^{g^{-1}}_{i}L^{g}_{i+1}- P_i^{(D_8)}\,.
%  \\
%  =&-\tfrac{1}{8}\sum_i\lb
% \bbI_{i}\bbI_{i+1}+R^{a}_{i}L^{a}_{i+1}+R^{b}_{i}L^{b}_{i+1}+
% R^{ab}_{i}L^{ab}_{i+1}+R^{c}_{i}L^{c}_{i+1}+R^{cab}_{i}L^{cab}_{i+1}+
%     R^{cb}_{i}L^{ca}_{i+1}+
%     R^{ca}_{i}L^{cb}_{i+1}\rb-\sum_i\bbI_i=\\
%     =&-\tfrac{1}{8}\sum_i
%     \bbI_{i}\bbI_{i+1}-\tfrac{1}{16}\sum_i\lbb
%     (\bbI+Z^{\I})X^{\II}+(\bbI-Z^{\I})X^{\III}\rbb_i\lbb X^{\II}\rbb_{i+1}-\tfrac{1}{16}\sum_i\lbb
%     (\bbI+Z^{\I})X^{\III}+(\bbI-Z^{\I})X^{\II}\rbb_i\lbb X^{\III}\rbb_{i+1}+\\
%     &-\tfrac{1}{8}\sum_i(X^{\II}X^{\III})_{i} (X^{\II}X^{\III})_{i+1}-\tfrac{1}{8}\sum_iX^{\I}_{i} X_{i+1}^{\I}(\Swap)_{i+1}^{\II,\III}-\tfrac{1}{8}\sum_i\lb X^{\I}X^{\II}X^{\III}\rb_{i}
%     (X^{\I}X^{\II}X^{\III})_{i+1}(\Swap)_{i+1}^{\II,\III}\\
%     &-\tfrac{1}{16}\sum_i\lbb(\bbI+ Z^{\I}) X^{\I}X^{\III}+
%     (\bbI -Z^{\I}) X^{\I}X^{\II}\rbb_{i}\lbb X^{\I}X^{\III}(\Swap)^{\II,\III}\rbb_{i+1}+\\
%     &-\tfrac{1}{16}\sum_i\lbb(\bbI+ Z^{\I}) X^{\I}X^{\II}+
%     (\bbI -Z^{\I}) X^{\I}X^{\III}\rbb_{i}\lbb X^{\I}X^{\II}(\Swap)^{\II,\III}\rbb_{i+1}-\sum_i\bbI_{i}\,.
\ea\ee
The space of ground states is spanned by terms with holonomy in the same conjugacy class, whose list can be found in eq. \eqref{eq:D8_conjclasses}
\be
\ba
    \ket{\Psi_{[h]}}\equiv\tfrac{1}{\sqrt{8^L}}\sum_{\{\vec{g}\;|\;g\in[h]\}}\ket{\vec{g},g}\,.
\ea
\ee
Collecting them according to $\Rep(D_8)$ symmetry, they are:
\be \label{eq:GS_D8}
\ba
    \ket{\GS,1}&= \ket{\Psi_{[1]}}+\ket{\Psi_{[ab]}}+\ket{\Psi_{[ca]}}+\ket{\Psi_{[c]}}+\ket{\Psi_{[a]}},\\
    \ket{\GS,2}&= \ket{\Psi_{[1]}}+\ket{\Psi_{[ab]}}-\ket{\Psi_{[ca]}}+\ket{\Psi_{[c]}}-\ket{\Psi_{[a]}},\\
    \ket{\GS,3}&= \ket{\Psi_{[1]}}+\ket{\Psi_{[ab]}}-\ket{\Psi_{[ca]}}-\ket{\Psi_{[c]}}+\ket{\Psi_{[a]}},\\
    \ket{\GS,4}&= \ket{\Psi_{[1]}}+\ket{\Psi_{[ab]}}+\ket{\Psi_{[ca]}}-\ket{\Psi_{[c]}}-\ket{\Psi_{[a]}},\\
    \ket{\GS,5}&= 2\ket{\Psi_{[1]}}-2\ket{\Psi_{[ab]}}\,.\\
\ea
\ee
The action of $\Rep(D_8)$ on the ground states is:
\be
\ba
     \cS_{1_c}: \quad \ket{\GS,1}&\leftrightarrow\ket{\GS,2},\quad \ket{\GS,3}\leftrightarrow\ket{\GS,4}, \\
     \cS_{1_a}: \quad \ket{\GS,1}&\leftrightarrow\ket{\GS,3},\quad \ket{\GS,2}\leftrightarrow\ket{\GS,4}, \\
     \cS_{1_{ca}}: \quad \ket{\GS,1}&\leftrightarrow\ket{\GS,4},\quad \ket{\GS,2}\leftrightarrow\ket{\GS,3}, \\
     \cS_{1_k}\ket{\GS,5}&=\ket{\GS,5} \qquad \forall\; 1_k\in\{1_c,1_a,1_{ca}\},\\
    \cS_{E}\ket{\GS,p}&=\ket{\GS,5} \qquad \forall\;p\in\{1,2,3,4\},\\
    \cS_{E}\ket{\GS,5}&=\sum_{p=1}^4\ket{\GS,p}\,,\\
\ea
\ee
from which we see that $\Rep(D_8)$ is fully spontaneously broken.
The order parameters are all the charges in $\Bsym=\cA_{33}$: we denote by $O^{[g]}$ the operators $L^g$ or $R^g$ for $g \in [g] $ and compute
\be
\ba
    \bra{\GS,p}O^{ab}_{i}\ket{\GS,p}&=+1 \quad \forall\;p\in\{1,2,3,4\}\cr 
    \bra{\GS,5}O^{ab}_{i}\ket{\GS,5}&=-1, \\
    \bra{\GS,p}O^{[c]}_{i}\ket{\GS,p}&=+1 \quad p\in\{1,2\} \cr 
    \bra{\GS,p}O^{[c]}_{i}\ket{\GS,p}&=-1 \quad p\in\{3,4\} \\
     \bra{\GS,p}O^{[a]}_{i}\ket{\GS,p}&=+1 \quad p\in\{1,3\}\cr 
    \bra{\GS,p}O^{[a]}_{i}\ket{\GS,p}&=-1 \quad p\in\{2,4\} \\
    \bra{\GS,p}O^{[ca]}_{i}\ket{\GS,p}&=+1 \quad p\in\{1,4\}\cr 
    \bra{\GS,p}O^{[ca]}_{i}\ket{\GS,p}&=-1 \quad p\in\{2,3\} \\
    \bra{\GS,5}O^{[g]}_{i}\ket{\GS,5}&=0 \quad [g]\in\{[a],[c],[ca]\}\,.
\ea
\ee
One can write the $\Rep(D_8)$ SSB ground states \eqref{eq:GS_D8}, as follows:
\be
\ba
    \ket{\GS,1}&=\bigotimes_i\sum_{g\in D_8}\tfrac{1}{\sqrt{8}}\ket{g}_i\,,\\
    \ket{\GS,2}&=\cS_{1_c}\ket{\GS,1}\,,\quad
    \;\,\ket{\GS,3}=\cS_{1_a}\ket{\GS,1}\,,\\
    \ket{\GS,4}&=\cS_{1_{ca}}\ket{\GS,1}\,,\quad \ket{\GS,5}=\cS_E\ket{\GS,1}\\
\ea
\ee
$\ket{\GS,p}$ for $p=1,2,3,4$ are therefore tensor-product states, whereas we numerically computed that $\ket{\GS,5}$ has non-zero entanglement.

\subsubsection{$\Z_2$ SSB phase for $F^\beta=D_8^-$} \label{sec:D8-}
We now consider $F=D_8$ as the previous case, but now with the non-identity
\be
    \beta\in H^2(D_8,U(1))=\Z_2\,.
\ee
As a representative of $\beta$, we choose:
\be \label{eq:betaD8}
\ba
    \beta(c^{c_1}a^{a_1}b^{b_1},c^{c_2}a^{a_2}b^{b_2})=(-1)^{\lb(1-c_2)a_1+c_2b_1\rb b_2+a_1b_1c_2}\,.
\ea
\ee
The Hamiltonian for this gapped phase is:
\be \label{eq:HD8-}
 H_{D_8^-}\simeq-\tfrac{1}{2}\sum_i \left[ (R^a_\beta)_i(L^a_\beta)_{i+1}+(R^c_\beta)_i(L^c_\beta)_{i+1}\right]-\sum_i\bbI_i
\ee
Here we define
\begin{equation}
     R_\beta^{f^{-1}}=Z_{(\beta,R)}^f R^{f^{-1}},\qquad L_\beta^f=L^{f}Z_{(\beta,L)}^f\,,
\end{equation}
with the following non-trivial twists, which follow from \eqref{eq:betaD8}:
%Z_{(\beta,R)}^f=\beta^{-1}(\cdot,\,f)\,, \qquad  Z_{(\beta,L)}^f=\beta(f,\,\cdot)\,,
\be \label{eq:D8-beta}
\ba
   Z_{(\beta,L)}^{a}&=\tfrac{1}{2}\lbb\bbI-Z^{\I}+Z^{\III}+Z^{\I}Z^{\III}\rbb\,,  \\
   Z_{(\beta,R)}^{c}&=\tfrac{1}{2}\lbb\bbI+Z^{\II}+Z^{\III}-Z^{\II}Z^{\III}\rbb\,,
    %Z_{(\beta,L)}^{a}=Z_{(\beta,L)}^{ca}&=\tfrac{1}{2}\lbb\bbI-Z^{\I}+Z^{\III}+Z^{\I}Z^{\III}\rbb\,, \\ 
    %Z_{(\beta,L)}^{b}=Z_{(\beta,L)}^{cb}&=\tfrac{1}{2}\lbb \bbI+Z^{\I}+Z^{\II}-Z^{\I}Z^{\II}\rbb \,,\\
     %Z_{(\beta,L)}^{ab}=Z_{(\beta,L)}^{cab}&=\tfrac{1}{2}\lbb -Z^{\II}+Z^{\III}+Z^{\I}(Z^{\II}+Z^{\III})\rbb \,,\\
     %Z_{(\beta,R)}^{ab}=Z_{(\beta,R)}^{b}&=\tfrac{1}{2}\lbb Z^{\II}+Z^{\III}+Z^{\I}(Z^{\II}-Z^{\III})\rbb\,, \\
     %Z_{(\beta,R)}^{c}=Z_{(\beta,R)}^{ca}&=\tfrac{1}{2}\lbb\bbI+Z^{\II}+Z^{\III}-Z^{\II}Z^{\III}\rbb\,,\\
     %Z_{(\beta,R)}^{cab}=Z_{(\beta,R)}^{cb}&=\tfrac{1}{2}\lbb \bbI-Z^{\I}(Z^{\II}-Z^{\III})+Z^{\II}Z^{\III}\rbb\,.
\ea
\ee
%zR[ca]=1/2(+id8+z2+z3-z2z3)
%zL[ca]=1/2(+id8-z1+z3+z1z3)
explicitly:
\be
\ba
 H_{D_8^-}\simeq -\sum_i\bbI_i-\tfrac{1}{4}\sum_i \left[(\bbI+Z^\I)\, X^{\II}+(\bbI-Z^\I)\, X^{\III}\right]_i X^\II_{i+1}\\[-3mm]
 \times \tfrac{1}{2}\lbb\bbI-Z^{\I}+Z^{\III}+Z^{\I}Z^{\III}\rbb_{i+1}\\
 -\tfrac{1}{2}\sum_i\lbb\bbI+Z^{\II}+Z^{\III}-Z^{\II}Z^{\III}\rbb_i\\[-2mm]
 \times\tfrac{1}{4}X^\I_i\left[(\bbI+Z^\II Z^\III)\, X^{\I}+(\bbI-Z^\II Z^\III)\, X^{\I}X^\II X^\III\right]_{i+1}
\ea
\ee
One can diagonalize this Hamiltonian and determine that the ground state space is spanned by $\Psi^\beta_{[1]}$ and $\Psi^\beta_{[ca]}$ with holonomy respectively in the $[1]$ and $[ca]$ conjugacy class. We thus have:
\be \label{eq:GS_D8-}
\ba
\ket{\GS,+}&\equiv \ket{\Psi^\beta_{[1]}}+\ket{\Psi^\beta_{[ca]}}\,,\\
\ket{\GS,-}&\equiv \ket{\Psi^\beta_{[1]}}-\ket{\Psi^\beta_{[ca]}}\,.
\ea
\ee
The action of the $\Rep(D_8)$ generators \eqref{eq:SGamma} on the ground states \eqref{eq:GS_D8-} is:
\be
\ba
    \cS_1\ket{\GS,\pm}=\;\cS_{1_{ca}}\ket{\GS,\pm}&=\ket{\GS,\pm}\,,\\
    \cS_{1_c}\ket{\GS,\pm}=\cS_{1_a}\ket{\GS,\pm}&=\ket{\GS,\mp}\,,\\
    \cS_{E}\ket{\GS,\pm}&=\ket{\GS,+}+\ket{\GS,-}\,.
\ea
\ee
This is therefore a $\Z_2$ SSB phase for $\Rep(D_8)$: the $\Z_2$ symmetries exchanging the two ground states are $\cS_{1_c}$ and $\cS_{1_a}$. The local order parameter is:
\be
\ba
    \Q_{([ca],1)}^\beta&\sim\lb\bbI+Z^{\III}-Z^{\I}Z^{\II}(\bbI-Z^{\III})\rb X^{\I}X^{\II}+\\
    &+\lb\bbI+Z^{\II}+Z^{\I}Z^{\III}(\bbI-Z^{\II})\rb X^{\I}X^{\III}\,,
\ea
\ee
for which
\be \bra{\GS,\pm}\Q_{([ca],1)}^\beta\ket{\GS,\pm}=\pm 1\,.
\ee

\subsection{$\Rep(D_8)$ Phase Transitions}
\label{app:gapless}
We can find lattice models for second-order phase transitions between the gapped phases discussed above. This can be done using the general approach of \cite{Bhardwaj:2024kvy}, which takes as an input a lattice model for a known phase transition, like the critical point of Ising model, and embeds it into a larger lattice model to produce the desired transition. 
%These transformations can be viewed as generalized versions of Kennedy-Tasaki (KT) transformations \cite{Kennedy:1992ifl, kennedy1992hidden}. 
For $\Rep (D_8)$ symmetry the critical models, describing second order phase transitions between two gapped phases were determined in \cite{Bhardwaj:2024qrf} (table IV, v3). 
These continuum predictions can be complemented by lattice models as shown in \cite{Bhardwaj:2024kvy} taking the input from the SymTFT club sandwich construction. An alternative, less systematic way to see the phase transitions in the lattice models is to construct the  Hamiltonian $H$ describing the desired phase transition as 
\be
H_{(l,m)}(\lambda)=\lambda H_{(l)}+(1-\lambda) H_{(m)}
\ee
for some $\lambda\in[0,1]$ where $H_{(l)}$ and $H_{(m)}$ are the commuting projector Hamiltonians for the gapped phases lying on the two sides of the phase transition. 
\\ 

\medskip \noindent{\bf Trivial to $\Z_2$ SSB Transition.} 
The ansatz for the interpolating Hamiltonian is 
\be
   H_{(1,\,\Z_2^a)}(\lambda)=\lambda H_{1}+(1-\lambda)H_{\Z_2^a} \,.
\ee
The low-energy physics is restricted to the $\Z_2^a$ subspace, i.e. qubit $\II$, on each site. We can thus write a simplified Hamiltonian, by restricting to $\Z_2^a$:
\begin{equation}
    H_{(1,\Z_2^a)}\approx -\tfrac{1}{2}\sum_i\lbb\lambda\, Z^{\II}_i+(1-\lambda)X^{\II}_{i}X^{\II}_{i+1}\rbb\,.
\end{equation}
Setting $\lambda=\tfrac{1}{2}$ we obtain the transverse-field critical Ising Hamiltonian, describing the phase transition corresponding to a gSPT phase. This type of analysis can be repeated for all gapped phases. We will consider a few more examples to illustrate this. \\

\medskip \noindent{\bf Trivial to $\Rep (D_8)/\Z_2 \times \Z_2$ SSB Transition.}
Similarly taking the Hamiltonians $H_{1}$ and $H_{\Z_2^{ab}}$, there are common projectors  $(1+Z^\I)/2$ and we can project first onto the ground state of $-(1+ Z^{\II})(1+Z^{\III})/4$ and then restrict 
$(1+ Z^{\II}Z^{\III})/2$ to that. The resulting Hamiltonian restricted to that subspace is 
\be
\ba
H_{(1,\Z_2^{ab})} \approx  
-\tfrac{1}{2}\sum_i&[ \lambda\, (Z^{\II} + Z^{\III})_i+
 \cr 
 &(1-\lambda)X^{\II}_{i}X^{\III}_{i}X^{\II}_{i+1}X^{\III}_{i+1}]  \,.
\ea
\ee
This reduces again to the critical Ising model, for the diagonal combination $\Z_2^{ab}$.

\medskip \noindent{\bf Trivial to  $\Z_2\times \Z_2$ SSB Transition.} 
Here we can restrict to the $\II$ and $\III$ qubit subspace only. The two $XX$ terms from the $\Z_2\times\Z_2$ SSB result in two Ising chains, giving an Ising$\times$Ising transition.
\be
\ba
H_{(1,\Z_2^a\times\Z_2^b)^+} \approx  
-\tfrac{1}{2}\sum_i&[ \lambda\, (Z^{\II} + Z^{\III})_i+
 \cr 
 &(1-\lambda)(X^{\II}_{i}X^{\II}_{i+1}+X^{\III}_{i}X^{\III}_{i+1})]  \,.
\ea
\ee

\medskip \noindent{\bf Trivial to SPT Transition.} 
Similarly to the previous case, we can restrict to the $\II$ and $\III$ qubit subspace only. 
\be
\ba
H_{(1,\Z_2^a\times\Z_2^b)^-} \approx  
-\tfrac{1}{2}\sum_i[ \lambda\, (Z^{\II} + Z^{\III})_i+
 \cr 
 (1-\lambda)(X^{\II}_{i}(X^{\II}Z^{\III})_{i+1}+(Z^{\II}X^{\III})_{i}X^{\III}_{i+1})]  \,.
\ea
\ee
By applying a (unitary) Kennedy-Tasaki (KT) transformation \cite{Kennedy:1992ifl,kennedy1992hidden}, the SPT Hamiltonian can be mapped to the $\Z_2\times\Z_2$ SSB Hamiltonian, while the trivial Hamiltonian remains unchanged. Therefore the universality class of this transition is the same as the previous case, i.e. Ising$\times$Ising.

%\medskip \noindent{\bf $\Rep (D_8)/\Z_2 \times \Z_2$ SSB to $\Z_2\times \Z_2$ SSB Transition.}
%Similarly here we can restrict to the qubits $\II$ and $\III$. Restricting the $XX$ term of the $\Rep (D_8)/\Z_2 \times \Z_2$ SSB   to the eigenstates of the $XX$ term in the $\Z_2 \times \Z_2$ SSB gives again two decoupled Ising chains, consistent with the Ising$\oplus$Ising  gapless phase.  

\medskip \noindent{\bf $\Rep (D_8)/\Z_2 \times \Z_2$ SSB to SPT Transition.}
% A22
The phase transition between the SPT labeled as $(F)^{\beta}=(\Z_2^a\times \Z_2^b)^{-}$ and the $\Rep (D_8)/\Z_2 \times \Z_2$ SSB labeled as $(F)^{\beta}=\Z_2^{ab}$ takes place in the projected subspace of $Z^{\I}=1$.
In this restricted subspace
\begin{equation}
\cS_E\Big|_{\Z^{\I}=1}=\cS_{\II}+\cS_{\III}\,, 
\end{equation}
and 
\begin{equation}
  \cS_{1_a}=\bbI\,, \qquad
  \cS_{1_c}=\cS_{1_{ca}}=\cS_{\II}\cS_{\III}
\end{equation}
where 
\begin{equation}    \cS_{\alpha}=\prod_{i=1}^{L} \lb Z^{\alpha}\rb_{i} \quad \alpha\in\{\II\,,\III\}\,. 
\end{equation}
Therefore the effective symmetry in this subspace is $\Z^{\II}_2\times \Z^{\III}_2$ generated by $\cS_{\II}$ and $\cS_{\III}$.
$\Rep (D_8)/\Z_2 \times \Z_2$ SSB in this restricted space realizes the $\Z^{\II}_2\times \Z^{\III}_2\to \Z_2^{\rm diag}$ SSB phase, while the SPT is indeed the $\Z_2^{\II}\times \Z_2^{\III} $ SPT. 
The transition takes place at $\lambda=1/2$ along the following line in parameter space
\begin{equation}\label{eqref:trans}
    \wt{H}(\lambda)= \lambda\wt{H}_{\rm SPT} + (1-\lambda) \wt{H}_{\Rep(D_8)/\Z_2^{2}}\,,
\end{equation}
where $\wt{H}$ denotes the restriction of $H$ to the $Z^{\I}=1$ space.
In this space, the Hamiltonians have the form 
\begin{equation}
    \begin{split}
      \wt{H}_{\rm SPT}&= -\tfrac{1}{4}\sum_i\lbb\bbI_i\bbI_{i+1}+X^{\II}_{i}(X^{\II}Z^{\III})_{i+1}\rbb \\[-3mm]
      &
      \qquad \qquad \ \times \lbb\bbI_i\bbI_{i+1}+(Z^{\II}X^{\III})_{i}X^{\III}_{i+1}\rbb\,,
    \\[2mm]
     \wt{H}_{\Rep(D_8)/\Z_2^2}&= -\tfrac{1}{4}\sum_i\lbb\bbI_i\bbI_{i+1}+(X^{\II}X^{\III})_{i}(X^{\II}X^{\III})_{i+1}\rbb \\[-3mm]
     &
      \qquad \qquad \ \times \lbb\bbI_i\bbI_{i+1}+Z^{\II}_{i}Z^{\III}_{i}\rbb\,,
    \end{split}
\end{equation}
We claim that the transition at $\lambda=1/2$ in \eqref{eqref:trans} lies in the Ising universality class.
To see this, we will map the transition to a more familiar form in two steps.
In the first step we define a unitary that has the following action on the operators generating the $\Z_2^{\II}\times \Z_2^{\III}$ bond algebra \cite{Moradi:2023dan}
\begin{equation}
    U_{1}:
    \begin{pmatrix}
        Z^{\II}_j\\
        Z^{\III}_j\\
        X^{\II}_{j}X^{\II}_{j+1} \\
        X^{\III}_{j}X^{\III}_{j+1} \\
    \end{pmatrix}
    \longmapsto
    \begin{pmatrix}
        Z^{\II}_j\\
        (Z^{\II}Z^{\III})_j\\
    (X^{\II}X^{\III})_{j}(X^{\II}X^{\III})_{j+1} \\
        X^{\III}_{j}X^{\III}_{j+1} \\
    \end{pmatrix}
\end{equation}
This transition implements an automorphism of $\Z_2^{\II}\times \Z_2^{\III}$ symmetry. 
It can be easily seen that this unitary preserves the bond algebra. 
This unitary maps leaves the SPT Hamiltonian unchanged while it transforms the $\Rep(D_8)/\Z_2^2$ Hamiltonian into
\begin{equation}\label{eq:psb}
\begin{split}
    -\tfrac{1}{4}\sum_i
\lbb 1 +X^{\II}_{i}X^{\II}_{i+1}\rbb \times [1+Z^{\III}_i]\,,
\end{split}
\end{equation}
which realizes the phase $\Z_2^{\II}\times \Z_2^{\III}\to \Z_{2}^{\III}$ SSB.
Then in the second step, we implement a (SPT entangler) unitary that realizes the following map on the bond algebra
\begin{equation}
    U_{2}:
    \begin{pmatrix}
        Z^{\II}_j\\
        Z^{\III}_j\\
        X^{\II}_{j}X^{\II}_{j+1} \\
        X^{\III}_{j}X^{\III}_{j+1} \\
    \end{pmatrix}
    \longmapsto
    \begin{pmatrix}
(Z^{\II}_jX^{\III})_{j}X^{\III}_{j+1}\\
        X^{\II}_{j-1}(X^{\II}Z^{\III})_j\\
    X^{\II}_{j}X^{\II}_{j+1} \\
        X^{\III}_{j}X^{\III}_{j+1} \\
    \end{pmatrix}
\end{equation}
This unitary leaves the partial symmetry breaking Hamiltonian \eqref{eq:psb} unaltered but maps the SPT Hamiltonian to
\begin{equation}\label{eq:triv}
\begin{split}
    -\tfrac{1}{4}\sum_i
\lbb 1 +Z^{\II}_i\rbb \times [1+Z^{\III}_i]\,,
\end{split}
\end{equation}
which realizes the $\Z_2^{\II}\times \Z_2^{\III}$ disordered phase.
Clearly the transition between \eqref{eq:psb} and \eqref{eq:triv} lies in the Ising universality class.
Since these transformations were unitary, they preserve the spectrum and \eqref{eqref:trans} too lies in the Ising universality class wherein the $\Z_2^{\II}$ symmetry is broken.
Although the two unitaries have the effect of performing an automorphism on the group $\Z_2^{\II}\times \Z_2^{\III}$ and subsequently pasting an SPT which modifies the charges of the various twisted sectors without affecting the untwisted sectors.
In terms of the SymTFT, both these unitaries can be thought of as descending from braided autoequivalences of the SymTFT (the doubled Toric Code with lines generated by $e^{\II}\,, m^{\II}\,, e^{\III}\,, m^{\III}$) that leave the symmetry boundary (which is the $e^{\II}$ and $e^{\III}$ condensed boundary) invariant.
These unitaries implement 
\begin{equation}
\begin{split}
U_1:    \{e^{\II}\,, m^{\II}\,, e^{\III}\,, m^{\III}\}& \mapsto 
\{e^{\II}e^{\III}\,, m^{\II}\,, e^{\III}\,, m^{\II}m^{\III}\}
\\
U_2:    \{e^{\II}\,, m^{\II}\,, e^{\III}\,, m^{\III}\}& \mapsto
\{e^{\II}\,, m^{\II}e^{\III}\,, e^{\III}\,, m^{\III}e^{\II}\}
\\
\end{split}
\end{equation}
See \cite{Moradi:2022lqp} for more details on how the twisted sectors map.

\medskip \noindent{\bf $\Z_2\times \Z_2$ SSB to $\Rep(D_8)$ SSB Transition.}
The transition from the $\Z_2\times\Z_2$ SSB phase with $F^\beta=(\Z_2^a\times\Z_2^b)^+$ to the $\Rep(D_8)$ SSB goes out of the $s^\I=0$ subspace, so one must interpolate between the unprojected Hamiltonian in \eqref{eq:Ham_Z2aZ2b} and the $\Rep(D_8)$ SSB Hamiltonian \eqref{eq:Ham_D8}. The critical Hamiltonian, at $\lambda=\tfrac{1}{2}$, is:
\begin{align}
    H_{\text{crit}}=&-\tfrac{1}{8}\sum_i(\bbI+Z^\I)_i(X^\III_iX^\III_{i+1}+2X^\II_iX^\II_{i+1}+2)\nn\\
    &-\tfrac{1}{8}\sum_i(\bbI-Z^\I)_i(X^\II_iX^\III_{i+1}+2X^\III_iX^\II_{i+1})\nn\\
    &-\tfrac{1}{8}\sum_iX^\I_i[(\bbI+Z^\II Z^\III)X^\I]_{i+1}-\tfrac{1}{2}\sum_i\bbI_i\nn\\
    &-\tfrac{1}{8}\sum_iX^\I_i[(\bbI-Z^\II Z^\III)X^\I X^\II X^\III]_{i+1}
\end{align}
The first line dominates over the second (because of the $+2$), therefore qubits $\II$ and qubits $\III$ are independently aligned. The third lines reveals that on the local Hilbert space spanned by $\Z_2^c\times\Z_2^{ab}=\{\ket{1}=\ket{000},\ket{c}=\ket{100},\ket{ab}=\ket{011},\ket{cab}=\ket{111}\}$, we have the $\Ising$ Hamiltonian for the first qubit $H_{\Ising}=-\tfrac{1}{2}\sum_i\lb X^\I_i X^\I_{i+1}+Z^\I\rb$: indeed the $\cS_{1_a}=\prod_i Z^\I$ symmetry gets spontaneously broken along this transition. The last line implies that on the remaining states the $\II$ and $\III$ qubits are aligned eigenstates, giving a total of 3 gapless $\Ising$ `universes'.
\be
\begin{tikzpicture}[baseline]
\begin{scope}[shift={(-1,2.8)}]
\node (5) at (2.5,-3) {$\Ising_0$};
\node at (3.5,-3) {$\oplus$};
\node (3) at (4.5,-3) {$\Ising_+$};
\node (3b) at (4.55,-3.2) {};
\node at (5.5,-3) {$\oplus$};
\node (4) at (6.5,-3) {$\Ising_-$};
\node (4b) at (6.55,-3.2) {};
\draw[blue,stealth-stealth] (3) [bend right=30] to (5);
\draw[blue,stealth-stealth] (4) [bend right=30] to (5);
\draw[ForestGreen,-stealth] (3) edge [in=70, out=110,looseness=5] (3);
\draw[ForestGreen,-stealth] (4) edge [in=70, out=110,looseness=5] (4);
\draw[ForestGreen,-stealth] (5) edge [in=70, out=110,looseness=5] (5);
\draw[red,-stealth] (5) edge [in=-70, out=-110,looseness=5] (5);
\draw[red,stealth-stealth,transform canvas={xshift=0mm,yshift=-0.5mm}] (3) [bend right=40] to (4);
% \draw[red,stealth-stealth,transform canvas={xshift=3.5mm,yshift=0mm}] (1) [bend right=30] to (2);
% \draw[red,stealth-stealth,transform canvas={xshift=-3.5mm,yshift=0mm}] (3) [bend right=30] to (4);
%\oplus\ket{\GS,--}\oplus\ket{\GS,+-}\oplus\ket{\GS,-+}$};
\end{scope}
\end{tikzpicture} 
\ee
The critical point is therefore a gapless SSB phase, where the spontaneously broken summery is the $\Ising$ category symmetry: the invertible $\Z_2$ corresponds to $\cS_{1_c}$ (shown in red, which exchanges two of the 3 universes), and the non-invertible symmetry to $E$ (shown in blue). There is $\sqrt{2}$ `relative Euler term' between $\Ising_0$ and each of $\Ising_\pm$, since $E$ has quantum dimension 2 whereas the $\Ising$ non-invertible symmetry has quantum dimension $\sqrt{2}$. Each copy of $\Ising$ is invariant under $\cS_{1_a}$ (shown in green). This is an intrinsically gapless SSB because it can only flow to gapped phases with a larger number of vacua (i.e. 4 or 5): indeed $\Rep(D_8)$ does not admit any gapped phases with 3 vacua.
This phase was predicted from the continuum SymTFT approach in \cite{Bhardwaj:2024qrf}. Our current work enabled us to verify numerically that indeed the lowest three eigenstates of $H_{\text{crit}}$ are degenerate and compatible with central charge $c=\tfrac{1}{2}$. They can be written as:
\be
\ba
    \Ising_0&\equiv\Ising^\I\otimes(\sqrt{2}\Psi_{[1]}-\sqrt{2}\Psi_{[ab]})\,,\\
    \Ising_+&\equiv\Ising^\I\otimes(\Psi_{[1]}+\Psi_{[ab]}+\Psi_{[a]})\,,\\
    \Ising_-&\equiv\Ising^\I\otimes(\Psi_{[1]}+\Psi_{[ab]}-\Psi_{[a]})\,.\\
\ea
\ee

\section{Realization in Rydberg atom arrays}\label{app:appB}

\subsection{Elementary quantum gates}
\begin{figure}
	\centering
	\includegraphics[width= \linewidth]{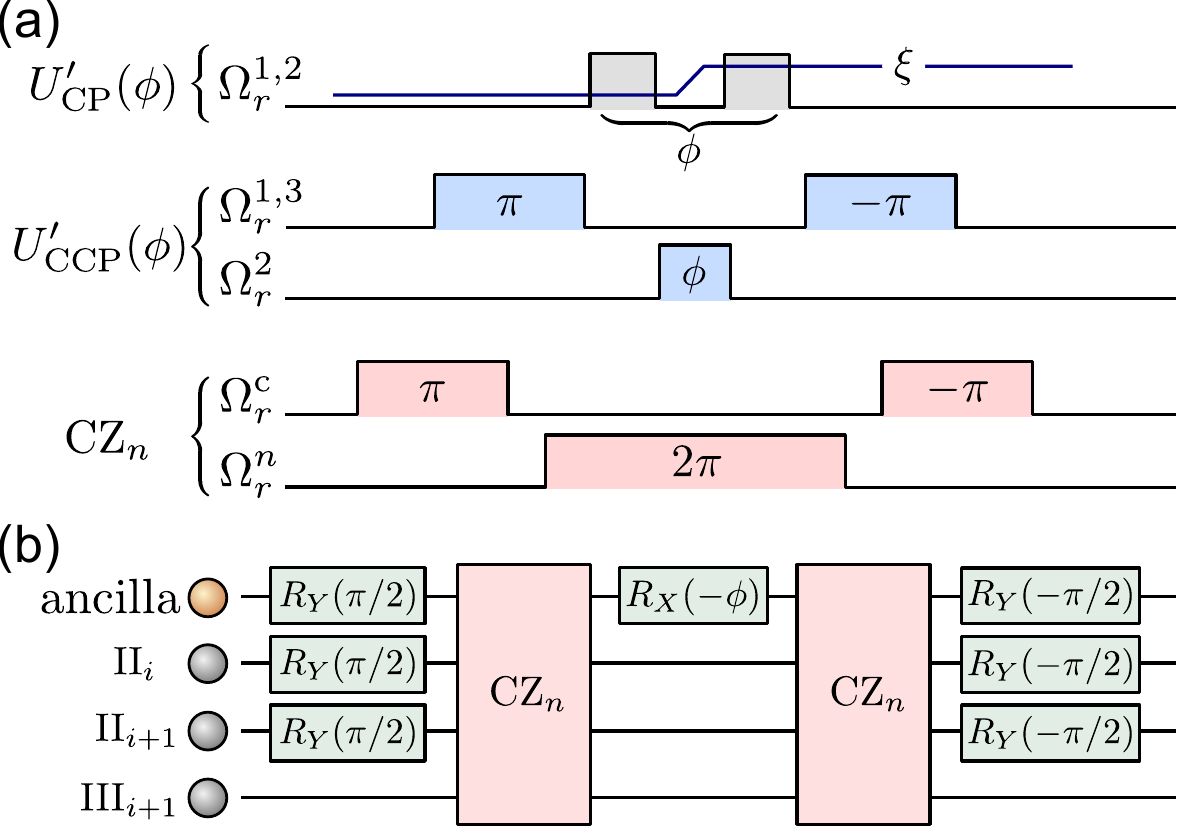}% Here is how to import EPS art
	\caption{(a) Rydberg pulse sequence for the elementary quantum gates. (b) Gate sequence for simulating the three-body plaquette evolution $U_\square(\phi)=\exp(-\mathrm{i}\phi X_i^\mathrm{II}X_{i+1}^\mathrm{II}Z_{i+1}^\mathrm{III})$.} \label{fig:fig_sequence}
\end{figure}

With the atom-array setup illustrated in the main text, the unitary evolution governed by arbitrary combinations of the Hamiltonians $\{H_1,H_{\Z_2^{a}},H_{\Z_2^{ab}},H_{(\Z_2^a\times\Z_2^b)^\pm}\}$ can be realized via Trotterization by four types of elementary quantum gates. Here, we describe in detail the physical implementation of these elementary gates. The corresponding pulse sequence is illustrated in Fig.~\ref{fig:fig_sequence}(a).

First, we note that single-qubit rotations $R_\mathcal{N}(\phi)=\exp(-\mathrm{i}\phi\mathcal{N}/2)$ along different axes $\mathcal{N}=X,Y,Z$ can be straightforwardly realized. The rotation along the $x$- and the $y$-axis can be achieved by a two-photon Raman transition between qubit states, i.e., $({\Omega}\ketbra{0}{1}+\mathrm{H.c.})/2$ with $\phi=|\Omega| t$. The rotation along the $z$-axis can be realized by off-resonantly coupling states $\ket{0},\ket{1}$ to an intermediate state $|e\rangle$ to acquire an additional light shift $\Delta_\mathrm{LS}Z$ with $\phi=\Delta_\mathrm{LS}t$. These single-qubit operations are carried out in the ground-state manifold and can thus be executed in a fast and accurate manner, making them efficient building blocks in the multiqubit gate sequence.

(i) \textit{Single-qubit phase gate}.---The single-qubit phase gate $U_\mathrm{P}(\phi)=e^{-\mathrm{i}\phi Q}$ is equivalent to the single-qubit rotation
$R_{Z}(\phi)=\exp(-\mathrm{i}Z\phi/2)$ up to a global phase factor since $Q=\ketbra{0}{0}=(\mathbb{I}+Z)/2$.

(ii) \textit{Two-qubit controlled phase gate}.---The controlled phase gate $U_\mathrm{CP}(\phi)=e^{-\mathrm{i}\phi Q_1 Q_2}$ between qubits 1 and 2 can be achieved by a modified version of the gate discussed in Ref.~\cite{levine2019parallel}, which only requires global addressing of two atoms. Specifically, we consider two atoms within the blockade radius, described by
\begin{align}
H_\mathrm{Ryd}= \tfrac{\Omega_r}{2}&\big[(\mathbb{I}-\sigma_1^{rr})(e^{i\xi}\sigma_2^{0r}+\mathrm{H.c.}) +\nonumber\\
&+(e^{i\xi}\sigma_1^{0r}+\mathrm{H.c.})(\mathbb{I}-\sigma_2^{rr})\big]-\Delta_r(\sigma_0^{rr}+\sigma_1^{rr}),
\end{align}
where $\sigma_i^{\alpha\beta}=\ketbra{\alpha}{\beta}_i$. The basis state $\ket{11}$ is invariant under $H_\mathrm{Ryd}$, while states $\ket{01}$ and $\ket{10}$ execute Rabi oscillations at frequency $\Omega$. The Rydberg blockade will affect the state $\ket{00}$, which executes a collective Rabi oscillation at an enhanced frequency $\sqrt{2}\Omega$. To ensure that all basis states return to the ground-state manifold, a sudden phase jump $\xi\neq0$ is applied at time $t_1=2\pi/\sqrt{2\Omega_r^2+\Delta_r^2}$, when the collective Rabi oscillation experiences a cycle. A careful choice of $\xi$ will make the single-particle Rabi oscillation follow a symmetric trajectory on the Bloch sphere, i.e., $\ket{00}$ also returns to the initial state at time $t=2t_1$. The full action is then described by $U_\mathrm{CP}^\prime(\phi)=\exp[i\theta (Q_1+Q_2)-i\phi Q_1Q_2]$, which, after single-qubit rotations $R_{Z_2}(\theta)R_{Z_1}(\theta)U_\mathrm{CP}^\prime(\phi)$ is equivalent to the standard form $U_\mathrm{CP}(\phi)$ up to a global phase factor. The strength $\phi$ is adjustable by tuning the ratio $\Omega_r/\Delta_r$. Different from the original gate design of Ref.~\cite{levine2019parallel}, here we only need a small Trotter step $\phi\ll\pi$, which further reduces the time consumption of the gate operation.

(iii) \textit{Three-qubit controlled phase gate}.---To realize the controlled phase gate $U_\mathrm{CCP}(\phi)=e^{-\mathrm{i}\phi Q_1Q_2Q_3}$ between qubits 1, 2, and 3, we consider a chain of three atoms, where only nearest-neighbor atoms are within the blockade radius. Then, one performs the sequential operation: (a) Apply a Rydberg $\pi$ pulse to the edge atoms 1 and 3, transferring $|0\rangle$ to $|r\rangle$; (b) Off-resonantly driving the central atom 2, which experiences a detuned Rabi cycle and acquires a phase $\phi$; (c) Apply a Rydberg $-\pi$ pulse to the edge atoms, transferring the Rydberg population back to $|0\rangle$. Such a sequence realizes an operation $U_\mathrm{CCP}^\prime(\phi)=\exp[-i\phi (\mathbb{I}-Q_1)Q_2(\mathbb{I}-Q_3)]$, because the phase accumulation is possible only when both edge atoms are in the state $\ket{1}$. The operation can be conveniently transformed to $U_\mathrm{CCP}(\phi)=R_{X_3}(\pi)R_{X_1}(\pi)U_\mathrm{CCP}^\prime(\phi)R_{X_1}(\pi)R_{X_3}(\pi)$ by single-qubit rotations.

Noting that the on-site projector $P_i$ in the Hamiltonian $\{H_1,H_{\Z_2^{a}},H_{\Z_2^{ab}},H_{(\Z_2^a\times\Z_2^b)^\pm}\}$ takes a general form $P_i=aQ_i^\mathrm{I}+Q_i^\mathrm{I}(bQ_i^\mathrm{II}+cQ_i^\mathrm{III})+dQ_i^\mathrm{I}Q_i^\mathrm{II}Q_i^\mathrm{III}$, the elementary gates (i-iii) are complete for implementing all possible on-site unitary evolutions.

(iv) \textit{Multi-qubit controlled-$Z$ gate}.---For a system containing $n$ data qubits and one ancillary qubit, the multi-qubit gate operation of interest is
\begin{equation}
\mathrm{CZ}_n=|0\rangle\langle 0|_\mathrm{c}\otimes\mathbb{I}+(-1)^n|1\rangle\langle 1|_\mathrm{c}\otimes\prod_{i=1}^n{Z_i}.
\end{equation}
Such an operation can be easily accomplished under the scenario where the data qubits do not interact with each other but can only be blockaded by the ancillary qubit, e.g., the dual-species plaquette configuration considered in the main text. Specifically, $\mathrm{CZ}_n$ is realized by the sequence: (a) Apply a Rydberg $\pi$ pulse on the ancillary atom, transferring $|0\rangle_\mathrm{c}$ to $|r\rangle_\mathrm{c}$; (b) Apply a $2\pi$ pulse simultaneously on all data atoms, letting $|0\rangle_i$ do a complete Rabi cycle; (c) Apply a Rydberg $-\pi$ pulse on the ancillary atom, transferring $|r\rangle_\mathrm{c}$ back to $\ket{0}_\mathrm{c}$.

The multi-qubit gate (iv) can be used to compose the unitary evolution caused by multi-body interactions between data qubits. For example, the unitary operation $U_n^Z(\phi)=\exp(-\mathrm{i}\phi \prod_{i=1}^nZ_i)$ can be realized by $R_{Y_\mathrm{c}}(-\pi/2)(\mathrm{CZ}_n) R_{X_\mathrm{c}}[(-1)^n2\phi](\mathrm{CZ}_n)R_{Y_\mathrm{c}}(\pi/2)$ with the ancillary qubit prepared in $\ket{0}_\mathrm{c}$. To understand such a decomposition (e.g., $n=4$), one can image the action on an initial state $\ket{\psi_0}=(c_+\ket{+}+c_-\ket{-})$, where $\ket{\pm}$ denotes the eigenstate of the multi-body interaction $B = \prod_{i=1}^4{Z_i}$, i.e., $B\ket{\pm}=\pm\ket{\pm}$. Then, the first rotation $R_{Y_\mathrm{c}}(\pi/2)$ transforms $\ket{\psi_0}\otimes\ket{0}_\mathrm{c}$ into $(c_+\ket{+}+c_-\ket{-})\otimes(\ket{0}_\mathrm{c}+\ket{1}_\mathrm{c})$, which under the action of $\mathrm{CZ}_n$ becomes $c_+\ket{+}\otimes(\ket{0}_\mathrm{c}+\ket{1}_\mathrm{c})+c_-\ket{-}\otimes(\ket{0}_\mathrm{c}-\ket{1}_\mathrm{c})$. Since $(\ket{0}_\mathrm{c}\pm\ket{1}_\mathrm{c})$ are eigenstates of $X_\mathrm{c}$, the rotation $R_{X_\mathrm{c}}(\phi)$ will make $|\pm\rangle$ pick up a phase $e^{\mp i\phi}$, respectively. The rest of the operation will transform the control qubit back to $\ket{0}_\mathrm{c}$, while the data qubits evolve into $c_+e^{i\phi}\ket{+}+c_-e^{-i\phi}\ket{-}=U_4^Z(\phi)|\psi_0\rangle$. The fact that all Pauli operators are transformable in terms of single-qubit rotations, e.g., $X = R_Y(-\pi/2)ZR_Y(\pi/2)$, implies that arbitrary $U_n(\phi)=\exp(-\mathrm{i}\phi \prod_{i=1}^n\mathcal{O}_i)$ ($\mathcal{O}_i=X_i,Y_i,Z_i$) can be achieved. For the array configuration considered in the main text, the above discussed gate manifests as a plaquette evolution $U_\square(\phi)=\exp(-\mathrm{i}\phi \mathcal{O}_i^\mathrm{II}\mathcal{O}_{i+1}^\mathrm{II}\mathcal{O}_{i+1}^\mathrm{III}\mathcal{O}_{i}^\mathrm{III})$ with $\mathcal{O}_j^\alpha\in\{\mathbb{I},X,Y,Z\}_j^\alpha$, which implements each multi-body evolution term in the inter-site evolution $e^{-\mathrm{i}V_{i,i+1}\tau}$, e.g., $U_\square(\phi)=\exp(-\mathrm{i}\phi X_i^\mathrm{II}X_{i+1}^\mathrm{II}Z_{i+1}^\mathrm{III})$ required for the SPT phase [see Fig.~\ref{fig:fig_sequence}(b)].

\subsection{Probing the quantum phase transition}
To probe the quantum phase transition, one should apply a quantum annealing, in which the Trotter sequence should implement the unitary evolution governed by the time-dependent Hamiltonian $H_{(l,m)}[\lambda(t)]=\lambda(t) H_{(l)}+[1-\lambda(t)]H_{(m)}$. For simplicity, we consider a linear ramping with $\lambda(t)=t/T$. To ensure a good adiabaticity, the annealing time $T$ should be comparable to the inverse of the energy gap $\Delta E$ at the critical point. 
\begin{figure}[b]
	\centering
	\includegraphics[width=\linewidth]{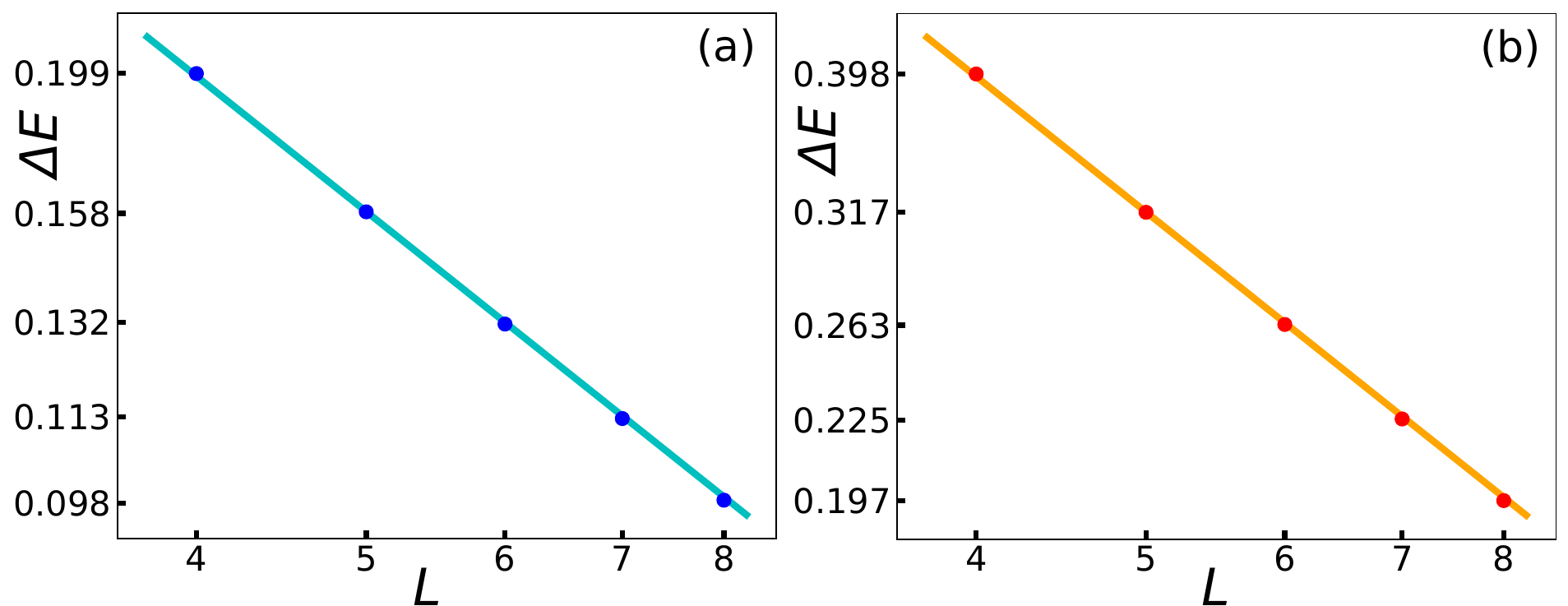}
	\caption{(a) and (b) show the scaling of the energy gap $\Delta E=E_1-E_0$, where $E_0$ is the energy of the ground state and $E_1$ of the first excited state, as a function of the number of lattice sites $L$ (each containing 3 data qubits), plotted in log-log scale. (a) and (b) refer to the $\mathrm{Trivial}\rightarrow \Rep(D_8)/(\Z_2\times\Z_2)$ SSB and and $\mathrm{Trivial}\rightarrow\mathrm{SPT}$ transitions respectively. We rescaled the Hamiltonians at the critical point to remove the overall factor of $1/2$, for consistency with the conventions of \cite{CARDY1986200,oshikawa2019universalfinitesizegapscaling}. The scaling is polynomial, of the form $\Delta E={2\pi \beta}/{L}$, as expected for (1+1)d CFTs, where $\beta$ is compatible with $\beta_\Ising={1}/{8}$ in (a) and with $\beta_{\Ising\times \Ising}={1}/{4}$ in (b).} \label{fig:fig_E1-E0}
\end{figure}

The quantum phase transitions in this work can be described by (1+1)d CFTs, for which the energy gap $\Delta E$ scales  polynomially with system size $L$ \cite{CARDY1986200,oshikawa2019universalfinitesizegapscaling}
\begin{equation}
     \Delta E={2\pi \beta}/{L},
\end{equation}
where $\beta$ is a constant determined by the universality class of the CFT. This relation is confirmed by the numerical results shown in Fig.~\ref{fig:fig_E1-E0}, which considers the phase transitions illustrated in Fig.~\ref{fig:fig_tebd} of the main text. The $\Ising$ igSSB transition, whose gap closing plot is shown in Fig.~\ref{fig:fig_igssb} of the main text, has three degenerate gapless ground states (also called ``universes''): we verified numerically that each one exhibits the same scaling as the $\Ising$ CFT, i.e. the same values as those in Fig.~\ref{fig:fig_E1-E0}(a). As a result, the annealing time $T\sim 1/\Delta E\propto L$ increases linearly with the system size, implying a good scalability of our scheme.

To gain insights into the universal property of the quantum phase transition, one can utilize the Kibble-Zurek mechanism by measuring the growth of the spatial correlations. Specifically, by decreasing the ramping speed $s=1/T$ of the control parameter $\lambda$, the correlation length $\xi$ of the system grows as $\xi=\xi_0(s_0/s)^{\nu/(1+\nu z)}$, where $\nu$ and $z$ are critical exponents of the system. Such a universal behavior of the quantum correlation allows for an efficient extraction of the critical exponents by data collapse at different ramping speeds \cite{keeslingQuantumKibbleZurek2019b}. 

\begin{figure}
	\centering
	\includegraphics[width= \linewidth]{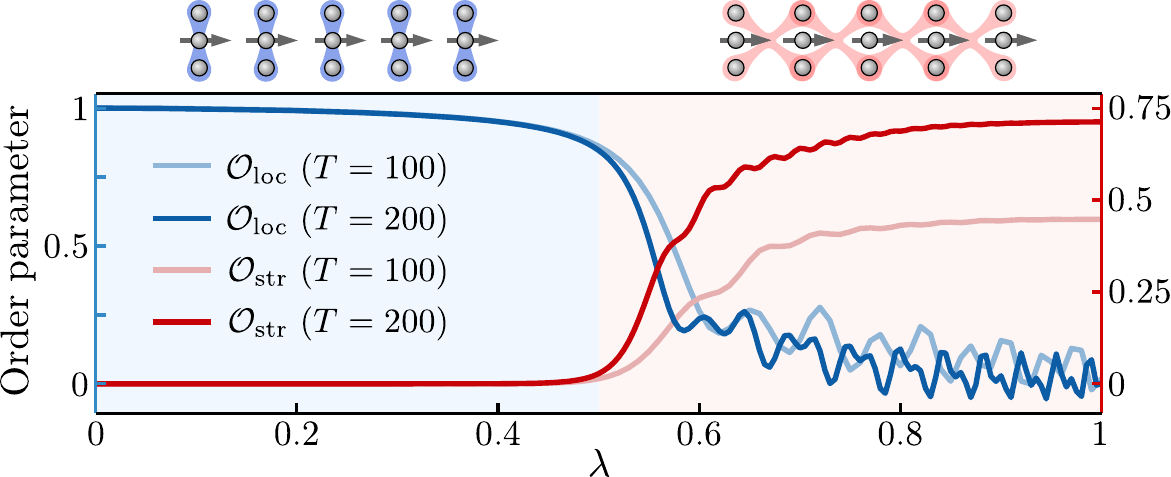}% Here is how to import EPS art
	\caption{Probing the transition from the $\mathsf{Rep}(D_8)/(\mathbb{Z}_2\times \mathbb{Z}_2)$ SSB phase to the SPT phase. The parameters are the same as in Fig.~2(a) of the main text ($54$ data qubits and $\tau=1$).} \label{fig:fig_transi}
\end{figure}
In the main text, we illustrate our scheme for the transition from the trivial phase to the nontrivial ones. However, it is also possible to probe the transition between the two nontrivial phases, for which one needs to prepare the ground state of one of the nontrivial phases, e.g., by a first annealing, and then apply the second annealing to study the phase transition of interest. For certain nontrivial phases, their ground states can be prepared more efficiently by local quantum gates, such as the $\mathsf{Rep}(D_8)/(\mathbb{Z}_2\times \mathbb{Z}_2)$ SSB phase, whose ground states are locally entangled Bell pairs. Figure~\ref{fig:fig_transi} illustrates the transition from such an SSB phase to the SPT phase, characterized by the decay of the local order parameter $\mathcal{O}_\mathrm{loc}$ and the growth of the string order parameter $\mathcal{O}_\mathrm{str}$.

In a realistic Rydberg system, experimental imperfections bring in unwanted evolutions for each Trotter step, causing extensive errors in the final state fidelity. However, many observables of interest, such as the order parameters and the correlation functions, remain stable in the presence of errors \cite{Trivedi:2022lsi}. To demonstrate the robustness of our scheme, we consider here a simplified error model with coherent and incoherent bit (phase)-flip errors, while a more accurate error model can be developed in the same manner by considering dominant error sources in a specific experimental setup \cite{de2018analysis}.

\begin{figure}[b]
	\centering
	\includegraphics[width= \linewidth]{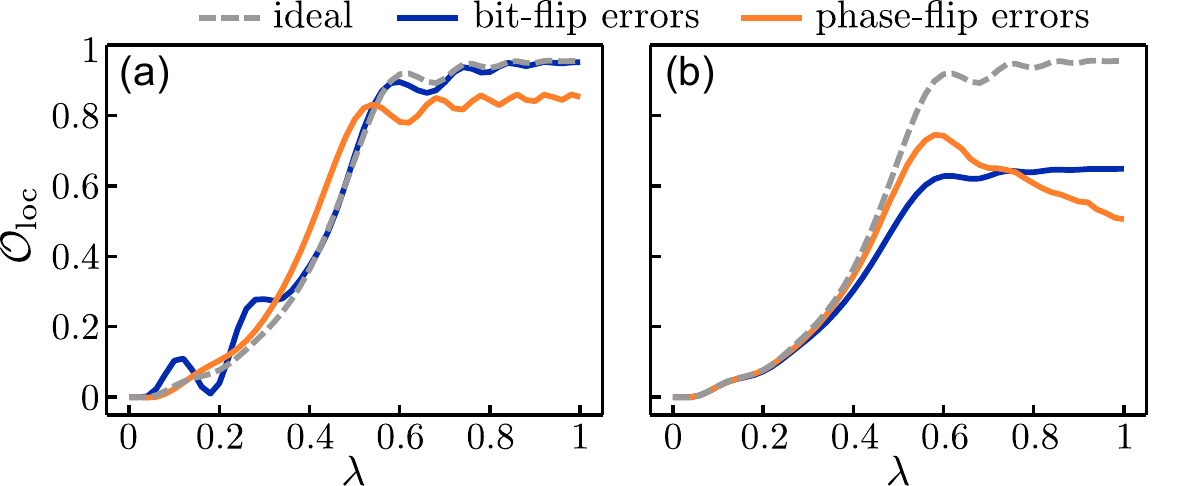}% Here is how to import EPS art
	\caption{Quantum annealing dynamics under coherent (a) and incoherent (b) gate errors. The considered phase transition is from the trivial phase to the $\mathsf{Rep}(D_8)/(\mathbb{Z}_2\times \mathbb{Z}_2)$ SSB phase for 18 data qubits with $\tau=1$ and $T=50$.} \label{fig:fig_error}
\end{figure}

To describe the coherent errors, we introduce an additional spin rotation after each Trotter step:
\begin{equation}
U_\mathcal{K}=\exp[-\mathrm{i}\sum_{j=1}^{L}\sqrt{\epsilon/2}(\mathcal{K}_j^\mathrm{II}+\mathcal{K}_j^\mathrm{III})],   
\end{equation}
where $\mathcal{K}=X$ and $\mathcal{K}=Z$ represent bit-flip and phase-flip errors, respectively. For the incoherent errors, we assume that the density matrix $\rho$ of the system is mapped to 
\begin{equation}
\mathcal{E}_\mathcal{K}(\rho)=(1-\epsilon L)\rho+(\epsilon/2)\sum_{j=1}^L (\mathcal{K}_j^\mathrm{II}\rho\mathcal{K}_j^\mathrm{II}+\mathcal{K}_j^\mathrm{III}\rho\mathcal{K}_j^\mathrm{III}), 
\end{equation}
and simulate the dynamics by sampling 100 trajectories. In current Rydberg array setups, the fidelity of the entangling gate has reached a high value $>99.5\%$ \cite{evered2023high}, so we assume a reasonable error rate $\epsilon=0.5\%$ and study the phase transition: $\mathrm{Trivial}\rightarrow \Rep(D_8)/(\Z_2\times\Z_2)$ in an intermediately large system size containing 18 data qubits. As shown in Fig.~\ref{fig:fig_error}, we find that the key signature of the spontaneous symmetry breaking, i.e., establishment of the local order parameter, is preserved in the presence of both types of errors, while a better tolerance to coherent errors is observed. We further notice that, by applying erasure conversions \cite{ma2023high},
it is possible to mitigate certain types of errors and improve the performance of the simulation.

%%%%%%%%%%%%%%%%%%%%%%%%%%%%%%%%%%%%%%%%%%%%%%%%%

\end{document}